\documentclass{elsart}

\usepackage{graphicx}

\usepackage{amssymb}

\begin{document}

\begin{frontmatter}

\title{Three-Phase Traffic Theory and Highway Capacity}

\author{Boris S.~Kerner
}

\address{Daimler Chrysler AG, RIC/TS, T729, 70546 Stuttgart, Germany \\
}

\maketitle              

\begin{abstract} 
Hypotheses and some results of the three-phase
traffic theory by the author are compared with results of the fundamental diagram approach
to traffic flow theory. A critical
discussion of model results about congested pattern features
which have been derived within the
fundamental diagram approach to traffic flow theory and modelling is made.
The empirical basis of the three-phase traffic theory is discussed and some new
spatial-temporal features of the traffic phase "synchronized flow" are considered. 
A probabilistic theory of highway capacity is presented which
 is based on the three-phase
traffic theory. In the frame of this theory, the probabilistic nature of highway capacity in free flow is linked to
 an occurrence of 
the first order local phase transition from the traffic phase
"free flow" to the traffic phase "synchronized flow". 
A  numerical study of congested pattern highway capacity based on simulations of
a  KKW cellular automata model  within the three-phase traffic theory
is presented.
A congested pattern highway capacity which
depends on features of congested spatial-temporal
patterns upstream of a bottleneck is  studied.
\end{abstract}
\end{frontmatter}

\section{Introduction}

Real traffic is 
a dynamical process which occurs both in space and  time.
This spatial-temporal process shows  very complex 
dynamical behavior. In particular, highway traffic  can be either free or congested.
Congested traffic states can be defined as the traffic states where
the average vehicle speed is lower than the minimum possible average
speed in free flow (e.g.,~\cite{Koshi}).  It is well known that in
contrast to free traffic flow, in congested traffic a collective
behavior of vehicles plays an important role (the collective flow by
Prigogine and Herman~\cite{Prigogine}), and a synchronization of
vehicle speeds across different highway lanes usually
occurs~\cite{Koshi}. In congested traffic, complex spatial-temporal
patterns are observed, in particular a sequence of moving traffic
jams, the so called "stop-and-go" phenomenon (e.g., the classical
works by Treiterer~\cite{Tr} and Koshi {\it et al.}~\cite{Koshi}).
Recall that {\it a moving jam} is a  moving localized structure. The moving jam
is spatially restricted by two upstream moving jam fronts where the vehicle speed 
and the density change
sharply. The vehicle speed is low (as low as zero) and the density is high inside 
the moving jam.

 Congested traffic  usually occurs at a  highway bottleneck, e.g., at the bottleneck due to an on-ramp.
In empirical investigations, the onset  of congested traffic is 
accompanied by the breakdown
phenomenon, i.e., by a sharp decrease
in the vehicle speed at the 
bottleneck (see e.g., papers by 
Athol and Bullen~\cite{Athol}, 
Banks~\cite{Banks1990,Banks1991}, 
 Hall {\it et al.}~\cite{Hall1991,Hall1992},  Elefteriadou {\it et al.}~\cite{El},
 Kerner and Rehborn~\cite{KR1997}
and by Persaud {\it et al.}~\cite{Persaud}).
It
has been found  that the breakdown phenomenon
has a probabilistic nature, i.e., the probability of the speed breakdown is an increasing
function of the flow rate in free flow at the bottleneck~\cite{El,Persaud}. 
Besides, it has been found  that the capacity of congested bottleneck, i.e., highway capacity 
after the breakdown phenomenon
at the bottleneck  has occurred
is usually lower than the capacity in free flow before - the so called phenomenon
"capacity drop"~\cite{Banks1990,Banks1991,Hall1991,Hall1992}.

Concerning the important role of highway bottlenecks it should be
noted that although congested traffic can occur away from
bottlenecks~\cite{Kerner1999A}, they have an important impact just like
 defects in physical systems which can play an important role for the
phase transitions and for the formation of spatial-temporal
patterns. The role of the bottlenecks in traffic flow is as follows:
Congested traffic occurs most frequently at highway bottlenecks 
(e.g.,~\cite{Da,May}). The bottlenecks can  result from
for example due to road works, on and off ramps, a decrease in the
number of highway lanes, road curves and road gradients.

Although the complexity of traffic is linked to the occurrence of
spatial-temporal patterns, some of the traffic features can be
understood if {\it average} traffic characteristics are
considered. Thus, important empirical methods in traffic
science are empirical flow-density and speed-density relationships
which are related to measurements of some {\it average} traffic
variables at a highway location, in particular at a highway bottleneck.
The empirical relationship of the {\it average} vehicle speed on the
vehicle density must be related to an obvious result observed in real
traffic flow: the higher the vehicle density, the lower the average
vehicle speed.  When the flow rate, which is the product of the vehicle
density and the average vehicle speed, is plotted as a function of the
vehicle density one gets what is known as the empirical fundamental
diagram.
It must be noted that  the empirical fundamental
diagram is successfully used for different important applications
 where some average
traffic flow characteristics should be determined
(e.g.,~\cite{May,Manual}).

Over the past 80 years scientists have developed a wide range of different mathematical models 
of traffic flow to understand these complex non-linear traffic phenomena (see the books by Leuzbach~\cite{Leu},
by May~\cite{May}, by Daganzo~\cite{Da}, by Prigogine and Herman~\cite{Prigogine},
 by Wiedemann~\cite{Wid}, by Whitham~\cite{Wh2}, by Cremer~\cite{Cremer}, by Newell~\cite{New},
 the 
reviews by Chowdhury {\it et al.}~\cite{Sch},  Helbing~\cite{Helbing2001},
Nagatani~\cite{Nagatani_R}, Nagel {\it et al.}~\cite{Nagel2003A} and the conference
 proceedings~\cite{Lesort,Ceder,Taylor,SW1,SW2,SW3,SW4}). 
Clearly these models must be based on the real behavior 
of drivers in traffic, and their solutions should show phenomena observed in real traffic.

\subsection{Fundamental Diagram Approach to Traffic Flow Theory and Modelling
\label{FD_Sunsec}}

Up to now by a development of a mathematical traffic flow model which should
explain empirical spatial-temporal congested patterns, it has been self-evident that 
hypothetical {\it steady} state solutions
of the  model
should belong to a curve in the
flow-density plane (see,
e.g.,~\cite{LW,GH,New2,Prigogine1961,Prigogine,SW1,SW2,SW3,NS,Nagel1995,Schreckenberg,Wh,KK1994,B1995A,B1995B,Su,Herrmann,Kra,Bar,Nagatani,Nagatani2,Hel,Trei,Mahnke,Wh2,Cremer,Leu,KKK1997,KKS1995,Lee1998,Helbing1998,Helbing1999,Lee1999,Lee2000A,Lee2000B,Helbing2000,Treiber,Tomer2000,Knospe,Knospe2001}
and the recent reviews~\cite{Sch,Helbing2001,Nagatani_R,Nagel2003A}). 
The above term {\it steady state} designates
 the hypothetical model solution where
vehicles move at the same distances to one  another with the same
time-independent vehicle speed. Therefore, steady states are hypothetical
spatially homogeneous and time-independent
traffic states (steady states are
also often called  "homogeneous" or
"equilibrium" model solutions; we will use in the article for these 
hypothetical model traffic states
the term "steady" states or "steady speed" states). 
The curve in the flow-density plane for steady  state
model solutions
 goes
through the origin and has at least one maximum. This curve
 is called {\it the 
fundamental diagram} for traffic flow. 

The postulate about the
fundamental diagram
underlies almost all 
traffic flow modeling approaches  up to now 
(see reviews~\cite{Sch,Helbing2001,Nagatani_R,Nagel2003A}) in the sense
that the models are constructed such that in the {\it unperturbed, noiseless limit}
they have a fundamental diagram of steady states, i.e., the
steady states form a curve in the flow-density plane. The fundamental diagram is
either a result of the model (e.g., for the models by Gazis, Herman, Rothery~\cite{GH}, Gipps~\cite{Gipps},
Nagel-Schreckenberg 
cellular automata (CA for short)~\cite{NS,Nagel1995,Schreckenberg,Bar,Knospe},  Krau{\ss} {\it et al.}~\cite{Kra}, 
Helbing, Treiber and co-workers~\cite{Trei,Helbing2000,Helbing2002},
Tomer, Halvin and co-worker~\cite{Tomer2000,Tomer2002}) or the fundamental diagram
is hypothesized in a model (e.g.,  the optimal velocity models by Newell~\cite{New2},
Whitham~\cite{Wh}, Bando, Sugiyama and co-workers~\cite{B1995A,B1995B} and the macroscopic model
by Payne~\cite{Pa}).
Moving jams which are calculated from these models for a
homogeneous road (i.e., a road without bottlenecks)
are due to the instability of steady states of the fundamental
diagram within some range of vehicle densities
 (see reviews~\cite{Sch,Helbing2001,Nagatani_R,Nagel2003A}). This is 
one of the reasons why we
find it helpful to classify these models as belonging to what 
we call the "fundamental diagram approach''.

In 1955 
Lighthill and Whitham~\cite{LW} wrote in their classical work
(see page 319 in~\cite{LW}):
"... The fundamental hypothesis of the theory is that at any point of the road the flow  (vehicles per hour) 
is a function of the concentration  (vehicles per mile)...".
Apparently the empirical fundamental diagram was the reason that the
fundamental diagram approach has already been introduced in the first
traffic flow models derived by Lighthill and Whitham~\cite{LW}, by
Gasis, Herman and Rothery~\cite{GH}, and by Newell~\cite{New2}.

Concerning the  theoretical
fundamental diagram, it must be noted
that in real congested traffic 
complex spatial-temporal traffic patterns are observed 
(e.g.,~\cite{Koshi,Tr}).
These patterns are  spatially non-homogeneous.
This spatial behavior of congested patterns is a complex function of time.
An averaging of traffic variables related to congested patterns
over long enough time intervals gives a relation between different {\it averaged} vehicle speeds and densities. 
Thus, the empirical fundamental diagram 
is related to  
{\it averaged} characteristics of  spatial-temporal congested  patterns measured at a highway location
rather than to features of the hypothetical steady states of congested traffic
on the theoretical fundamental diagram.
  For this reason, the existence of the theoretical fundamental diagram is only  {\it a hypothesis}.

It has recently been found
that   empirical features of the phase transitions in traffic flow and
most of empirical
spatial-temporal pattern
features~\cite{Kerner1998B,Kerner2000C,Kerner2002B} are qualitatively
different from those which follow from mathematical traffic flow
models in the fundamental diagram
approach which is considered in the 
reviews~\cite{Sch,Helbing2001,Nagatani_R,Nagel2003A}.

\subsection{Three-Phase Traffic Theory}

For this reason in 1996-2000 the author, based on an empirical traffic
flow analysis, introduced a concept called "synchronized flow" and
the related three-phase traffic
theory~\cite{Kerner1999A,Kerner1998B,Kerner2000C,KR1996B,Kerner1998C,Kerner2000A,Kerner1999B,Kerner2000B,Kerner2001A,Kerner1998A,Kerner1999C}.

\subsubsection{The Concept "Synchronized Flow" \label{Concept_Objective}}

In the concept "synchronized flow", 
there are two qualitatively different phases, the traffic phase
called "synchronized flow" and the traffic phase called "wide moving
jam", which should be distinguished in congested 
traffic~\cite{Kerner1998B,Kerner2000C,KR1996B,Kerner1998C,Kerner2000A,Kerner1999B}. This
distinguishing is based on qualitatively different {\it empirical
spatial-temporal} features of these phases. Traffic consists of free flow and congested traffic.
Congested traffic consists of two traffic phases.
Thus, there are three traffic phases:
\begin{enumerate}
\item  Free flow. 
\item  Synchronized flow. 
\item  Wide moving jam.
\end{enumerate}
In the three-phase traffic theory, 
features of spatial-temporal congested patterns are explained
based on the phase transitions between these three traffic phases.

Objective criteria to distinguish between the traffic phase
 "synchronized flow" and the traffic phase
 "wide moving jam"  
are based on qualitative different {\it empirical spatial-temporal} features
of these two traffic phases. 
A wide moving jam is a moving jam which possesses the following characteristic feature. Let us consider
the 
downstream front of the wide moving jam where vehicles 
accelerate escaping from a standstill inside  the wide moving
jam. This downstream jam front propagates 
on a highway
{\it keeping} the mean velocity of this front.
As long as a moving jam is a wide moving jam this 
characteristic effect -
the keeping of the velocity of the downstream jam front -
 remains even if the wide moving jam propagates through any complex traffic states 
and through any highway  bottlenecks.
In contrast, the downstream front of the  traffic phase "synchronized flow" 
(where vehicles accelerate escaping from synchronized flow to free flow)
is usually fixed at the bottleneck. Corresponding to the definition of the
traffic phase "wide moving jam" and to the concept "synchronized flow" where
there are only two traffic phases in congested traffic,
any state of congested traffic which does not possess
the above characteristic feature of a wide moving jam is related to the traffic state "synchronized flow".
A more detailed consideration of the objective criteria of
traffic phases and empirical examples of the application of these objective criteria for the
determination of the phase "synchronized flow" and the phase "wide
moving jam" in congested traffic can be found in~\cite{Kerner2002B}.

It should be stressed that the concept "synchronized flow" and the
related methodology of the congested pattern study, which has been
used for the definition of the three traffic phases below, is based on
an analysis of {\it empirical spatial-temporal} features of congested
patterns~\cite{Kerner1998B,KR1996B,Kerner1998C} 
rather than on a dynamical analysis of data (e.g., in the
flow-density plane) which is measured at only one highway
location. First, a
spatial-temporal study of traffic must be made. Only after the
traffic phases "synchronized flow" and "wide moving jam" have already
been distinguished, based on this spatial-temporal data analysis, some
of the pattern features can further be studied in the flow-density
plane. In particular, this procedure has already been used in~\cite{KR1996B}:
At the first step, a spatial-temporal analysis of empirical data has
been made and the phases "synchronized flow" and "wide moving jam"
have been identified. At the next step, the traffic phase
"synchronized flow" has been plotted in the flow-density plane without
any wide moving jams and some of the features of "synchronized flow" have
 been studied.  Measured
data on sections
of the highway 
A5 in Germany
which have been used 
in~\cite{KR1996B,Kerner1998C,Kerner1998A,Kerner1998B,Kerner1999B,Kerner1999A,Kerner1999C,Kerner2000A,Kerner2000B,Kerner2000C,Kerner2001A,Kerner2002B} 
also comprise the information
about vehicle types (number of vehicles and long vehicles)
and individual vehicle speeds passing
the detector during each one minute interval of averaging.
Using the latter information
in addition to the spatial-temporal data analysis of one minute averaged data,
the determination of the type of  synchronized flow in empirical studies of 
synchronized flow 
(the type (i), or (ii) or else (iii)~\cite{KR1996B}) have been made. Besides, the individual vehicle speeds 
allow us to
answer the question of whether there are  
narrow moving jams in synchronized flow or  not. 
All these steps of data analysis
have been made 
in~\cite{KR1996B,Kerner1998C,Kerner1998A,Kerner1998B,Kerner1999B,Kerner1999A,Kerner1999C,Kerner2000A,Kerner2000B,Kerner2000C,Kerner2001A,Kerner2002B}.
We will  illustrate this in Sect.~\ref{concept} where an empirical example of a synchronized flow pattern
will be studied.

\subsubsection{The Fundamental Hypothesis of The Three-Phase Traffic Theory
\label{Fun_Hyp} }

Another claim of the concept of
"synchronized flow" is the hypothesis about steady states of synchronized flow.
This is the fundamental hypothesis of the three-phase traffic theory.
The fundamental  hypothesis of the three-phase traffic 
theory reads as follows~\cite{Kerner1998B,Kerner1998C,Kerner2000A,Kerner1999B}:

{\it Hypothetical steady  states of synchronized flow cover a
two-dimensional region in the flow-density plane} (Fig.~\ref{Diagram}). This means that in
these hypothetical steady  states of synchronized flow, where all
vehicles move at the same distance to one  another and with the same
time-independent speed, a given steady vehicle speed is related to an
infinite multitude of different vehicle densities and a given vehicle
density is related to an infinite multitude of different steady vehicle
speeds.  This hypothesis means that there is {\it no} fundamental
diagram for hypothetical steady speed states of synchronized flow.
This hypothesis has recently been used in a microscopic
three-phase traffic flow theory~\cite{KKl,KKW,KKl2003A}. It occurs
that this theory, which is also based on other
hypotheses of the three-phase traffic theory (Sect.~\ref{Com_Sec}),
explains and predicts main features of empirical phase
transitions and spatial-temporal congested patterns found in~\cite{Kerner1998B,Kerner2002B}.

The fundamental hypothesis of the three-phase traffic theory
is therefore in contradiction with the 
hypothesis about the existence of the fundamental diagram for
hypothetical steady  states of mathematical models and theories
in the fundamental diagram approach. An explanation of the fundamental hypothesis of
the three-phase traffic theory
will be done
in Sect.~\ref{Exp_H_Sec}.

\subsubsection{Explanation of  The Terms "Synchronized Flow"
and "Wide Moving Jam"}

The 
 term "synchronized flow" 
 should reflect the
following  features of this traffic phase: (i) It is a non-interrupted traffic flow
rather than a long enough standstill 
as it usually  occurs  inside a wide moving jam. The word "flow" should reflect this feature. 
(ii) There is a {\it tendency} to a synchronization of vehicle speeds across different lanes on a multi-lane
road in this flow. Besides, there is 
a {\it tendency} to
a synchronization of vehicle
speeds on each of the road lanes (a bunching of the vehicles)
in synchronized flow
due to a relatively low mean probability of passing in synchronized flow. 
The word "synchronized"  should reflect these speed synchronization effects.

The term "wide moving jam" should reflect the characteristic feature of the jam 
to propagate through any other states of traffic flow and through any bottlenecks keeping
the velocity of the downstream jam front.
The word combination  "moving jam" should reflect the feature of the jam {\it propagation} as a whole 
localized structure on a road. 
If the width of a moving jam is considerably higher than the widths of the jam fronts
and  the speed inside the jam is zero then the moving jam possesses this characteristic feature.
The word "wide" (the jam width in the longitudinal direction) should reflect this characteristic feature 
of the jam propagation {\it keeping} the velocity of the downstream jam front.

This article is organized as  follows. In Sect.~\ref{concept},
firstly an overview of known features of synchronized flow is made (Sect.~\ref{Syn_Overview})
 and then new empirical results about spatial-temporal
features of synchronized flow are presented. These empirical results should help to understand some of 
the hypotheses to the three-phase traffic theory. 
A critical analysis of the application of the
 fundamental diagram approach for a description of phase transitions and of
spatial-temporal  features of congested patterns will be made in Sect.~\ref{Discussion}.
In Sect.~\ref{Com_Sec},
a comparison of already known hypotheses to the author's three-phase
traffic 
theory
with results of the
fundamental diagram approach to traffic flow theory will be considered. 
In this section, we will also consider some new hypotheses about a Z- and 
{\it double} Z-shaped characteristics of traffic flow. The double Z-characteristic should explain
phase transitions which are responsible for the
wide moving jam emergence in real traffic flow.
Other new results are presented in Sect.~\ref{Cap_Sec} where
a probabilistic theory of highway capacity which is based on
 the three-phase traffic theory is considered. This general theory 
will be illustrated and confirmed by  new numerical results of a study of
a KKW cellular automata model within
the three-phase traffic theory.

\section{Empirical Features of Synchronized Flow
\label{concept} }

In this section some new empirical features of synchronized flow will be considered. These features 
will be used below for an explanation of the three-phase traffic theory. In particular, these results allow us
to give the  empirical basis for the hypothesis about a Z-shape of the
probability of passing in traffic flow as a function of the density.
However, firstly a brief overview
of empirical features of synchronized flow is made.

\subsection{Main Empirical Features of Phase Transitions, Onset of Congestion 
and Congested Patterns at Bottlenecks (Overview) \label{Syn_Overview}}

Moving jams do not emerge in free flow, if synchronized flow is not
hindered~\cite{Kerner2000C}. Instead, the moving jams emerge due to a
sequence of two first order phase transitions~\cite{Kerner1998B}:
First the transition from free flow to synchronized flow occurs (it
is called the F$\rightarrow$S transition) and only later and
usually at a different highway location wide moving jams emerge in the synchronized
flow (the latter transition is called the
S$\rightarrow$J transition and the sequence of both transitions is called the
F$\rightarrow$S$\rightarrow$J transitions).

In particular, the onset of congestion at a bottleneck, i.e.,
 the well-known breakdown phenomenon in free flow is
linked to  the F$\rightarrow$S transition in an initial free flow
rather than to the wide moving jam emergence (the F$\rightarrow$J transition)~\cite{Kerner1998B,Kerner2002B}.
This means that
at the same density in free flow at the bottleneck
 the probability of the  F$\rightarrow$S transition should be considerably higher than 
the probability of the F$\rightarrow$J transition.

Empirical investigations show~\cite{Kerner2002B,Kerner2002}
that there are two main types of congested patterns at an
isolated bottleneck\footnote{Note that an isolated bottleneck is a highway bottleneck which
 is far enough away from other
bottlenecks where congested patterns can emerge.}:
{\it The general pattern} or {\it GP} for short: The GP is the
congested pattern at the isolated bottleneck where synchronized flow
occurs upstream of the bottleneck and wide moving jams spontaneously
emerge in that synchronized flow. Thus the GP consists of both traffic
phases in congested traffic: "synchronized flow" and "wide moving
jam".  
{\it The synchronized flow pattern} or {\it SP} for short: The
SP consists of synchronized flow upstream of the isolated bottleneck
{\it only}, i.e., {\it no} wide moving jams emerge in that
synchronized flow.

However, dependent on the bottleneck features and on traffic demand,
the GP and the SP show a diverse variety of special cases.
In particular, there are three main different types of 
synchronized flow patterns (SP) at the isolated bottleneck:
The localized SP (LSP), i.e., the SP whose width is spatially limited 
over time\footnote{LSP 
 can also be called a "shortening" SP. 
In the article, following the empirical congested pattern study~\cite{Kerner2002B} we will use the term LSP.}, 
the widening SP (WSP), i.e., the SP whose width is continuously widening over time and
the moving SP (MSP), i.e., the SP which propagates
as a whole localized pattern 
on the road. 

The usual scenario of the GP emergence is the following~\cite{Kerner1998B,Kerner2002B}:
First, the F$\rightarrow$S transition occurs at a highway bottleneck. The downstream front of the
synchronized flow (the front where drivers escape from the synchronized flow upstream to 
the free flow downstream) is fixed at the bottleneck. The upstream front of the synchronized flow 
 (this front separates free flow upstream from the synchronized flow downstream)
propagates upstream. Later in this synchronized flow upstream of the bottleneck narrow moving jams
spontaneously emerge. Narrow moving jams propagate upstream and 
self-grow\footnote{A narrow moving jam is a moving jam which  does not
possess the characteristic feature of a wide moving jam to keep the downstream jam front velocity
propagating through any state of traffic and through any bottlenecks.
Narrow moving jams belong to congested traffic. Corresponding to
the objective criteria of
traffic phases in congested traffic any state of congested traffic which does not possess
this characteristic feature of a wide moving jam is related to the traffic state "synchronized flow".
Thus, narrow moving jams are states of the traffic phase
 "synchronized flow"~\cite{Kerner2002B}. \label{fn_n_jam}}.
Finally, the growing narrow moving jams (or only a part of them) transform
into wide moving jams. This transformation is related to the S$\rightarrow$J transition.

It must be noted that the well-known and very old  term 
$\lq\lq$stop-and-go$\lq\lq$ traffic which is related to
a sequence of moving traffic
jams   will not be
used
in this article. This is linked to the following:
For a traffic observer,  both a sequence of narrow moving jams and a sequence of wide moving jams is
 $\lq\lq$stop-and-go$\lq\lq$ traffic. However, as it has already  been mentioned
(see  footnote~\ref{fn_n_jam}) narrow moving jams 
belong to the traffic phase
$\lq\lq$synchronized flow$\lq\lq$ whereas wide moving jams belong to the qualitatively different
traffic phase $\lq\lq$wide moving jam$\lq\lq$.

It has been found 
that moving jams are most to emerge in dense synchronized flow of lower
average vehicle speed~\cite{Kerner1998B,Kerner2002B}. In other words, the
average speed in synchronized flow of a GP where a
moving jam emerges is usually considerably lower than the speed
in synchronized flow of a SP. The dense synchronized flow in GP
appears due to the pinch effect in the synchronized flow,
i.e., due to a strong compression of the synchronized flow upstream of the bottleneck.
Besides,
the lower the average speed in the pinch region of  synchronized flow in a GP is the higher  the frequency of 
the moving jam emergence in this synchronized flow.  In particular, the average speed 
in the pinch region of synchronized flow is decreasing upstream 
of an on-ramp if the flow rate to the on-ramp is increasing. Consequently the frequency of the moving jam
emergence increases when the flow rate to the on-ramp  increases. This is because 
the average speed in synchronized flow upstream
of the on-ramp decreases.
Even at the highest observed flow rate to the on-ramp moving jams emerge
(with the highest frequency) in synchronized flow upstream of the on-ramp.

In contrast, empirical investigations of synchronized flow allow us to suggest
that in synchronized flow of higher vehicle speed moving jams do not necessarily
emerge~\cite{Kerner1998B,Kerner2002B}. 
In this case, a
SP  can occur where no wide moving jams emerge~\cite{Kerner2002B,KKl}.
 One of the SPs is a
 WSP. 
 The WSP should occur when a bottleneck introduces
a relatively small disturbance for traffic flow.

This
case which can be observed at a bottleneck due to off-ramps will be considered below.
It will be shown, that in this case
synchronized flow in a WSP can indeed exist upstream of the bottleneck on a long
stretch of the highway (about 4.5 km) during a long time (more than 60
min) without wide moving jam emergence in that synchronized flow.
This confirms the above suggestion made  that 
in synchronized flow of higher vehicle speed moving jams do not necessarily
emerge.

The GP and the SP which have briefly been discussed above appear at an isolated bottleneck.
As a result, an influence of the other bottlenecks  on the pattern formation at the isolated bottleneck
and on the pattern features should be negligible.

On real highways there are a lot of bottlenecks where different
congested patterns almost simultaneously can emerge.
If two (or more) such bottlenecks exist, then an expanded congested pattern (EP) can be formed.
In the EP, synchronized flow covers at least two bottlenecks.
For example, let us consider two bottlenecks which are close to one another.
The bottleneck downstream will be called "downstream bottleneck" and the bottleneck upstream 
will be called "upstream bottleneck". If the F$\rightarrow$S transition occurs at the
downstream  bottleneck and later narrow moving jams emerge in the synchronized flow, then
narrow moving jams can reach the upstream bottleneck before they have transformed
into wide moving jams. In this case, the synchronized flow propagates upstream of the
upstream bottleneck covering  both bottlenecks: An EP appears.

Another empirical example which will be considered below is the following:
If a WSP occurs at the downstream bottleneck,
then due to the continuous upstream widening of  synchronized flow in the WSP
 the upstream front of the synchronized flow  always at some time reaches the upstream
bottlneck. After the synchronized flow is upstream of the upstream bottleneck, the 
synchronized flow covers
both bottlenecks, i.e., the initial WSP transforms into an 
EP. From this example we may  conclude
that the congested pattern at
the downstream bottleneck can be considered as a WSP in only a finite time interval as long as
the synchronized flow of the WSP does not reach the upstream bottleneck.
After the upstream front of synchronized flow of the WSP has reached the upstream bottleneck, 
an EP appears. Synchronized flow 
of this EP consists of 
the synchronized flow of the initial WSP and the synchronized flow at the upstream bottleneck.
The upstream bottleneck may make a great influence on the synchronized flow of the
initial WSP, e.g., this bottleneck can lead to the wide moving jam formation in the synchronized flow. This occurs
in the example of this WSP.

There can be a lot of different types of EPs. As it has been shown in~\cite{Kerner2002B}, 
 EPs can be explained within the three-phase traffic theory.
This conclusion is linked to the empirical fact~\cite{Kerner2002B}
that  congested traffic of all known EP types consists of 
the traffic phase "wide moving jam" and the traffic phase "synchronized flow" only. 
This result of the empirical study~\cite{Kerner2002B} is illustrated
in Fig.~\ref{Expanded} where an  example of an EP is shown~\cite{FOTO2003A}.
The synchronized flow inside the EP covers at least four bottlenecks
(the locations of the bottlenecks are marked $B_{1}$, $B_{2}$, $B_{3}$ and $B_{1}$
in Fig.~\ref{Expanded} (b)).
We can see in Fig.~\ref{Expanded} (b) that congested traffic in the EP consists of 
the traffic phase "wide moving jam" and the traffic phase "synchronized flow".
The empirical results are
 also confirmed by a microscopic three-phase traffic theory of EPs
which has recently been developed in~\cite{KKl2003A}.

\subsection{Empirical Example of WSP \label{Wide_Sec} }

The empirical study made
in~\cite{Kerner2002B} allows us to assume that WSP can
occur at off-ramps. The importance of the WSP analysis is linked to
the possibility to show empirical features of synchronized flow of a
relatively high vehicle speed. 

An example of a WSP which occurs upstream of the off-ramp D25-off on the
section of the highway A5-North in Germany (the State "Hessen") is shown in
Figs.~\ref{Wide1},~\ref{Wide2},~\ref{Wide3} and
~\ref{Wide4}. The section of the highway A5-North
(Fig.~\ref{Wide1} (a)) has already been described
in~\cite{Kerner2002B} (see Fig. 3 (c) in this paper). A WSP occurs as a
result of the F$\rightarrow$S transition upstream of the off-ramp
(also discussed in~\cite{Kerner2002B}). It can be seen in
Figs.~\ref{Wide1} (b) and~\ref{Wide2} that the vehicle speeds slowly
decrease within a WSP in the upstream direction whereas the flow rate
does not change considerably when a WSP occurs. This is a peculiarity of
synchronized flow.

When the vehicle speed in a WSP decreases in the upstream direction,
some narrow moving jams emerge in this synchronized flow of low
vehicle speed (D17 and D16, Fig.~\ref{Wide2}). However, D16 is already
at the on-ramp (D15-on). For this reason, the synchronized flow
propagating upstream, covers this upstream bottleneck at the
on-ramp. As a result, an expanded congested pattern (EP)
occurs~\cite{Kerner2002B} (this is not
shown in Fig.~\ref{Wide2}). Thus, WSP upstream of the off-ramp at D25-off
and downstream of the on-ramp at D16 is only a part of this EP. 
 Nevertheless, the consideration of this WSP
allows us to come to some important conclusions about  features
of the traffic phase "synchronized flow".

\subsection{Overlapping of States of Free Flow and Synchronized Flow in Density}

In particular, when free flow (black quadrates in Fig.~\ref{Wide3})
and synchronized flow inside WSP (circles in Fig.~\ref{Wide3}) are
shown in the flow-density-plane it can be seen, that at least at
detectors D20-D18 states of synchronized flow partially overlap with
free flow in the density.

The same conclusion can be made if the vehicle speed as a function of
the density is drawn (Fig.~\ref{Wide4} (a)). This means that at the
same density either a state of synchronized flow or a state of free
flow is possible. 

 If now the average absolute values of the vehicle speed difference
between the left lane and the middle lane $\Delta v$ for free flow (curve $F$
in Fig.~\ref{Wide4} (b)) and for synchronized flow (curve $S$ in
Fig.~\ref{Wide4} (b)) are shown, then we see that there is 
overlapping of the speed difference in density.
This overlapping can lead to a hysteresis between the
free flow states and the synchronized flow states.
It can be assumed that this overlapping
is related to  a Z-shaped form of the dependence $\Delta v$
of the density.\footnote{The assumption that the overlapping
of the speed difference $\Delta v$ in the density should be related to a Z-shaped
characteristic $\Delta v(\rho)$ is made in an analogy to  a
huge number of physical, chemical and biological systems where
states of two different  phases overlap in a control system parameter.
Naturally, there is usually a hysteresis effect due to this overlapping.
 In  the theory of these spatially distributed systems, dependent of the
form of this overlapping   it is usually assumed
that there is one of the N-, S- or Z-shaped characteristics of the 
system~\cite{Scholl1987,Vasilev,Michailov1,KO}.
The middle branch of these
characteristics cannot usually be observed in experiments. This is because this branch should
be related to unstable states of the system~\cite{Scholl1987,Vasilev,Michailov1,KO}.
The  assumption under consideration can be considered as
 correct if the theory predicts features of spatial-temporal patterns which are observed in 
experiments~\cite{Scholl1987,Vasilev,Michailov1,KO}.
The three-phase traffic theory (where the Z-shaped
traffic flow characteristic is used~\cite{KKl2003A})
predicts main features of empirical spatial-temporal congested traffic 
patterns~\cite{KKl,KKW,KKl2003A}. For this reason we will use the
 assumption about the Z-shaped traffic flow characteristic
made above in the further consideration.}
Such Z-shaped characteristic  is in agreement with the
hypothesis about the Z-shape of the mean probability for
passing~\cite{Kerner1999A,Kerner1999B,Kerner1999C}
(see Sect.~\ref{Z_FS_Sec}). 

The physical meaning of the result in Fig.~\ref{Wide4} (b)
is the following: In free flow the difference in the average vehicle
speed on German highways between the left (passing) highway lane and the
middle lane due to the high mean probability of passing is
considerably higher than that in synchronized flow. However, at the
same density in a limited range (e.g., at D19 from 18 vehicles/km to
26 vehicles/km) either states of free flow or synchronized flow can
exist. This may lead to a nearly Z-form of the dependency $\Delta v$ on
the vehicle density. The lower the vehicle speed in synchronized flow
is the less  the density range of the overlapping of the curves $F$
and $S$ (Fig.~\ref{Wide4}(b), D18). This overlapping disappears fully
 if the vehicle speed in synchronized flow further
decreases.

\subsection{Analysis of Individual Vehicle Speeds \label{Ind_Sec}}

To see the difference between free flow and synchronized flow 
and features of synchronized flow more
clearly, distributions of the number of vehicles as a function of the
individual vehicle speed for synchronized flow (Fig.~\ref{Wide4} (c))
and for free flow (Fig.~\ref{Wide4} (d)) are shown.  This is possible
because the types of vehicles and their individual vehicle speed during
each of the one minute intervals
are also available.

Firstly, it can be seen
that in synchronized flow the mean vehicle speed of vehicles and long
vehicles are almost the same for different highway lanes whereas for
free flow these mean values are strongly shifted to one another.

Secondly, we see that at the detectors D19 during
 121 min of the observation  individual vehicle speeds
in synchronized flow   were not lower than 40 km/h (Fig.~\ref{Wide4} (c), left).
At the detectors D18 during 105 min of the observations individual speeds of 6181 vehicles 
which passed the detectors were measured (Fig.~\ref{Wide4} (c), right). Among these 6181 vehicles
 there were no vehicles which had the speed below 20 km/h,
there were only 9 vehicles which had 
 individual speeds between 20 and 30 km/h and 59 vehicles which had  individual speeds
between 30 and 40 km/h. All other 6113 vehicles had  individual speeds
higher than 40 km/h. Thus, there were no narrow moving jams 
in  synchronized flow between D19 and D18.
Nevertheless, these states of synchronized flow cover 2D regions in the flow-density plane
(Fig.~\ref{Wide3}, D19-left, D18-left).

\section{Short Coming of the Fundamental Diagram Approach for 
Description of Traffic Congestion \label{Discussion} }

Different explanations of empirical features of wide moving jams
 and synchronized flow~\cite{KR1996B,Kerner1998B,Kerner1999A,Kerner1999B,Kerner2000A,Kerner2000B}
 are up to now being discussed between different
 groups (e.g.,~\cite{Lee1998,Helbing1998,Helbing1999,Lee1999,Lee2000A,Lee2000B,Helbing2000,Treiber,Tomer2000,Knospe,Knospe2001,Lubashevsky,Lubashevsky2001,Nelson,Wagner2002A,Helbing2002,Wagner2002B,Tomer2002,KKl,Fukui,Helbing2002B,KKW}
 and the review~\cite{Helbing2001}).

Due to the effort of different scientific groups (see
e.g.,~\cite{SW1,SW2,SW3,NS,Nagel1995,Schreckenberg,KK1994,B1995A,B1995B,Su,Herrmann,Kra,Bar,Nagatani,Nagatani2,Hel,Trei,Mahnke,KKK1997,KKS1995,Lee1998,Helbing1998,Helbing1999,Lee1999,Lee2000A,Lee2000B,Helbing2000,Treiber,Tomer2000,Knospe,Knospe2001,Nishinari}
and the reviews~\cite{Sch,Helbing2001})  considerable progress has
been made in the understanding of the theoretical spatial-temporal congested
patterns in different traffic flow models within the fundamental
diagram approach.  In particular, in this approach two main classes of
traffic flow models may be distinguished which claim to show moving jams and other
congested
patterns upstream from an on-ramp:
\begin{enumerate}
\item [(i)] Models where at a sufficiently high initial flow rate on the main road,
upstream from an on-ramp, moving jams spontaneously occur if the flow
rate at the on-ramp beginning from zero gradually is increasing. However, the range of the flow rate to the
on-ramp where  moving jams spontaneously occur is
limited. Beginning at a high enough flow rate to the on-ramp spatial
homogeneous states of traffic flow which have been called "homogeneous
congested traffic" (HCT)~\cite{Helbing1999} occur upstream of the
on-ramp where no moving jams spontaneously emerge
(e.g.,~\cite{KKS1995,Helbing1999,Lee1999,Lee2000A,Lee2000B,Helbing2001}).
\item [(ii)] Models where as well as in the models of item (i),
beginning at  some flow rate to the on-ramp, moving jams
spontaneously occur upstream of the on-ramp. However, no HCT occurs
in these models.
\end{enumerate}

How does the traffic phase "synchronized flow" and the traffic phase
"wide moving jam" emerge in an initially free traffic flow at an
isolated bottleneck (i.e., the bottleneck is far away from other
effective bottlenecks), e.g., at a bottleneck due to an on-ramp?
Empirical observations in~\cite{Kerner1998B,Kerner2002B} and discussed above
allow us to conclude that the following scenarios
are responsible for the phase transitions and for the pattern
evolution in traffic flow at the
on-ramp:
\begin{enumerate}
\item
Moving jams do {\it not} emerge in an initial free flow at the on-ramp
when the flow rate at the on-ramp is gradually increasing. Rather than
moving jams the phase transition from free flow to synchronized flow
occurs at the on-ramp.
\item
At a low enough flow rate to the on-ramp the vehicle speed in
synchronized flow which has occurred upstream of the on-ramp is
relatively high. Moving jams do {\it not} necessarily emerge in that
synchronized flow. If the flow rate to the on-ramp is high, then the
vehicle speed in the synchronized flow is low and moving jams, in
particular wide moving jams, emerge in that synchronized flow.
\item 
The lower the average vehicle speed in synchronized flow upstream of
the on-ramp, the higher the frequency of the moving jam emergence in
that synchronized flow. This means that the moving jam emergence goes
on up to the highest possible values of the flow rate to the on-ramp:
Traffic states of high density and low vehicle speed, where moving
jams do not emerge, are {\it not} observed in synchronized flow
upstream of the on-ramp.
\end{enumerate}

The empirical results in item (1) and (2) are  qualitatively in contradiction
with  both model classes (i) and (ii) in the fundamental diagram
approach. In these models, at high enough initial flow rates on the main road
upstream
of the on-ramp moving jams must emerge in an initial free flow if the
flow rate to the on-ramp beginning from zero is gradually
increased~\cite{Helbing2001}. The last empirical result in item 3
means that on average the higher the vehicle density, the lower 
the stability of traffic flow with respect to the moving jam emergence
in that flow. This result of observations which seems to be intuitively
obvious for each driver is in a qualitative contradiction with the
models of class (i) in the fundamental diagram approach where HCT,
i.e., homogeneous congested traffic of high density and low vehicle
speed must occur where moving jams do not emerge~\cite{Helbing2001}.

The features of the theoretical
diagram~\cite{Helbing1999,Lee1999,Helbing2000,Helbing2001} given a
high enough initial flow rate on a highway upstream of the on-ramp may
be illustrated with the following simple {\it theoretical scheme 1}:
\begin{itemize}
\item
A low flow rate to the on-ramp $\rightarrow$ different kinds of moving jams must emerge.
\item
A high enough flow rate to the on-ramp $\rightarrow$ HCT where
the density is high and the speed is very low and no moving jams spontaneously emerge
must occur.
\end{itemize}
In contrast to this theoretical 
result~\cite{Helbing1999,Lee1999,Helbing2000,Helbing2001}, in empirical observations~\cite{Kerner2002B}
the following  {\it empirical scheme 2} is observed:
\begin{itemize}
\item
A low flow rate to the on-ramp $\rightarrow$ synchronized flow
where the density is relatively low and the speed is relatively
high occurs where
moving jams do not necessarily emerge.
\item
A high enough flow rate to the on-ramp $\rightarrow$ moving jams must spontaneously emerge
in synchronized flow upstream of the on-ramp at any high flow rate to the on-ramp.
\end{itemize} 
Thus, these two schemas (the theoretical scheme
1~\cite{Helbing1999,Lee1999,Helbing2000,Helbing2001} and the empirical
scheme 2~\cite{Kerner2002B}) are in contradiction with one  another.

It should be noted that in models with the fundamental diagram for steady states
which should explain empirical spatial-temporal
congested traffic patterns beyond the instability of some of these steady states,
fluctuations
and instabilities let the system evolve in time
through a 2D region 
in the flow-density plane 
as well. This 2D region is related to
some   "dynamical" model solutions.
Examples of
the 2D region in the flow-density plane for the "dynamical" model states 
 are the Nagel-Schreckenberg CA models
(e.g.,~\cite{Knospe}), the models by Tomer, Havlin and co-workers~\cite{Tomer2002},
Helbing, Treiber and co-workers~\cite{Helbing2002,Treiber2003A}.
However, in these traffic flow models,
steady model states are related to a {\it one-dimensional}
region in the flow-density plane, i.e., to
the fundamental diagram.

If, in accordance with the three-phase traffic theory,
{\it steady states} of synchronized flow form a 2D  region in
the flow-density plane,
the model dynamics is fundamentally different~\cite{KKl,KKW,KKl2003A}. 
This leads also to qualitative differences between the patterns of congested
traffic obtained in the three-phase traffic theory or in the fundamental
diagram approach, respectively.

It must be noted that this critical consideration of the application
of the fundamental diagram approach for the
description of congested traffic  does not concern
some important mathematical  ideas 
which  have been introduced and developed in  models and theories
within the fundamental diagram
approach with the aim to describe the traffic flow dynamics,
 for example the ideas
about the modelling of vehicle safety conditions,
 fluctuations, vehicle acceleration and deceleration, different vehicle time delays
 and other  important 
effects.
In particular, the related  pioneer mathematical  ideas have been introduced in
 models and theories  within the fundamental diagram approach by 
Lighthill and Whitham~\cite{LW}, Richards~\cite{Rich},  
 Prigogine~\cite{Prigogine1961},
Gazis,  Herman, Rothery, Montroll,   Potts~\cite{GH}, 
Komentani and Sasaki~\cite{KS},
 Newell~\cite{New2},  Whitham~\cite{Wh},    
 Bando, Sugiyama and colleagues~\cite{B1995A},  Payne~\cite{Pa},
  Gipps~\cite{Gipps},  
Wiedemann~\cite{Wid},
   Nagel  
and  Schreckenberg and co-workers~\cite{NS,Nagel1995,Schreckenberg,Bar,Knospe}, 
 Takayasu and Takayasu~\cite{TaTa},
  Krau{\ss} 
{\it et al.}~\cite{Kra}, 
 Mahnke, K\"uhne {\it et al.}~\cite{Mahnke,Mahnke1997,Kuehne2002},  Helbing, 
Treiber and co-workers~\cite{Trei,Helbing2000,Treiber2003A},
 Nishinari and Takahashi~\cite{Nishinari},
 Fukui {\it et al.}~\cite{Fukui,FI},  Havlin, Tomer  and co-workers~\cite{Tomer2000,Tomer2002},
 Nagatani and Nakanishi~\cite{NaNa}
and by many other groups
(see  references in
the reviews by Chowdhury {\it et al.}~\cite{Sch},  Helbing~\cite{Helbing2001},
Nagatani~\cite{Nagatani_R}, Nagel {\it et al.}~\cite{Nagel2003A}, Wolf~\cite{Wolf}). 
These mathematical ideas  
 are also very important elements of the  three-phase traffic 
theory~\cite{Kerner1999A,Kerner1999B,Kerner1998A}
and of  microscopic models within this theory~\cite{KKl,KKW}. 
The main feature of the  three-phase traffic  theory 
is that this theory {\it rejects} the 
basic hypothesis about the
fundamental diagram of  earlier traffic flow theories and models. The three-phase traffic theory
introduces the new phase of traffic flow, synchronized flow, whose steady
states  cover 2D region in the flow density plane. 
 This allows us to overcome the above problems of the fundamental diagram approach
and
to explain empirical spatial-temporal congested pattern features~\cite{Kerner1998B,Kerner2002B}.

\section{Comparison of Hypotheses to  Three-Phase Traffic Theory with some
Results of  Fundamental Diagram Approach \label{Com_Sec}}

\subsection{Steady Speed States~\protect\cite{Kerner1998C,Kerner1999A,Kerner1999B}}

\subsubsection{Explanation of  Fundamental Hypothesis of  The Three-Phase Traffic Theory
\label{Exp_H_Sec}}

\begin{figure}
\begin{center}
\includegraphics[width=1.0\textwidth]{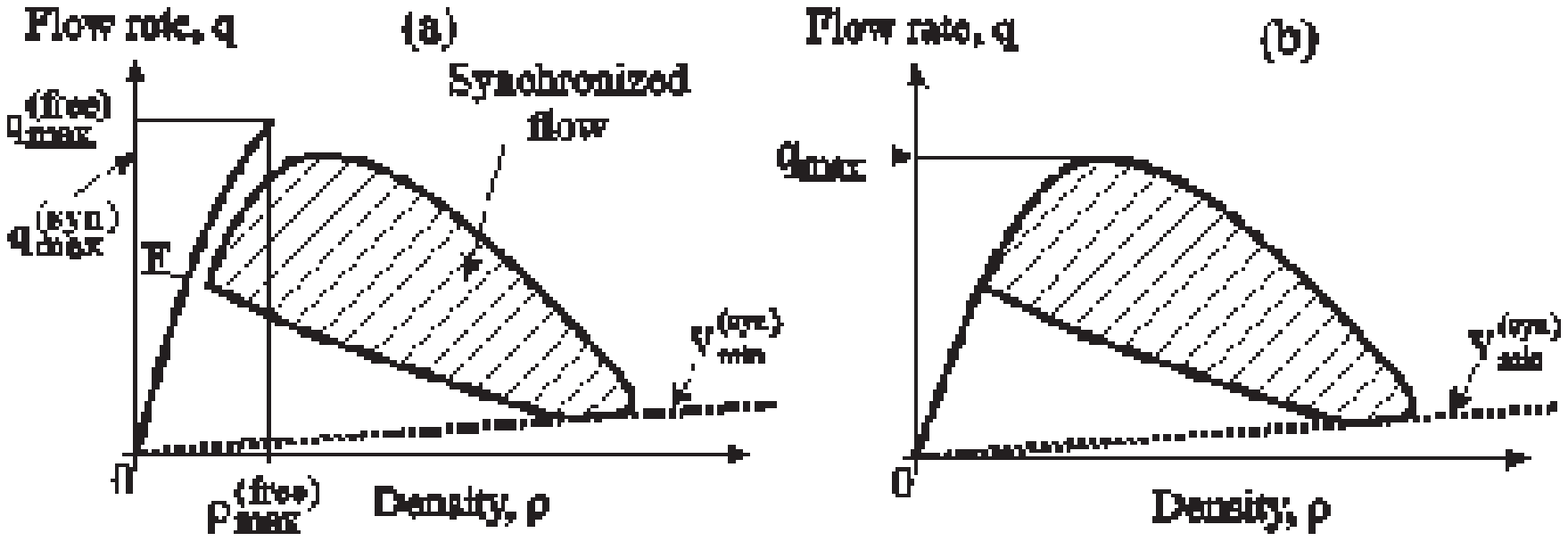}
\end{center}
\caption[]{The fundamental hypotheses of the three-phase traffic theory~\protect\cite{Kerner1998B,Kerner1998C}:
States of free flow (curve $F$) and spatially homogeneous 
and time-independent states (steady states) of synchronized flow (dashed region)
in the flow density plane for a multi-lane (in one direction) homogeneous
(without bottlenecks) road in the three-phase traffic theory (a). 
For a comparison, in (b) steady states
for a homogeneous
(without bottlenecks)
one-lane road are shown.
The dotted line is related to a  minimum possible speed in steady states of synchronized flow 
$v^{\rm (syn)}_{\rm min}$~\protect\cite{Kerner2000C,Kerner2001A}. 
}
\label{Diagram}
\end{figure}

\begin{figure}[]
\begin{center}
\includegraphics[width=0.9\textwidth]{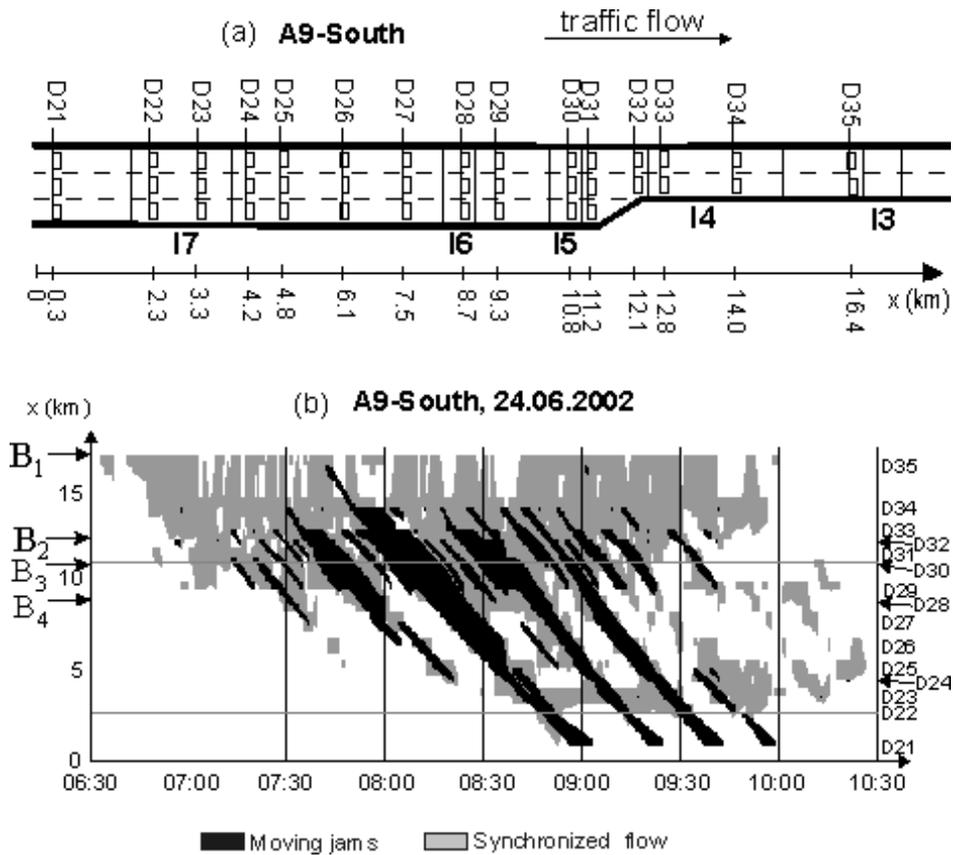}
\end{center}
\caption[]{Example of expanded congested pattern (EP): (a) - Scheme
of the  infrastructure of the section of the
freeway 
A9-South  near Munich, Germany. Highway bottlenecks are linked to the
merge of three-lane section into two-lane section
in the vicinity of  the detectors D32 and to on- and off-ramps in the intersections
I5, I6, and I7 with other freeways.
(b) - The graph
 of synchronized flow and wide moving jams in space and time in the EP
which has been reconstructed by the models ASDA and FOTO
based on measurements of the average vehicle
speed and the flow rate at different detectors (D35-D21). Taken from~\cite{FOTO2003A}.   \label{Expanded} }
\end{figure}

\begin{figure}[]
\begin{center}
\includegraphics[width=0.9\textwidth]{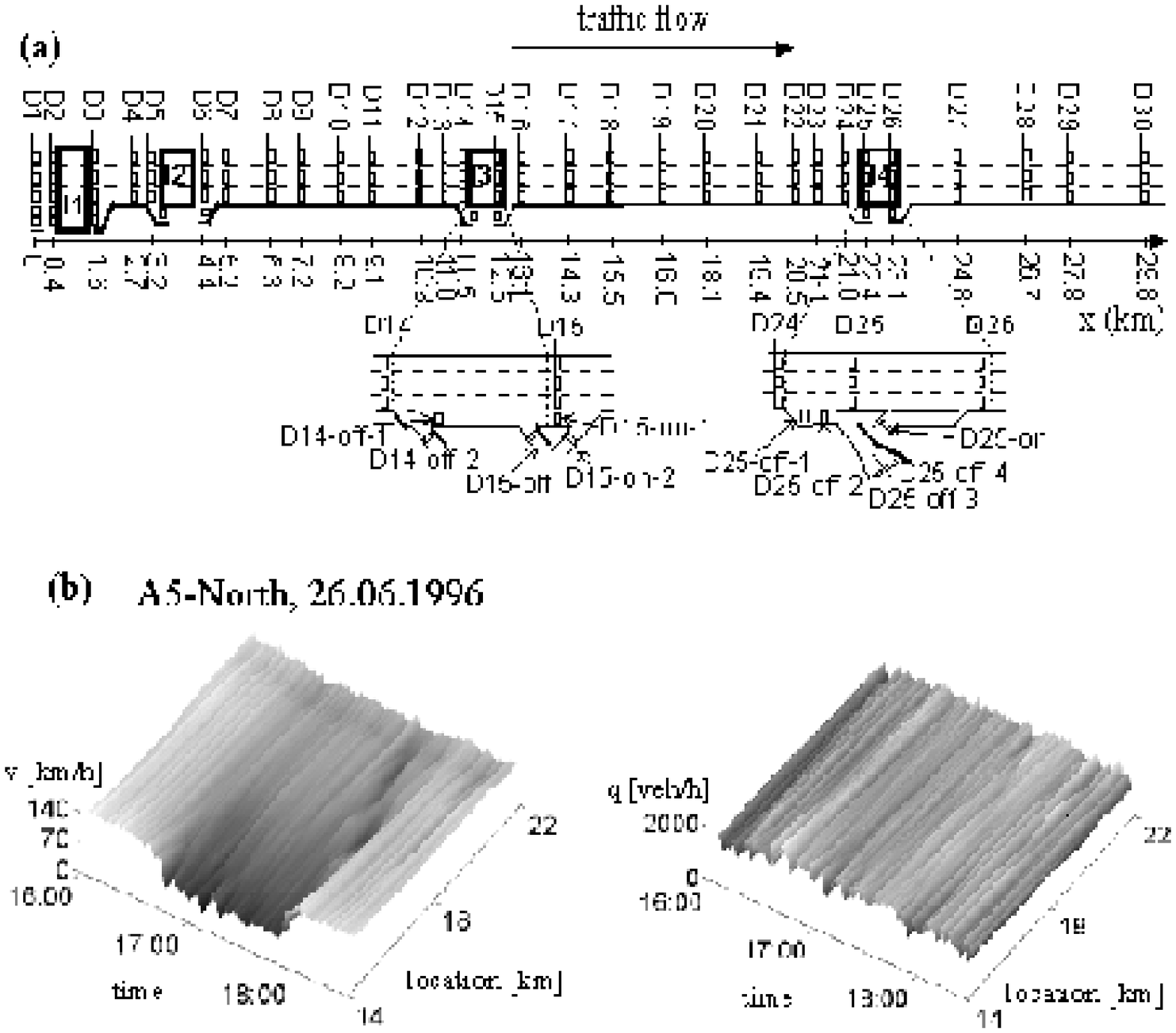}
\end{center}
\caption[]{The widening synchronized flow pattern (WSP): (a) - Scheme
of the highway infrastructure and local measurements at the section of
the highway A5-North~\protect\cite{Kerner2002B}; (b) - the average
speed and the flow rate in WSP as functions of time and
location~\cite{Kerner2002E}.  \label{Wide1} }
\end{figure}

\begin{figure}[]
\begin{center}
\includegraphics[width=0.9\textwidth]{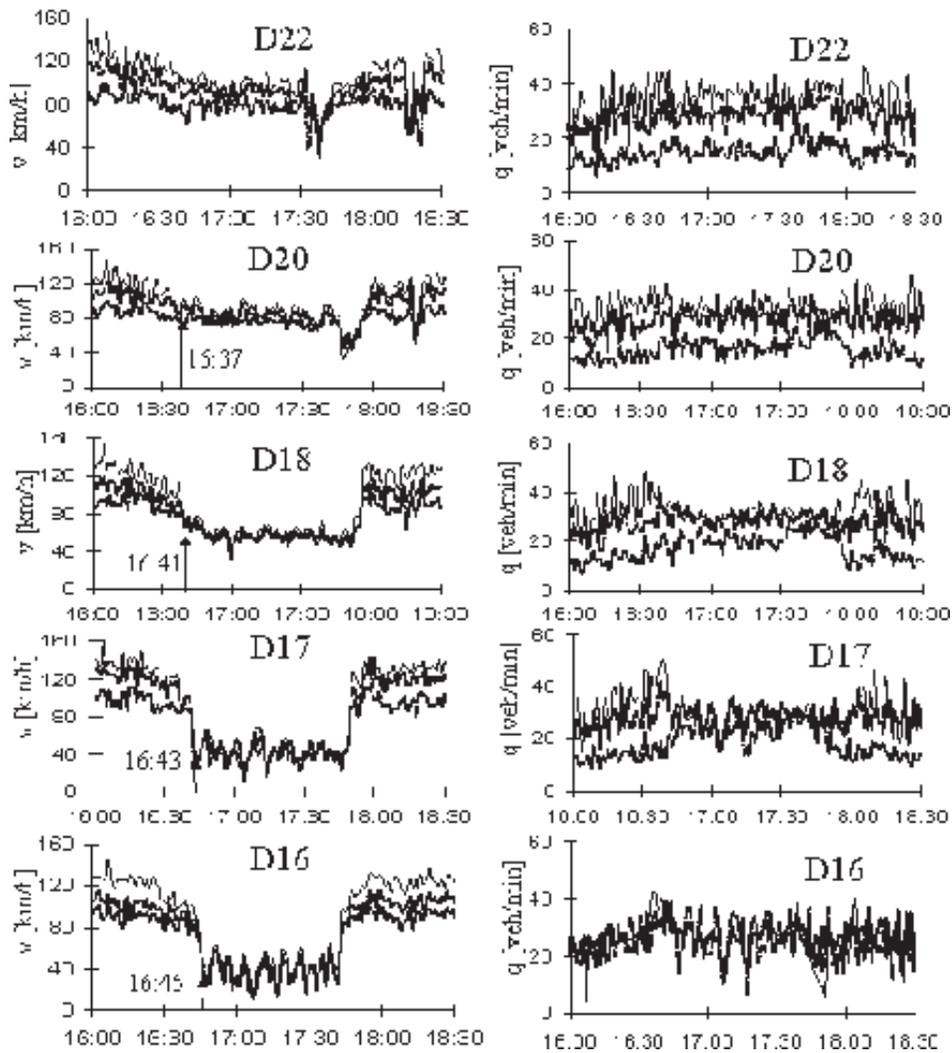}
\end{center}
\caption[]{The widening synchronized flow pattern (WSP): Time series
of the vehicle speed (left) and flow rate (right) for different
highway lanes for detectors D22 - D16. The F$\rightarrow$S transitions
at the related detectors leading to the WSP formation are marked with
up arrows.  \label{Wide2} }
\end{figure}

\begin{figure}[]
\begin{center}
\includegraphics[width=0.9\textwidth]{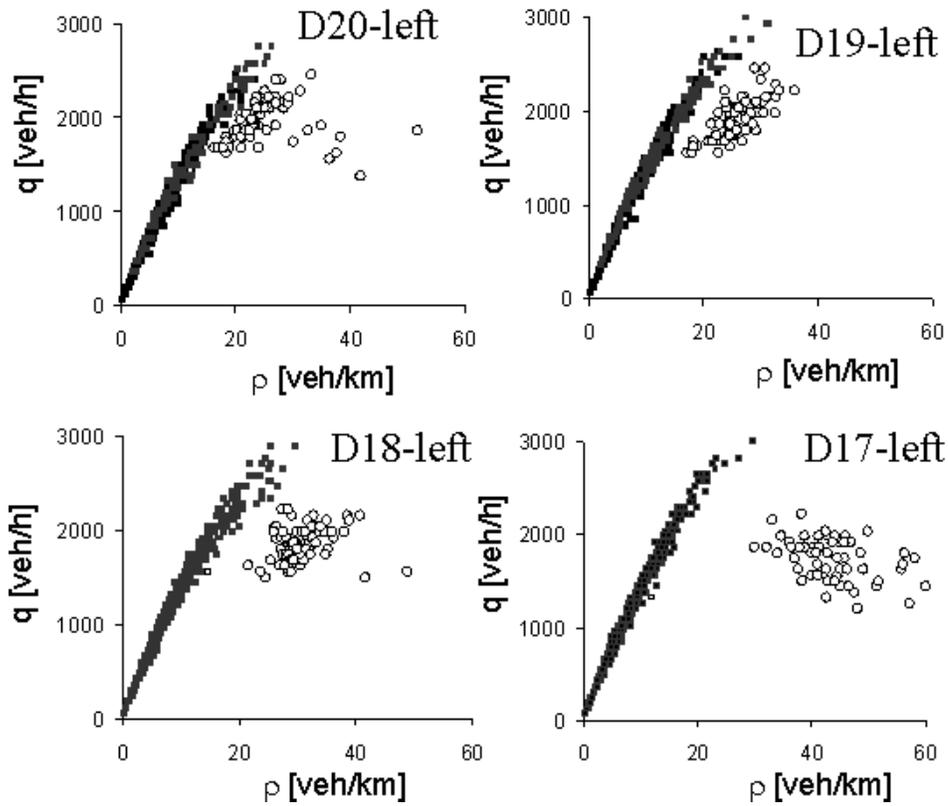}
\end{center}
\caption[]{The measurement points in the flow-density plane for WSP
shown in Fig.~\ref{Wide1} at the detectors D20 - D17 (left line).
Free traffic is related to black quadrates, synchronized flow is
related to circles. Overlapping of states of free flow and
synchronized flow at the detectors D19 is in the density range from
about 18 to 36 vehicles/km.
\label{Wide3} }
\end{figure}

\begin{figure}[]
\begin{center}
\includegraphics[width=0.6\textwidth]{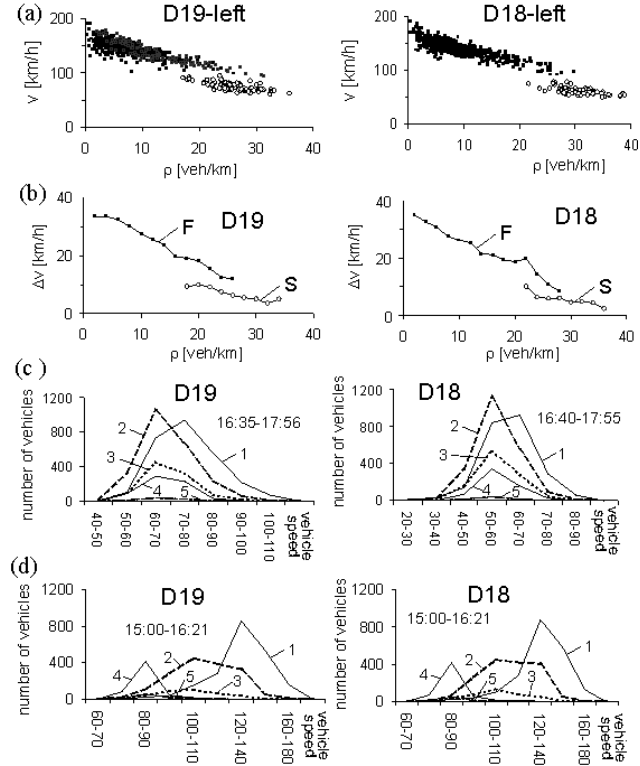}
\end{center}
\caption[]{Empirical features of synchronized flow at the detectors
D19 (left) and D18 (right) in the WSP shown in Figs.~\ref{Wide1} (b)
and~\ref{Wide2}: (a) - Measurement points in the
speed-density plane (free flow - black quadrates, synchronized flow -
circles; left lane), (b) - The average difference in the vehicle
speeds between left and middle highway lanes as a function of the
density, curve $F$ for free flow, curve $S$ for synchronized flow.
The speed differences in (b) are averaged for density intervals of 2
vehicles/km; (c, d) - Distribution of the number of vehicles as a
function of the different speed classes related to individual single
vehicle data for synchronized flow (c) and for free flow (d): The
curve 1 is related to vehicles on the left lane, 2 - vehicles on the
middle lane, 3 - vehicles on the right lane, 4 - long vehicles on the
right lane, 5 - long vehicles on the middle lane (long vehicles may
not move on the left (passing) lane of a three-lane (in one direction)
highway in Germany). In (c, d), measured single vehicle data are shown
whereby the number of vehicles in each of 15 different classes in
regard to the vehicle speed is used separately for vehicles and for long
vehicles.
\label{Wide4} }
\end{figure}

The fundamental hypothesis of the three-phase traffic theory has
already been formulated in 
Sect.~\ref{Fun_Hyp}~\cite{Kerner1998B,Kerner1998C,Kerner2000A,Kerner2000B,Kerner2001A,Kerner1998A,Kerner2002,Kerner2002C}:
Hypothetical spatially homogeneous and time-independent (stationary) states of synchronized flow,
 i.e., steady  states
of synchronized flow where vehicles move at the same distance to one  another with the same
time-independent vehicle speed
cover a 2D region in the flow-density
plane (Fig.~\ref{Diagram}).
These steady states are the same for  multi-lane and for  one-lane
roads. In other words, in the three-phase traffic theory there is
{\it no} fundamental diagram for steady states of synchronized flow.

This is not excluded by the empirical fact mentioned above, that a
given vehicle density determines the {\it average} vehicle speed.
Indeed, from empirical observations it may be concluded that at the
same distance between vehicles (at the same density) there may be a
continuum of different vehicle speeds within some finite range in
synchronized flow. Obviously the averaging of all these vehicle speeds
leads to one average value at the given density. 

The 2D region of steady states of synchronized flow (Fig.~\ref{Diagram}) is also not
ruled out by car following experiments, where a driver has the task to
follow a specific leading car and not lose contact with it
(e.g.,~\cite{Koshi}). In such a situation, the gap between the cars
will be biased towards the security gap depending on the speed of the
leading car. In synchronized flow the situation is different: The gap
between cars can be much larger than the security gap.

The hypothesis about steady states of synchronized flow which cover a
2D-region in the flow-density plane makes the three-phase traffic
theory~\cite{Kerner1998B,Kerner1998C,Kerner1998A} almost {\it
incompatible} with all classical traffic flow theories and present
models (see,
e.g.,
\cite{Leu,Wh2,Cremer,Sch,Helbing2001,SW1,SW2,SW3,LW,GH,New2,NS,Wh,KK1994,B1995A,B1995B,Herrmann,Kra,Bar,Hel,Trei,Mahnke,KKK1997,KKS1995,Lee1998,Helbing1998,Helbing1999,Lee1999,Helbing2000,Tomer2000,Knospe,Knospe2001})
which are based on the fundamental diagram approach.

This follows from the diagram of congested patterns
at bottlenecks which has recently been found  by the author~\cite{Kerner2002B,Kerner2002,Kerner2002D}. 
This diagram has been postulated on very general grounds within
the three-phase traffic theory~\cite{Kerner2002B,Kerner2002} (see Sect.~\ref{DCP})
and demonstrated for a microscopic traffic 
model by Kerner and Klenov~\cite{KKl}. In this model, the upper boundary of
a 2D-region of steady states of synchronized flow
in the flow-density plane is related to the well-known
dependence of the safe speed on the gap between vehicles.
The low boundary of these steady states is related to the dependence of a synchronization distance
$D$ on the speed. These characteristics of steady states of synchronized flow
have been used
by Kerner, Klenov and Wolf for the formulation of KKW cellular automata (CA)
models
to the three-phase traffic theory which may show qualitatively the same diagram of congested patterns~\cite{KKW}.
Recently Kerner and Klenov developed a miscoscopic
non-linear theory of spatial-temporal congested patterns at highway bottlenecks~\cite{KKl2003A}.
This theory which is based on the three-phase traffic theory
allows us to explain and predict  main empirical features of congested traffic
patterns~\cite{Kerner1998B,Kerner2002B}.

Note that the three-phase traffic theory is {\it a behavioral theory} of traffic flow.
This means that the hypotheses of this theory are based on  common behavioral
fundamental characteristics of drivers observed 
on highways. In particular, the fundamental hypothesis of the three-phase traffic 
theory (Sect.~\ref{concept})
is linked
to a driver's ability
to
recognize whether the distance to the vehicle ahead becomes  higher or lower over time~\cite{Kerner1998C}.
If gaps between vehicles are not very high, this driver's ability is true even if
the difference between vehicle speeds is negligible.

It has been shown~\cite{Kerner2002B,KKl,KKW,KKl2003A} that the main {\it qualitative} features of the 
congested pattern emergence observed in empirical investigations~\cite{Kerner1998B,Kerner2002B} can 
be shown in the three-phase traffic theory where
all drivers have the same characteristics and all
vehicles have the same parameters.  Obviously, in real traffic there are
differences in driver's characteristics and in vehicle parameters 
(e.g., different desired speeds
 and different safe speeds, 
 aggressive and timid driver's behavior, vehicles and long vehicles)
which may change some 
spatial-temporal congested pattern parameters
and conditions of the pattern emergence. However, these differences
 in driver's characteristics and in vehicle parameters
may first be neglected
when {\it fundamental} congested traffic pattern features are studied. 
This statement is confirmed in the microscopic theory of congested traffic patterns~\cite{KKl2003A}
which is able to explain and to predict main fundamental empirical congested pattern features.

\subsubsection{Hypothesis about Stability of Steady 
States~\protect\cite{Kerner1999A,Kerner1998C,Kerner1999B}}

In 1959  Herman,     Montroll,  Potts and 
Rothery
have introduced the concept of a driver's acceleration (deceleration)  delay time
in their car-following microscopic traffic flow model.
They found an instability in some of the steady states of traffic flow~\cite{GH1959}.
The existence of the delay times in the vehicle acceleration
and deceleration has been confirmed in a lot of empirical observations (e.g.,~\cite{Leu}). 
To study traffic phenomena beyond the instability, in
1961 Gazis,  Herman and Rothery~\cite{GH} and Newell~\cite{New2}
used 
the fundamental diagram for steady model states  (for the review see~\cite{Nagel2003A}). 
Later the driver's delay time and the fundamental diagram for steady model states
have been used in a huge number of other traffic flow models and theories  where  
the linear instability of steady states
should occurs 
 when the vehicle density  exceeds
 some {\it critical} vehicle density
(see references in the reviews~\cite{Sch,Helbing2001,Nagatani_R,Nagel2003A}).

In the three-phase traffic theory,  the driver's delay time
plays  a very important role
also,
however 
a different hypothesis is 
suggested~\cite{Kerner1998C,Kerner1998A,Kerner1998B,Kerner1999B,Kerner1999A,Kerner1999C}:
{\it Independent of the vehicle density in a steady state of synchronized flow infinitesimal 
perturbations of any traffic flow variables (e.g., the vehicle speed and/or the gap) 
do not grow}: In the whole possible 
density range steady states of synchronized flow can exist. In other words, in the whole 
possible density range (Fig.~\ref{Diagram}) there are no unstable steady states of 
synchronized flow with respect to infinitesimal perturbations of any traffic flow 
variables.

To explain this hypothesis from the driver's behavior,
let us consider  a small enough fluctuation in the
 braking of a vehicle in an initial synchronized flow
steady state which is related to one of the steady states inside  a 2D region
in the flow-density plane (dashed region in Fig.~\ref{Diagram}). This small braking  may lead to
a transition from the initial synchronized flow state to another state
with a lower gap. This lower gap  is equal to
a gap in another  state in a 2D region
in the flow-density plane. Thus, an occurrence of this fluctuation may cause a 
spatial-temporal transition to another state of synchronized flow.
 Therefore drivers should not immediately react 
on this transition. For this reason even after a time delay, which is due to a finite 
reaction time of drivers, the drivers upstream should not brake 
stronger than drivers in front of them to avoid an accident. As a result, a 
local perturbation of traffic variables (density or vehicle speed) of small 
enough amplitude does not grow. 

Already small amplitude fluctuations make
 real traffic flow  always  non-homogeneous and non-stationary. Thus,
as well as the theoretical fundamental diagram for steady states it is only a hypothesis
which cannot empirically be proven (Sect.~\ref{FD_Sunsec}),
 it is also  not possible to prove 
the fundamental hypothesis of the three-phase traffic theory and the hypothesis 
about stability of steady states
 based on a direct empirical measurement
of traffic  variables. As a  proof of the three-phase
traffic theory  mathematical results of this theory 
can be considered~\cite{KKl,KKW,KKl2003A}. These results allow us to overcome the
problems of the fundamental diagram approach
for the description of empirical features of the phase transitions and spatial-temporal
congested patterns.

\subsection{Phase Transitions on Homogeneous (without Bottlenecks) Roads}

\subsubsection{The Line J and Hypothesis about Emergence of Moving Jams~\cite{Kerner1998B}
\label{FJ_Hyp}}

In contrast to the traffic flow 
theories in the fundamental diagram approach~\cite{Sch,Helbing2001,Nagatani_R,Nagel2003A},
in the three-phase traffic theory,
there is {\it no}
critical density where traffic flow should become unstable with respect to the moving jam 
emergence~\cite{Kerner1998B,Kerner1998A,Kerner1998C}.
How is the moving jam emergence explained in the three-phase traffic theory?

To understand this, we consider first the process of the driver's escaping from a standstill inside a
wide moving jam. This process determines the velocity of the
upstream motion of the
downstream front of the wide moving jam.
In this process, each driver standing inside the 
wide moving jam
can start to escape from the jam after (i) the vehicle in front of the driver
has already escaped from the jam and (ii)
the distance between these vehicles has exceeded some "safety distance". Thus,
the driver begins to escape after a delay time in the vehicle acceleration. The related mean
delay time
$\tau^{\rm (a)}_{\rm del}$  determines the average time interval between two vehicles
following one another escaping from the wide moving jam.
Therefore, the velocity of the downstream jam front
$v_{\rm g}$ is:
\begin{equation}
v_{\rm g}=- \frac{1}{\rho_{\rm max}\tau^{\rm (a)}_{\rm del}},
\label{jam_velocity}
\end{equation}
where $\rho_{\rm max}$ is
the mean vehicle density inside the wide moving jam.

\begin{figure}
\begin{center}
\includegraphics*[width=.5\textwidth]{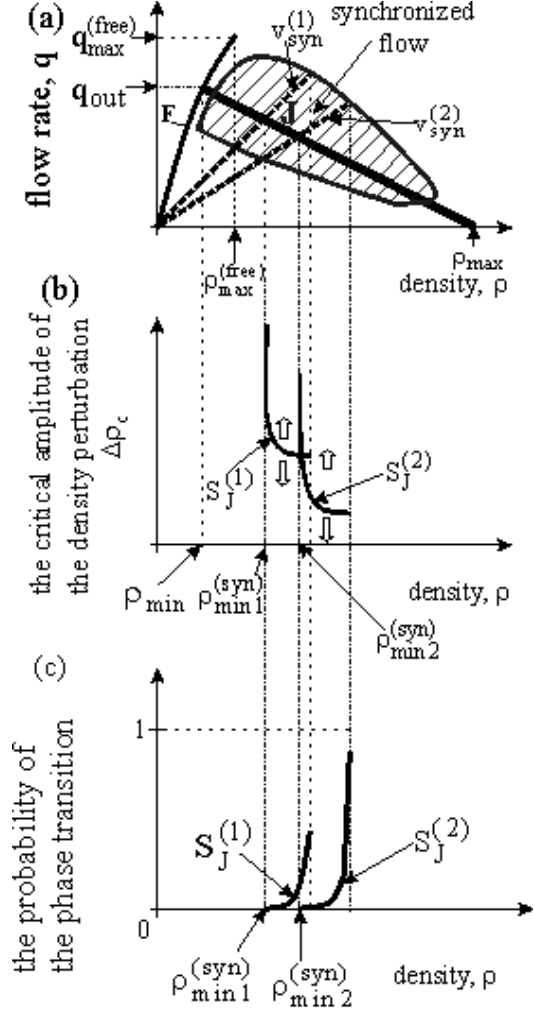}
\end{center}
\caption[]{Explanation of 
 the jam emergence in synchronized flow
in
the three-phase traffic theory~\protect\cite{Kerner1998B}: (a) -
the line $J$. 
States of free (curve $F$) and steady states of synchronized 
flow (hatched region)   are 
 the same as in Fig.~\protect\ref{Diagram} (a).
(b) - Qualitative dependences ($S^{\rm (1)}_{\rm J}$, $S^{\rm (2)}_{\rm J}$)
of the critical density amplitude of a local perturbation
$\Delta \rho_{\rm c}$
on the density in steady speed states. The curves $S^{\rm (1)}_{\rm J}(\rho)$ and $S^{\rm (2)}_{\rm J}(\rho)$
are related to two different constant speeds
$v^{\rm (1)}_{\rm syn}$ and $v^{\rm (2)}_{\rm syn}$, respectively ($v^{\rm (1)}_{\rm syn}>v^{\rm (2)}_{\rm syn}$).
The densities $\rho^{\rm (syn)}_{\rm min \ 1}$ and $\rho^{\rm (syn)}_{\rm min \ 2}$
are the threshold densities for the jam emergence for the vehicle speeds
$v^{\rm (1)}_{\rm syn}$ and $v^{\rm (2)}_{\rm syn}$,
respectively. 
}
\label{Hypote3}
\end{figure}

We propose that all vehicles are the same and all drivers have the same characteristics.
Thus, we can suggest that the mean parameters $\rho_{\rm max}$ and $\tau^{\rm (a)}_{\rm del}$ do not depend 
on time. This means that the velocity of the downstream wide moving jam front $v_{\rm g}$ (\ref{jam_velocity}) is 
also independent of time: The propagation of
the downstream wide moving jam  front is on average a stationary process. 
The  stationary motion of the downstream jam front can be presented
in the flow-density plane by the characteristic line $J$
(Fig.~\ref{Hypote3} (a)) which is called the line $J$~\cite{KR1996B}.
The slope of the line $J$ in the flow-density plane
is equal to the velocity of this downstream front $v_{\rm g}$.
If free flow is formed in the wide moving jam outflow, the flow rate in this jam outflow
is $q_{\rm out}$, the density is $\rho_{\rm min}$ and the speed is  $v_{\rm max}$.
As well as the velocity $v_{\rm g}$, 
the mean parameters of the jam outflow
$q_{\rm out}$, $\rho_{\rm min}$ and  $v_{\rm max}$
are  characteristic parameters which do not depend on initial conditions.
The stationary propagation of the downstream front of wide moving jams and the line $J$
have been found in empirical studies of wide moving jams~\cite{KR1996B,KR1996A}.

The following hypothesis of the three-phase traffic theory
is related to the wide moving jam emergence in the 2D steady states of 
synchronized flow~\cite{Kerner1998A,Kerner1998B}:

{\it All ({\bf an infinite number!}) 
steady states of traffic flow which are related to the line $J$ in the 
flow-density plane are {\bf threshold states} 
with respect to the wide moving jam emergence.}

This means that
the line $J$ separates all steady states of 
traffic flow into two qualitatively different classes:
\begin{enumerate}
\item In states which are related to points in the flow-density plane lying 
below (see axes in Fig.~\ref{Hypote3} (a)) the line $J$ {\it no} wide moving jams  can either
continue 
to 
exist or  be excited.
\item States which are related to points in the flow-density plane lying on and 
above the line $J$ are {\it metastable steady states} with respect to
the wide moving jam emergence (the F$\rightarrow
$J transition)
where the related nucleation effect can be realized. In a metastable steady state, a growth of an initial
 local perturbation
can  lead to the wide moving jam
formation {\it only if} the amplitude of this perturbation exceeds a critical amplitude. 
In contrast, if the amplitude of the perturbation is lower than the critical amplitude the 
initial perturbation does not lead to the wide moving jam formation.
In the latter case, a transition from one to another steady state in the 2D region
of steady states is however possible
rather than the wide moving jam formation.
\end{enumerate} 
This hypothesis has been confirmed by numerical simulations of the
model by Kerner and Klenov in the frame of the three-phase
traffic theory (see Fig. 1 (f) in~\cite{KKl}).

\begin{figure}
\begin{center}
\includegraphics*[width=.95\textwidth]{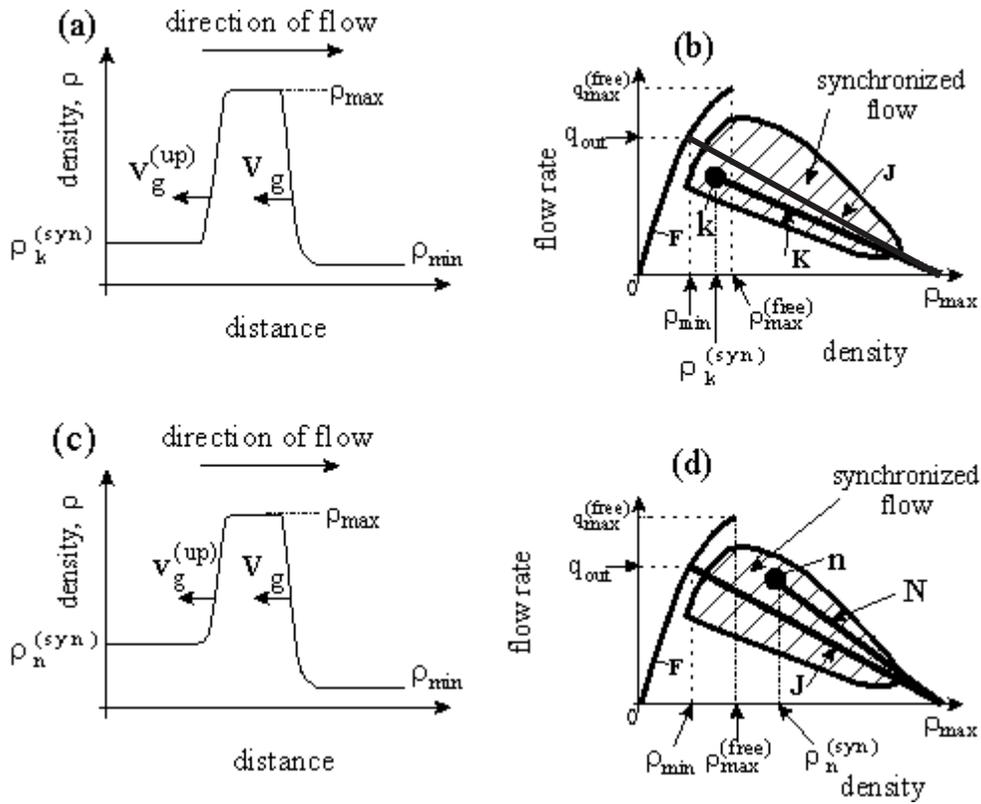}
\end{center}
\caption[]{Explanation of the wide moving
jam emergence in steady states of traffic flow
in~\protect\cite{Kerner1998B}. In (a, c) $q_{\rm out}$
$\rho_{\rm min}$ are the flow rate and  the density in the wide moving jam outflow,
respectively
(see Fig.~\ref{Hypote3}).
}
\label{Hypote4}
\end{figure}

To explain the hypothesis,
 note that  wide moving jams cannot be formed in any steady states of traffic flow situated below 
the line J. Indeed, let  the steady state of flow directly upstream of a wide moving jam is related to 
a point $k$ in the flow-density 
plane which is below the line J (Fig.~\ref{Hypote4} (a, b)). Because  the velocity 
of the upstream front 
of the wide moving jam  $v^{\rm (up)}_{\rm g}$
equals the slope of a line from $k$ to the point $[\rho_{\rm max}, 0]$, 
the related absolute value 
$\mid v^{\rm (up)}_{\rm g}\mid$
 is always lower than that of the downstream front $\mid v_{\rm g}\mid$
 which is determined  by the slope of the line $J$:
\begin{equation}
 \mid v^{\rm (up)}_{\rm g}\mid <\mid v_{\rm g}\mid.
\end{equation}
Therefore, the width of the wide moving jam is gradually decreasing. 
Otherwise, let us consider the case when
 a steady flow which is upstream from another wide moving jam 
is above the line $J$ (see a point $n$
 in Fig.~\ref{Hypote4} (d)).
In this case,  the velocity 
of the upstream front 
of the wide moving jam  $v^{\rm (up)}_{\rm g}$
equals the slope of a line from $n$ to the point $[\rho_{\rm max}, 0]$, i.e., 
the related absolute value 
$\mid v^{\rm (up)}_{\rm g}\mid$
 is always higher than that of the downstream front $\mid v_{\rm g}\mid$:
\begin{equation}
  \mid v^{\rm (up)}_{\rm g}\mid >\mid v_{\rm g}\mid.
\end{equation}
 Therefore,
 the width of the wide moving jam in Fig.~\ref{Hypote4} (d) should be gradually increasing. 
For these reasons, wide moving jams can be formed in steady states of traffic 
flow which lie on and above the line J. 

Recall that the line $J$ has first been introduced
by Kerner and Konh\"auser in 1994~\cite{KK1994}
in their theory of wide moving jams. The Kerner-Konh\"auser theory of wide moving jams
which has been derived within the fundamental diagram approach
 has been  further applied and developed
for a number of different traffic flow
models~\cite{B1995B,Kra,Bar,Trei,Mahnke,Helbing1999,Lee1999,Helbing2000,Knospe,Knospe2001,Sch,Helbing2001,Kuehne2002}.
In this theory~\cite{KK1994}, the line $J$ separates stable and metastable states of free flow
with respect to the wide moving jam emergence:
States of free flow which are on and above
the line $J$ (i.e., when the density and the flow rate in free flow
are either equal to or  higher than the
density $\rho_{\rm min}$
and the flow rate
$q_{\rm out}$  in the outflow from the
wide moving jam when free flow is formed downstream of the jam, respectively) are metastable states
with respect to the wide moving jam emergence (the  F$\rightarrow
$J transition).
The line $J$ and the characteristic wide moving
jam parameters following from the theory~\cite{KK1994}
have indeed been found in empirical
observations~\cite{KR1996A}.

However, in the fundamental diagram
approach at the limit point of this metastable region of free flow
($\rho^{\rm (free)}_{\rm max}$, $q^{\rm (free)}_{\rm max}$) (Fig.~\ref{Hypote3} (a))
 free flow becomes unstable with respect to
the moving jam emergence (the F$\rightarrow
$J-transition)~\cite{KK1994,B1995A,B1995B,Herrmann,Hel,Trei,KKK1997,KKS1995,Lee1998,Helbing1998,Helbing1999,Lee1999,Helbing2000,Tomer2000,Sch,Helbing2001,Nagatani_R,Nagel2003A,Kuehne2002}).
This common feature 
of all traffic flow models within the fundamental diagram
approach which claim to show moving jam emergence is in contradiction
with empirical observations: Moving jams do not
spontaneously emerge in real free flow.
In contrast, in the three-phase traffic theory at this limit point
of free flow  the F$\rightarrow
$S transition occurs rather than the F$\rightarrow
$J transition~\cite{Kerner1999A,Kerner1999B}.
This is in accordance with empirical results~\cite{Kerner1998B,Kerner2002B}
 (see for more detail Sect.~\ref{Limit}).

\subsubsection{Hypothesis about Continuous Spatial-Temporal
Transitions between 
Steady States of Synchronized Flow
on Homogeneous  Roads~\cite{Kerner1998C}
\label{Tran_Sec} }

If in a metastable steady state on and above the
line $J$ in the flow-density plane
a local perturbation occurs whose amplitude is lower
than the critical amplitude then
this perturbation does not lead to the moving jam emergence. However, 
the perturbation can nevertheless cause a local transition to
another  state of synchronized flow rather than traffic flow returning to the initial steady state.
The same case can also occur for steady states below the line $J$, i.e., for steady states
which are stable with respect to the moving jam emergence.
Thus, when a lot of different local perturbations
appear in traffic flow, then a very complex
spatial-temporal transitions between states of synchronized flow 
can occur. These synchronized flow
states can be close to  steady states.
This is the contents of 
the following hypothesis of the three-phase traffic 
theory~\cite{Kerner1998C,Kerner1998A,Kerner1998B,Kerner1999B,Kerner1999A,Kerner1999C}:

 {\it Local random perturbations in synchronized flow can cause continuous spatial-temporal 
transitions between different  states  
of synchronized flow} (a random "walking" between different 2D states in the
flow density plane).

Complex transformation between small amplitude
spatial-temporal states which are close to model steady states is  observed 
 in microscopic models~\cite{KKl,KKW} where steady speed states of synchronized flow
 cover a 2D-region
in the flow density plane. 
Complex spatial-temporal states of synchronized flow have also been observed
by Fukui {\it et al.}
in their cellular automata model~\cite{Fukui}.

Strictly speaking, already small amplitude random perturbations in synchronized flow
 destroy steady states: Rather than steady states some complex 
spatial-temporal states can appear. However, if the perturbation amplitude is low enough
these complex states
can be very close to  steady states. 
These small amplitude
spatial-temporal states  should possess the 
features of steady states  discussed
above and which will be consider in other hypotheses of the
three-phase traffic theory below.

\subsubsection{Two Kinds of Nucleation Effects and Phase Transitions in Free Flow
\label{TwoKind_Sub}}

How should the traffic phase "synchronized flow" and the traffic phase
"wide moving jam" emerge in an initial free traffic flow on a homogeneous multi-lane road
(curve $F$ in Fig.~\ref{Diagram} (a))?
The following  hypothesis of the three-phase traffic theory
answers these questions 
(Figs.~\ref{Hypote1} and~\ref{Hypote2})~\cite{Kerner1999B,Kerner1999A}:

{\it At the same density in free flow  there may be two qualitatively different nucleation effects 
and the related two qualitative different first order phase transitions in free flow:}
\begin{enumerate}
\item [(1)]
The nucleation effect which is responsible for the moving jam's emergence in free flow, i.e., 
for the phase transition from free flow
to a wide moving jam (the F$\rightarrow $J transition). 
\item [(2)]
The nucleation effect which is responsible for 
 the F$\rightarrow $S transition. 
\end{enumerate}

The F$\rightarrow $S transition is observed in real free flow: This transition is responsible for the
onset of congestion in free flow~\cite{Kerner2002B}. The onset of congestion is acompanied by the
speed breakdown. This is called {\it the breakdown phenomenon} in free  flow (e.g.,~\cite{Hall1992}).
Thus, in the three-phase traffic theory, the well-known breakdown phenomenon 
(the onset of congestion) is explained by
 the 
F$\rightarrow $S transition~\cite{Kerner1998A,Kerner1999A}.
The empirical breakdown phenomenon, i.e.,  the 
F$\rightarrow $S transition is accompanied by a hysteresis effect. The hysteresis effect
is an attribute of the first order phase transition.

Although the spontaneous
F$\rightarrow $J transition is not observed, nevertheless 
the 
F$\rightarrow $J transition can be {\it induced} in free flow. This is confirmed by two empirical facts: (i)
The induced F$\rightarrow$J transition  has been observed in~\cite{Kerner2000B}. (ii)
Empirical maximum flow rate in free flow $q^{\rm (free)}_{\rm max}$ satisfies the condition:
\begin{equation}
 q^{\rm (free)}_{\rm max}>q_{\rm out}.
\end{equation}
This means that there are states of free flow where
the flow rate  $q>q_{\rm out}$. These states are
 metastable with respect to the wide moving jam emergence.

Thus,  both phase transitions
are confirmed by empirical data. 
These both first order phase transitions are also found in a microscopic three-phase traffic
 theory~\cite{KKl,KKW,KKl2003A}.

\subsubsection{Priority of  F$\rightarrow $S Transition in Free Flow~\cite{Kerner1999A,Kerner1999B}
}

\begin{figure}
\begin{center}
\includegraphics*[width=.6\textwidth]{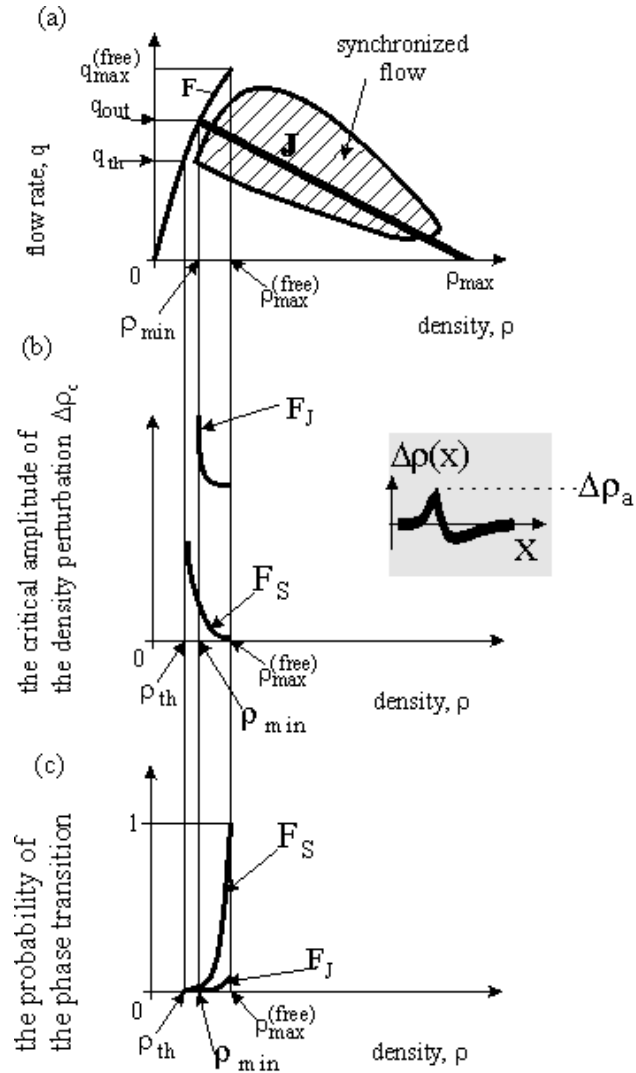}
\end{center}
\caption[]{Explanation of hypotheses to 
the three-phase traffic theory~\protect\cite{Kerner1999A,Kerner1999B}:
(a) - 
States of free (curve $F$) and synchronized 
flow (hatched region) which are 
 the same as in Fig.~\protect\ref{Diagram} (a).
(b) - Qualitative dependencies of the amplitude of the
critical density local perturbation  on the density.
(c) - Qualitative dependencies of the probability of the phase transitions
 In (b, c) the curve $F_{\rm S}$ is related to  the
F$\rightarrow$S transition and the curve
$F_{\rm J}$ is related to the
F$\rightarrow$J transition. 
In (b), right a form of a local density perturbation 
is schematically shown which grows if the amplitude of this
perturbation, $\Delta \rho_{\rm a}$, exceeds the critical 
amplitude $\Delta \rho_{\rm c}$ (links).
}
\label{Hypote1}
\end{figure}

The following hypothesis answers the question whether the F$\rightarrow $J transition
or the F$\rightarrow $S transition is more probable in free flow at the same density:

{\it At each given density in free flow
the critical amplitude of a local 
perturbation in the free flow 
which is needed 
for the 
 F$\rightarrow $S
transition (the curve $F_{\rm S}$     in 
Fig.~\ref{Hypote1} (b)) is considerably lower than the critical amplitude of a local 
perturbation which is needed for 
 the F$\rightarrow $J-transition 
(the curve $F_{\rm J}$     in Fig.~\ref{Hypote1} (b)).}

This hypothesis is confirmed by empirical observations: The onset of congestion in real free flow
is linked to the F$\rightarrow $S transition rather than to
 the F$\rightarrow $J transition~\cite{Kerner1998B,Kerner2002B}. 
This is also a result of a mathematical three-phase traffic theory
of the phase transitions in free flow
 (see Fig. 1 (b) in~\cite{KKl}).

Obviously, the higher the amplitude of a random local perturbation in a free flow, the lower
 the mean
probability (for a given time interval) of an occurrence of this
 perturbation. 
Thus,
at each given density 
in free flow 
the mean probability for a given time interval of an occurrence of the F$\rightarrow $S-transition, $P_{\rm FS}$
(the curve $F_{\rm S}$     in 
Fig.~\ref{Hypote1} (c)) should be 
considerably higher than the probability for the same time interval of 
the F$\rightarrow $J-transition, $P_{\rm FJ}$ (the curve $F_{\rm J}$     in Fig.~\ref{Hypote1} (c)).

\subsubsection{The Critical (Limit) Density in Free Flow~\cite{Kerner1999A,Kerner1999B}
\label{Limit} }

The following hypothesis explains the existence of the   critical (limit) point of free flow
$(\rho^{\rm (free)}_{\rm max}, \ q^{\rm (free)}_{\rm max})$ (Figs.~\ref{Diagram} (a) and~\ref{Hypote1} (a)):

{\it 
The existence of the critical (limit) point of free flow
$(\rho^{\rm (free)}_{\rm max}, \ q^{\rm (free)}_{\rm max})$  
is linked to the F$\rightarrow $S-transition rather than with the moving jam emergence.}
This hypothesis is related to the empirical result that in each free flow state, if 
the onset of congestion in this free flow occurs then this
is linked to the F$\rightarrow $S transition rather than to
 the F$\rightarrow $J transition~\cite{Kerner1998B,Kerner2002B}. The following hypotheses  also
result
from this empirical fact:

{\it  At the limit point of free flow
$(\rho^{\rm (free)}_{\rm max}, \ q^{\rm (free)}_{\rm max})$ the probability of the F$\rightarrow $S
transition, $P_{\rm FS}$ 
reaches one and 
 the critical amplitude
of the local perturbation 
for the F$\rightarrow $S-transition reaches zero}
(the curves $F_{\rm S}$ in Fig.~\ref{Hypote1} (b, c)).

{\it At the limit point of free flow
$(\rho^{\rm (free)}_{\rm max}, \ q^{\rm (free)}_{\rm max})$ the probability of the F$\rightarrow $J
transition  is very low
and consequently
 the critical amplitude of a local perturbation
 which is needed for the F$\rightarrow $J
transition is a relatively high finite value} 
(the curves $F_{\rm J}$ in Fig.~\ref{Hypote1} (b, c)).

These hypotheses are  also confirmed by a microscopic three-phase traffic theory~\cite{KKl,KKW,KKl2003A}
where the limit point of free flow is linked
to the  F$\rightarrow $S
transition rather than to the F$\rightarrow $J
transition.

In contrast to these results of the
three-phase traffic theory
and to results of empirical observations~\cite{Kerner2002B}, in the fundamental diagram approach
the breakdown phenomenon as well as
the limit point of free flow
$(\rho^{\rm (free)}_{\rm max}, \ q^{\rm (free)}_{\rm max})$ are explained by the moving jam emergence in free 
flow (see references in the reviews~\cite{Sch,Helbing2001,Nagatani_R,Nagel2003A}).

\subsection{Physics of Breakdown Phenomenon
\label{Z_FS_Sec}}

\subsubsection{Z-Shaped  Speed-Density Characteristic
\label{Z_FS}}

Because corresponding to the hypothesis discussed in
Sects.~\ref{TwoKind_Sub}-\ref{Limit} the empirical onset of congestion (the breakdown phenomenon)
in
an initial free flow is linked to
the F$\rightarrow $S
transition, we consider here the physics of this 
phase transition in more detail.
To do this, let us designate the critical amplitude of the critical local perturbation needed for the
F$\rightarrow$S transition as
$\Delta v^{\rm (FS)}_{\rm cr}$. For this critical amplitude we can write the formula
\begin{equation}
\Delta v^{\rm (FS)}_{\rm cr}=v^{\rm (free)}_{\rm initial}-v^{\rm (FS)}_{\rm cr},
\label{Am_Pert_Cr_Hom_For}
\end{equation}
where $v^{\rm (free)}_{\rm initial}$ is the speed in an initial free flow and
$v^{\rm (FS)}_{\rm cr}$ is the vehicle  speed inside the critical local perturbation.
At the critical (limit) point of free flow $(\rho^{\rm (free)}_{\rm max}, \ q^{\rm (free)}_{\rm max})$
the critical amplitude
\begin{equation}
\Delta v^{\rm (FS)}_{\rm cr}\mid_{\rho=\rho^{(free)}_{\rm max}}=0.
\label{Am_Pert_Cr_Hom_Cr_For}
\end{equation}
This means that at the limit point of free flow
the mean probability 
for the F$\rightarrow$S transition, $P_{\rm FS}$ is equal to
\begin{equation}
P_{\rm FS}\mid_{\rho=\rho^{(free)}_{\rm max}}=1.
\label{PFS1}
\end{equation}
This probability is defined as follows. We consider a large number of different realizations, $N_{\rm FS}$,
where the F$\rightarrow$S transition in an initial free flow are studied. Each of the realizations
should be performed at the same flow rate of free flow,   other intial conditions,
 and during the same time interval $T_{\rm ob}$
of  the observation of the spontaneous F$\rightarrow$S transition on a
chosen highway section of the length $L_{\rm ob}$. 
Let us assume that in $n_{\rm FS}$ of  these $N_{\rm FS}$ realizations
the F$\rightarrow$S transition occurs. Then, the mean probability of this transition
for the time interval $T_{\rm ob}$ and for the length of the road
$L_{\rm ob}$ is equal 
to\footnote{Strictly speaking the exact mean probability  of the spontaneous F$\rightarrow$S transition
 is determined when
$N_{\rm FS}\rightarrow \infty$.}
\begin{equation}
P_{\rm FS}=\frac{n_{\rm FS}}{N_{\rm FS}}.
\label{Prob_Def}
\end{equation}
 
When the density in free flow decreases, the critical amplitude
of the local perturbation for the
F$\rightarrow$S transition
 increases. There should be
a threshold point $(\rho_{\rm th}, \ q_{\rm th})$ for the
F$\rightarrow$S transition. This threshold point is defined as follows. Below the threshold point, i.e., at
$\rho<\rho_{\rm th}$ ($q<q_{\rm th}$) the mean probability of the
F$\rightarrow$S transition $P_{\rm FS}=0$. If the density in free flow is gradually increased
then the threshold point is related to the highest density 
$\rho=\rho_{\rm th}$ of free flow where this probability  still satisfies the condition:
\begin{equation}
P_{\rm FS}\mid_{\rho=\rho_{\rm th}}=0.
\label{TH}
\end{equation}
In this threshold point,
the critical amplitude of the local perturbation for the
F$\rightarrow$S transition
reaches a maximum value
\begin{equation}
\Delta v^{\rm (FS)}_{\rm cr}\mid_{\rho=\rho_{th}}=\Delta v^{\rm (FS)}_{\rm cr, \ max}.
\label{Am_Pert_Cr_Hom_th_For}
\end{equation}
This behavior of the critical amplitude of the local perturbation and of the
mean probability  of the
F$\rightarrow$S transition $P_{\rm FS}$ is shown by the curves $F_{\rm S}$ in Fig.~\ref{Hypote1} (b, c), respectively.

The curve $F$ for states of free flow, the critical branch $v^{\rm (FS)}_{\rm cr}$
(Fig.~\ref{Elementary_Z_FS_Hom})
which determines  the critical amplitude  (\ref{Am_Pert_Cr_Hom_For})
of 
local perturbation for the F$\rightarrow$S transition, and
 the 2D region for steady states of synchronized flow
 form  together 
a Z-shaped function of the speed of the density
(Fig.~\ref{Elementary_Z_FS_Hom} (b, c)).
This hypothesis about the Z-shaped traffic flow characteristic for the
breakdown phenomenon is confirmed in
the microscopic three-phase traffic flow theory (see Fig. 2 (a) in~\cite{KKl2003A}).

\subsubsection{Passing Probability}

This Z-shaped speed-density relationship
is qualitatively correlated with the hypothesis about the
mean probability of  passing $P$ on a multi-lane road
(Fig.~\ref{Hypote2}). In the three-phase traffic theory,
 the following hypothesis is 
valid~\cite{Kerner1999A,Kerner1999B,Kerner1999C}:

{\it The mean probability of  passing, $P$, on a multi-lane road is a Z-shaped function of  the
density} (Fig.~\ref{Hypote2} (b)).

At a very low vehicle density in free flow vehicles can freely pass. Thus, 
the probability of passing  $P$
in free flow should reach one at the density $\rho\rightarrow 0$
(the curve $P_{\rm F}$ in Fig.~\ref{Hypote2} (b)). To determine the probability of passing $P$,
a large number, $N_{\rm P}$,
 of different realizations (runs) for passing should be performed
at the same initial conditions and the same time interval 
 for passing. In each of these realizations, there should be
a driver who
moves with a speed higher than the speed of the vehicle ahead. 
When
approaching the vehicle ahead, the driver  should try to pass using a passing freeway lane.
It can occur that in some of the realizations the driver is able to pass, but
in the other realization the driver is not able to pass. The latter is
because the passing lane was occupied by other drivers.
If the number of realizations where the driver was able to pass is $n_{\rm P}$, then the probability of passing
is\footnote{Strictly speaking the exact probability of passing is determined when
$N_{\rm P}\rightarrow \infty$.}
\begin{equation}
P=\frac{n_{\rm P}}{N_{\rm P}}.
\end{equation}

To explain the  hypothesis about the Z-shaped form of the probability
of passing, note that in accordance with the fundamental 
hypothesis of the  three-phase traffic theory
 steady states of synchronized flow overlap
states of free flow in the vehicle density  
 within a density range (Fig.~\ref{Hypote2} (a))
\begin{equation} 
[\rho^{\rm (syn)}_{\rm max}, \ \rho^{\rm (free)}_{\rm max}].
\label{Overlap_Density_For}
\end{equation}

However, it is well-known that  the mean 
probability of passing in congested traffic (in our case, in one of the two
 traffic phases in congested traffic, $\lq\lq$synchronized flow$\lq\lq$) $P$ is considerably lower
than in free flow. Thus, in free flow the mean probability of passing, $P$
(the curve $P_{\rm F}$ in Fig.~\ref{Hypote2} (b))
should be higher than that in synchronized flow (the curve $P_{\rm S}$).
In the range of the density (\ref{Overlap_Density_For}), at the same given density there can be either a state 
of free flow where $P=P_{\rm F}$ is high or a steady state of synchronized flow
where $P=P_{\rm S}$ is low.
This leads to the Z-shaped function of  the mean probability of passing
in the three-phase traffic theory.

The Z-form of the dependence of the mean probability of passing  $P$
of the density
as well as the related Z-shape of the speed-density relationship
allows us to explain  the F$\rightarrow $S-transition on a multi-lane (in one direction) freeway.
If the density in free flow is gradually increasing then a drop
in the  mean probability of passing $P$ must occur
when the density reaches  the limit density for free flow
$\rho^{\rm (free)}_{\rm max}$   
(Fig.~\ref{Hypote2} (b)).
As a result of this drop the mean probability of passing $P$
decreases sharply  to low values of $P$
which are related to synchronized flow.
Thus, at the limit density for free flow
$\rho^{\rm (free)}_{\rm max}$ 
the spontaneous F$\rightarrow $S-transition must  occur.

The hypothesis about the Z-shaped mean probability of passing $P$
as a function of the vehicle density can also
be explained by the above empirical study of the WSP
where it has been shown that states of synchronized flow overlap with
states of free flow in the density (Figs.~\ref{Wide3} and~\ref{Wide4} (a); see also
Fig. 2 in~\cite{Kerner1999B}).
This means that at the
same density either a state of synchronized flow or a state of free
flow is possible. It is obvious that the mean probability of
passing is higher in free flow than in synchronized flow. Thus, the
empirical fact that states of free flow and synchronized flow overlap
in the vehicle density (Figs.~\ref{Wide3} and~\ref{Wide4} (a)) means
that the mean probability of passing should have a Z-shape. 
Indeed,
the lower the mean rate of passing,
the lower  
the mean probability of passing $P$ is. The mean rate of the passing
should decrease when the vehicle speed difference $\Delta v$
becomes lower.

However, it must be noted that this overlapping of states of free flow
and states of synchronized flow occurs in a narrow range of the
density in the vicinity of the limit density in free flow
$\rho^{\rm (free)}_{\rm max}$, i.e., when the average speed of synchronized
flow is relatively high (see Fig. 2 in~\cite{Kerner1999B} and Sect.~\ref{Wide_Sec}).
Therefore, if in some empirical data {\it only} synchronized flows of
a relatively low vehicle speed are observed then no overlapping of
states of free and synchronized flows could be found. As a result, some authors
make a conclusion that there is no overlapping of
states of free and synchronized flows, and as a result a
dependence of the mean probability of passing $P$ as a function of
the density should be a monotonous decreasing one (e.g.,~\cite{Helbing2002B}). However,
to make the correct conclusion about of whether an overlapping of
states of free and synchronized flows exists,  a more precise empirical study is necessary
(Sect.~\ref{Wide_Sec}).

\subsubsection{Competition between Over-Acceleration and Speed Adaptation}

The first order F$\rightarrow$S transition
may be explained by
 a {\it competition} between two contradicting tendencies
inside a
 local perturbation
 in an initial free flow:
\begin{description}
\item [(i)]
 A tendency to the initial free flow  due to  an $\lq\lq$over-acceleration$\lq\lq$. 
\item [(ii)] 
A tendency to a synchronized flow
due to the vehicle speed adaptation to the speed of the vehicle
ahead.
\end{description}
This hypothesis is confirmed by the microscopic
three-phase traffic theory~\cite{KKl2003A}.

The vehicle $\lq\lq$over-acceleration$\lq\lq$  can occur
due to
 passing:  For  passing the vehicle
has usually
to increase
its  speed.
The
tendency to synchronized flow
can occur
 due to the need in the adaptation of the vehicle speed to the speed 
of the leading vehicle when  passing is not possible or  is difficult. 

\begin{figure}
\begin{center}
\includegraphics*[width=.7\textwidth]{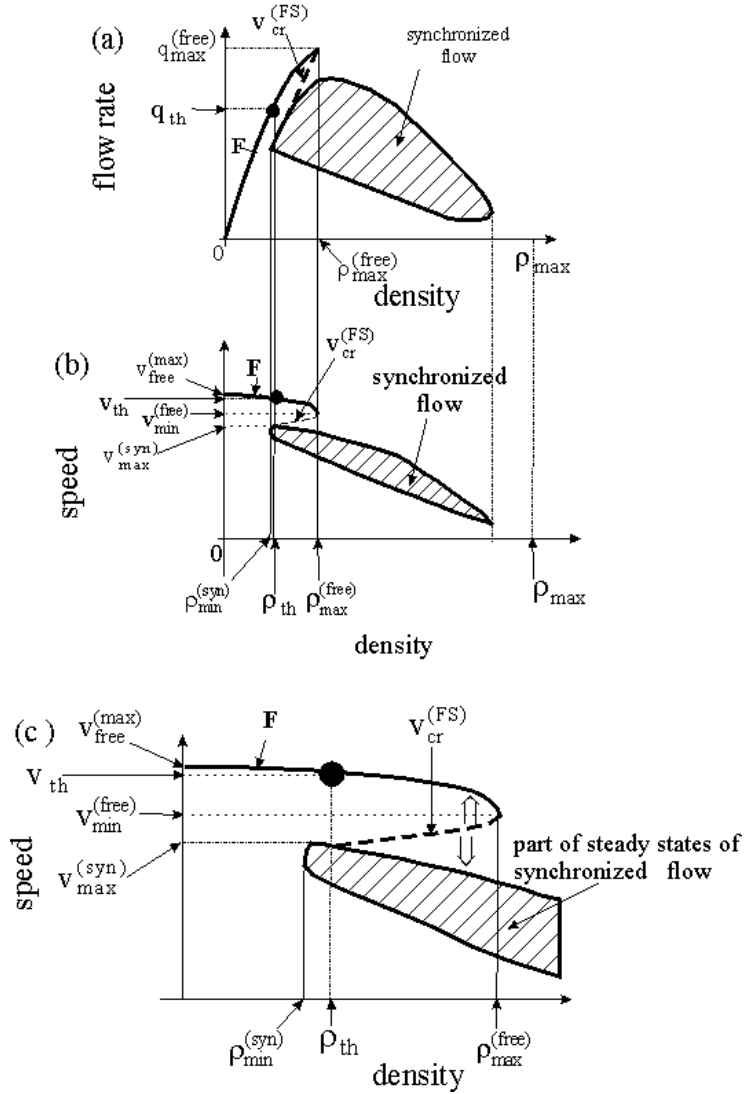}
\end{center}
\caption[]{Qualitative illustration of the F$\rightarrow$S transition
(the breakdown phenomenon):
The critical branch $v^{\rm  (FS)}_{\rm  cr}$ (dashed curve) gives the speed inside
the critical local perturbation,
 the threshold density $\rho_{\rm  th}$
and the threshold
 flow rate $q_{\rm  th}$ in free flow 
for the F$\rightarrow$S transition
on a homogeneous road in the flow-density
plane (a) and in the speed-density plane (b, c).
In (a)  states of free and synchronized flows
are taken from Fig.~\ref{Diagram} (a).
In (b) states of free and synchronized flows are related to (a).
In (c) a part of figure (b) for lower density range in a higher scale is shown.
The black point in (a-c) on the curve $F$ for free flow
shows the threshold point $\rho=\rho_{\rm th}$ for
the F$\rightarrow$S transition.
}
\label{Elementary_Z_FS_Hom}
\end{figure}

\begin{figure}[*]
\begin{center}
\includegraphics[width=.45\textwidth]{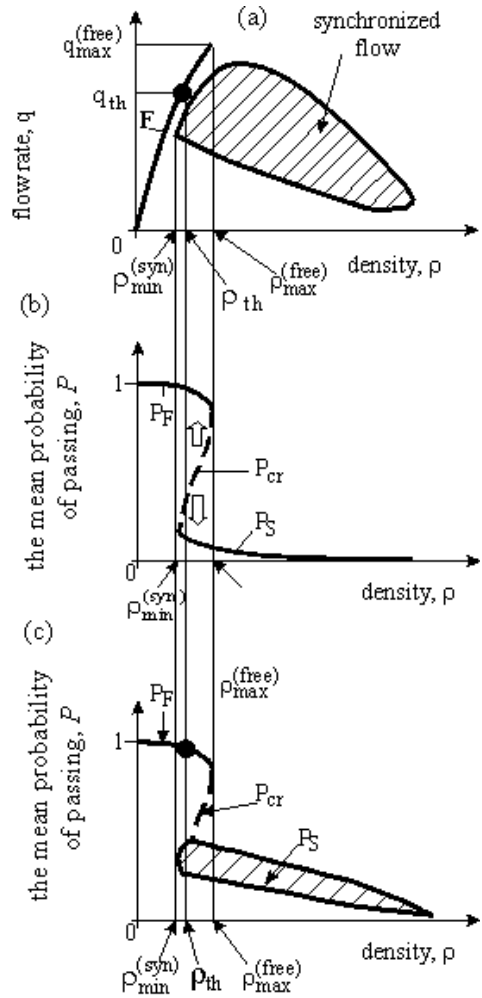}
\end{center}
\caption[]{Explanation of the hypothesis
about the Z-shaped dependence of the
 the mean probability of the passing $P$
in the three-phase traffic theory~\protect\cite{Kerner1999A,Kerner1999B}: 
(a) - 
States of free (curve $F$) and synchronized 
flow (hatched region) which are 
 the same as in Fig.~\ref{Diagram} (a).
(b) - Qualitative dependence of the mean probability of the passing  $P$
(which is averaged over all different steady states of synchronized flow
at a given density) as a function of the density.
(c) - Qualitative dependence of the mean probability of the passing $P$
as a function of the density without averaging of the mean probability for all different
steady states of synchronized flow at a given density.
The black point in (a, c) is related to the threshold point for the F$\rightarrow$S transition).
}
\label{Hypote2}
\end{figure}

The $\lq\lq$over-acceleration$\lq\lq$ is stronger at a
higher vehicle speed, exactly lower density.
In this case, an initial local perturbation in free flow decays
(up-arrow in Fig.~\ref{Elementary_Z_FS_Hom} (c)).
The over-acceleration is also responsible for the phase transition from synchronized flow to
free flow (the S$\rightarrow$F transition).
 
In contrast, the tendency to the speed adaptation 
is stronger at a lower speed, i.e., at a higher density. 
This causes a  decrease in the average vehicle speed  inside the initial
perturbation (down-arrow in Fig.~\ref{Elementary_Z_FS_Hom} (c)) 
and a self-maintenance of the emerging synchronized flow.

The curve $P_{\rm S}$ in Fig.~\ref{Hypote2} (b)
is
 related to some simplifications: These curves are  an averaging of all different
synchronized flow speeds at a given density to
one average speed.
If we consider all these different steady state synchronized flow speeds
(as it has been made in Fig.~\ref{Elementary_Z_FS_Hom} (b, c))
we naturally come to Fig.~\ref{Hypote2} (c) 
for the mean
probability of passing which reflects
the  2D region of steady states of synchronized flow.

\subsection{Why Moving Jams do not Emerge in Free Flow}

In the three-phase traffic theory, wide moving jams do not spontaneously emerge
in free flow (Sect.~\ref{TwoKind_Sub}). 
This hypothesis is related to results of empirical observations~\cite{Kerner1998B,Kerner2000B}
 and it is also 
confirmed in the mathematical three-phase traffic theory~\cite{KKl,KKW,KKl2003A}.

\begin{figure}
\begin{center}
\includegraphics*[width=.75\textwidth]{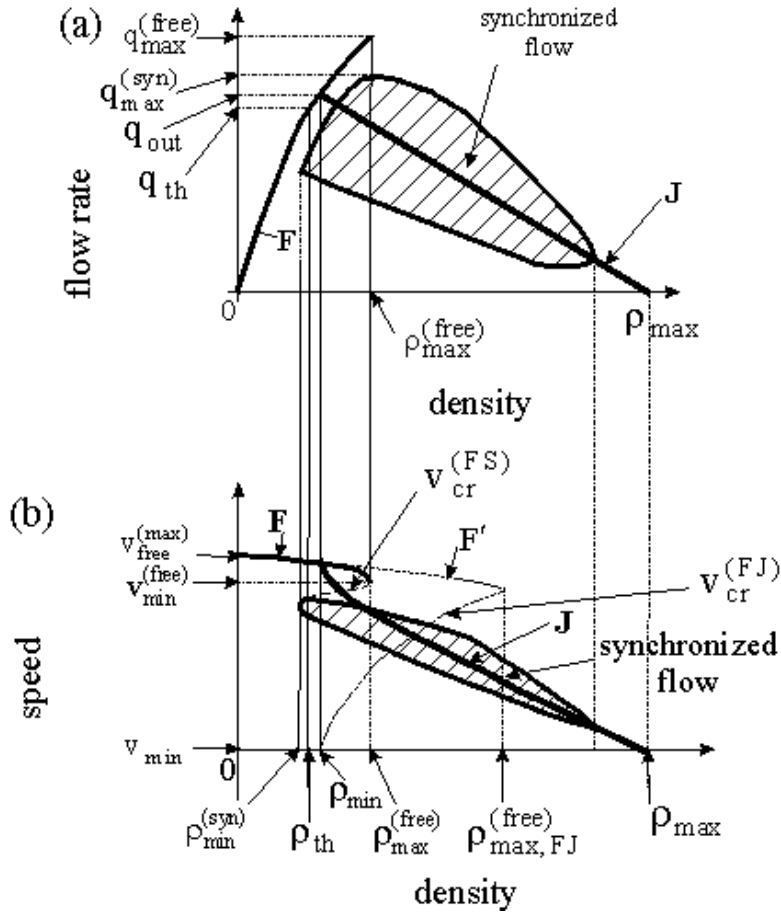}
\end{center}
\caption[]{Explanation of the  question why
  moving jams do not
emerge in free flow
in the
 three-phase traffic theory~\protect\cite{Kerner1998B}:
(a) - A qualitative concatenation of states of free flow (curve $F$), steady states of synchronized 
flow (hatched region), the critical  branch $v^{\rm (FS)}_{\rm  cr}$ with Line $J$ in the flow-density plane.
(b) -  states  in speed-density plane related to (a).
In (a, b)
states of free flow (curve $F$), steady states of synchronized 
flow (hatched region) and the critical  branch $v^{\rm (FS)}_{\rm  cr}$ are 
 taken from Fig.~\protect\ref{Elementary_Z_FS_Hom} (a) and
(b), respectively.
}
\label{Elementary_FJ_Hom}
\end{figure}

To explain this hypothesis of the three-phase traffic theory,
we
present  the line $J$ in the flow-density plane together with states of free flow and
steady states of synchronized flow as  is now shown in
Fig.~\ref{Elementary_FJ_Hom} (a).
The line $J$  in the flow-density plane (Fig.~\ref{Elementary_FJ_Hom} (a))
is related to
a parabolic function  in the speed-density plane 
(curve $J$ in Fig.~\ref{Elementary_FJ_Hom} (b)).

We suggest
that besides
the critical point of 
 the F$\rightarrow$S transition,  
$(\rho^{\rm (free)}_{\rm max}, \ q^{\rm (free)}_{\rm max})$,
there is another hypothetical 
critical point in free flow,
$(\rho^{\rm (free)}_{\rm max, \ FJ}, \ q^{\rm (free)}_{\rm max, \ FJ})$,
where
the critical amplitude of the critical perturbation
which is needed for  the F$\rightarrow$J transition
is
zero (Fig.~\ref{Elementary_FJ_Hom} (b)).
However, the density in free flow which is related to this 
critical density
$\rho^{\rm (free)}_{\rm max, \ FJ}$
is {\it much higher}
than the critical density
$\rho^{\rm (free)}_{\rm max}$
where
 the F$\rightarrow$S transition must occur.
This hypothesis is confirmed by the microscopic
three-phase traffic theory~\cite{KKl,KKl2003A}.

The density of
free flow $\rho^{\rm (free)}$, which is related to these
hypothetical free flow states
$\rho^{\rm (free)}_{\rm max}< \rho^{\rm (free)}\leq \rho^{\rm (free)}_{\rm max, \ FJ}$,
can not be reached in reality (states $F^{\prime}$, Fig.~\ref{Elementary_FJ_Hom} (b)). 
This is
because at the density 
$\rho^{\rm (free)}=\rho^{\rm (free)}_{\rm max}$
synchronized flow must occur.
 There should be
a critical branch
$v^{\rm (FJ)}_{\rm cr}$ in the speed-density plane
(Fig.~\ref{Elementary_FJ_Hom} (b)). 
The critical branch
$v^{\rm (FJ)}_{\rm cr}$ gives the speed inside the
critical perturbation, i.e., it determines  the
critical amplitude of the critical local perturbation $\Delta v^{\rm (FJ)}_{\rm cr}$
for the 
F$\rightarrow$J transition: 
$\Delta v^{\rm (FJ)}_{\rm cr}=v^{\rm (free)}-v^{\rm (FJ)}_{\rm cr}$.
Here $v^{\rm (free)}$ is the speed in free flow.

At the critical density
 $\rho^{\rm (free)}_{\rm max, \ FJ}$
the critical amplitude of the critical perturbation
for the F$\rightarrow$J transition
is
zero:
$\Delta v^{\rm (FJ)}_{\rm cr}=0$.
Thus, 
at the density $\rho^{\rm (free)}=\rho^{\rm (free)}_{\rm max, \ FJ}$
the branch
$v^{\rm (FJ)}_{\rm cr}$
should
merge
with
the branch  of states of free flow
$F^{\prime}$.
At lower density,
$\rho^{\rm (free)}<\rho^{\rm (free)}_{\rm max, \ FJ}$,
the critical amplitude of the critical perturbation
for the F$\rightarrow$J transition
should increase with the decrease in density.
At the threshold
point of free flow
for the F$\rightarrow$J transition, $(\rho_{\rm min}, \ q_{\rm out})$,
the critical local perturbation with the highest amplitude is needed where the speed is equal to
the speed inside a wide moving jam
$v_{\rm min}=0$
(Fig.~\ref{Elementary_FJ_Hom} (b)).

Therefore, we see that
the critical amplitude of the local perturbation
which is needed for the moving jam emergence in an initial free flow
$\Delta v^{\rm (FJ)}_{\rm cr}$
 at each density in free flow
is  higher than
the critical amplitude of the local perturbation
which is needed for the synchronized emergence in free flow
$\Delta v^{\rm (FS)}_{\rm cr}$ (\ref{Am_Pert_Cr_Hom_For}). 
This theoretical conclusion may explain why
the spontaneous emergence of moving jams is not observed in free flow~\cite{Kerner1998B,Kerner2002B}.

\subsection{Explanation of
F$\rightarrow $S$\rightarrow $J Transitions:
\protect\newline Double Z-Shaped   Traffic Flow Characteristics}
\label{Elementary_2Z_Sec}

\begin{figure}
\begin{center}
\includegraphics*[width=.85\textwidth]{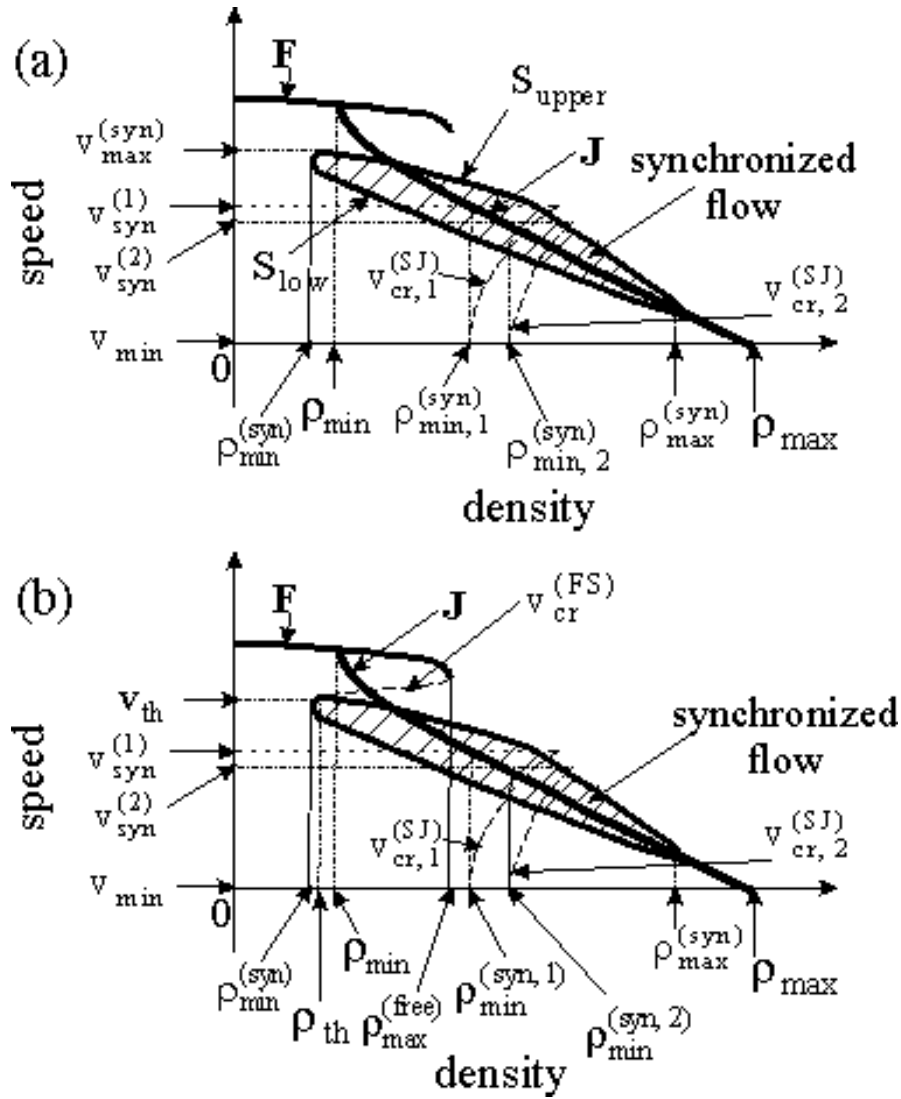}
\end{center}
\caption[]{Explanation of the  critical branches $v^{\rm (SJ)}_{\rm  cr, \ 1}$ and $v^{\rm (SJ)}_{\rm cr, \ 2}$
 for the S$\rightarrow $J-transition
for two different steady synchronized flow speeds
$v^{\rm (1)}_{\rm syn}$ and
$v^{\rm (2)}_{\rm syn}$, respectively.
States for free flow (curve $F$), steady states of synchronized flow
(dashed region) and the curve $J$ are taken from
Fig.~\ref{Elementary_FJ_Hom}b.
}
\label{Elementary_2Z_SJ_Hom}
\end{figure}

Let us first show that the S$\rightarrow $J-transition considered in Sect.~\ref{FJ_Hyp} 
is related to a Z-shaped function in the speed-density plane
(Fig.~\ref{Elementary_2Z_SJ_Hom} (a)). To explain this hypothesis, we consider
the speeds $v^{\rm (1)}_{\rm syn}$ and
$v^{\rm (2)}_{\rm syn}$ from 
Fig.~\ref{Hypote3}.
The threshold densities $\rho^{\rm (syn)}_{\rm min \ 1}$ and
$\rho^{\rm (syn)}_{\rm min \ 2}$ for
the S$\rightarrow $J-transition (Fig.~\ref{Hypote3})
correspond to the intersection
points between the horizontal line of the speeds $v^{\rm (1)}_{\rm syn}$ and
$v^{\rm (2)}_{\rm syn}$ with the curve $J$ in the speed-density plane
(Fig.~\ref{Elementary_2Z_SJ_Hom} (a)).
Because there is an infinite number of different speeds in
synchronized flow, there is also an  infinite number
of  threshold densities for these different
synchronized flow speeds. There is also  an infinite number of the related
critical branches
$v^{\rm (SJ)}_{\rm cr}$ each of then gives the speed inside
the critical local perturbation for S$\rightarrow $J-transition
(only two of them,
$v^{\rm (SJ)}_{\rm cr, \ 1}$ and
$v^{\rm (SJ)}_{\rm cr, \ 2}$, are shown in 
Fig.~\ref{Elementary_2Z_SJ_Hom} (a)
for the speeds $v^{\rm (1)}_{\rm syn}$ and
$v^{\rm (2)}_{\rm syn}$, respectively).
The  critical amplitude of the critical perturbation in the speed
for the S$\rightarrow$J transition is equal to
$\Delta v^{\rm (SJ)}_{\rm cr}=
v_{\rm syn}-v^{\rm (SJ)}_{\rm cr}$,
where $v_{\rm syn}$ is one of the infinite possible synchronized flow speeds.

At the threshold density
 the critical branch $v^{\rm (SJ)}_{\rm cr}$ should merge with the speed 
inside a wide moving jam $v_{\rm min}=0$ in the speed-density plane
(points $(\rho^{\rm (syn)}_{\rm min \ 1}, \ 0)$
and $(\rho^{\rm (syn)}_{\rm min \ 2}, \ 0)$ in Fig.~\ref{Elementary_2Z_SJ_Hom} (a)),
i.e., $\Delta v^{\rm (SJ)}_{\rm cr}=
v_{\rm syn}$.
At  critical points
for the S$\rightarrow$J transition where the critical amplitude
$\Delta v^{\rm (SJ)}_{\rm cr}=0$ the critical branch $v^{\rm (SJ)}_{\rm cr}$ merges with a
horizontal line in the speed-density plane. This horizontal line is related 
to a given
synchronized flow speed $v_{\rm syn}$
(e.g., the horizontal lines are related to the speeds
$v^{\rm (SJ)}_{\rm cr, \ 1}$ and
$v^{\rm (SJ)}_{\rm cr, \ 2}$ in 
Fig.~\ref{Elementary_2Z_SJ_Hom} (a)
for the critical branches
$v^{\rm (SJ)}_{\rm cr, \ 1}$ and $v^{\rm (SJ)}_{\rm cr, \ 2}$, respectively).
The synchronized flow states (dashed region in Fig.~\ref{Elementary_2Z_SJ_Hom} (a)), the critical branch
$v^{\rm (SJ)}_{\rm cr}$, and the line $v_{\rm min}=0$ for the speed inside a wide
moving jam give together a Z-shaped characteristic for
the S$\rightarrow$J transition (Fig.~\ref{Elementary_2Z_SJ_Hom} (a)).
This Z-characteristic is indeed found in the microscopic three-phase traffic theory~\cite{KKl2003A}.

If now
we  add
the critical branch for the
F$\rightarrow $S transition 
$v^{\rm (FS)}_{\rm cr}$
which has been considered in Sect.~\ref{Z_FS} 
 to
Fig.~\ref{Elementary_2Z_SJ_Hom} (a), we come to 
a
{\it double Z-shaped} characteristic of the speed on the
density (Fig.~\ref{Elementary_2Z_SJ_Hom} (b)).
This
explains the 
empirical result~\cite{Kerner1998B} that moving jam spontaneously emerge
due to a sequence
(cascade) of two phase transitions, first 
the F$\rightarrow $S transition, and later the S$\rightarrow $J
transition
(the F$\rightarrow $S$\rightarrow $J-transitions).
The 
double Z-shaped characteristic consists of states of free flow
$F$, the critical branch $v^{\rm (FS)}_{\rm cr}$ which gives the speed inside the critical perturbation for
the F$\rightarrow $S transition, the two-dimensional region of steady states of synchronized flow,
the infinite number of critical branches $v^{\rm (SJ)}_{\rm cr}$ which give the speeds inside
 the critical perturbations for
the S$\rightarrow $J transition for each of the synchronized flow speeds,
and the line $v_{\rm min}=0$ which gives the speed inside wide moving jams.
The double Z-characteristic is also confirmed in the microscopic
three-phase traffic theory~\cite{KKl2003A}.

\subsection{Phase Transitions and Patterns at Highway Bottlenecks  \label{Bottl_Sec} }

\subsubsection{Deterministic Perturbation and Local Breakdown at Bottlenecks}

Both the phase transitions and spatial-temporal patterns which occur
at highway bottlenecks
possess important  peculiarities in comparison with
a homogeneous (without bottlenecks) road considered above.
In particular, 
the following hypothesis
is related to this case~\cite{Kerner2000A}:

{\it The probability of the F$\rightarrow $S-transition, $P_{\rm FS}$ (for a highway section length and for
a time interval)
depends on the highway location.
This probability
has a maximum at  the bottleneck}
(see Fig. 1 (b) in~\cite{Kerner2000A}).
Note that the road location where this maximum is reached is called the effective location of the bottleneck 
(or the effective bottleneck
for short).

This hypothesis is related to the empirical result that
the onset of congestion (the F$\rightarrow $S-transition) is usually observed
at bottlenecks. This can also be  explained from driver's behavior.
Let us consider a road where only one bottleneck exists.
Upstream and downstream of the bottleneck the road
has the same characteristics and it is homogeneous. A feature of the bottleneck is that
in the vicinity of the bottleneck each driver should slow down, i.e., decrease the speed.
Thus, a {\it deterministic} local perturbation in the average speed appears at the bottleneck.
This perturbation is motionless and permanent because it is localized at the
highway bottleneck: The bottleneck forces all drivers permanently 
to slow down at approximately the same road location.

The deterministic perturbation is motionless.
Thus, the whole flow rate across the  road (when taking into account all
possible on- and off-ramps) 
 does not  depend on the highway location, i.e., this whole flow rate
remains the same inside the perturbation at the bottleneck  
and downstream of  the bottleneck.  Because inside the deterministic 
perturbation in free flow
the speed is lower than the speed away from the bottleneck but the flow rate does not change,
 the density must be higher
inside the perturbation.
This explains why the probability of the F$\rightarrow $S transition
as a function of freeway location should
have a maximum at  the bottleneck.
This hypothesis is also confirmed by  results of numerical simulations made in~\cite{KKl2003A}. 

Further we consider a bottleneck due to an on-ramp where the flow rate to the
on-ramp is $q_{\rm on}$ and the flow rate on the main road downstream of the on-ramp
is $q_{\rm on}$. Consequently, if free flow is at the bottleneck then the flow rate downstream of the
bottleneck is equal to
\begin{equation}
q_{\rm sum}=q_{\rm in}+q_{\rm on}.
\label{Sum_Flow_For}
\end{equation}

The higher the density in free flow, the higher the
amplitude of the
local deterministic perturbation.
 The growth of this  deterministic perturbation
should have a limit:
There should be some 
critical amplitude 
of the deterministic perturbation in the speed. 
When this  critical deterministic perturbation at the on-ramp is achieved the
{\it deterministic local speed breakdown} 
(the F$\rightarrow $S transition) spontaneously occurs:
The speed decreases  and the density increases avalanche-like at the location of the initial
deterministic perturbation.
This hypothesis is also confirmed by results of numerical simulations~\cite{KKl2003A}.
This deterministic 
F$\rightarrow$S transition
is realized even if no random perturbations
(fluctuations)
would be in free traffic flow
(the deterministic 
F$\rightarrow$S transition is symbolically marked by 
dotted arrows $K_{\rm determ}\rightarrow M_{\rm determ}$ in Fig.~\ref{Elementary_H_FS_Deter} (a, b)).
This phase transition together with the return S$\rightarrow$F transition 
(arrow $N\rightarrow P$ in Fig.~\ref{Elementary_H_FS_Deter} (a, b))
may explain the well-known hysteresis
in traffic at bottlenecks.

It must be stressed that
 the deterministic 
F$\rightarrow$S transition
occurs at a {\it lower} 
 flow rate $q_{\rm sum}=q^{\rm (B)}_{\rm cr, \ FS}$ 
downstream of the on-ramp
than the critical traffic variables
on a homogeneous road:
\begin{equation}
q^{\rm (B)}_{\rm cr, \ FS}<q^{\rm (free)}_{\rm max}
\quad (\rho^{\rm (B)}_{\rm cr, \ FS}<\rho^{\rm (free)}_{\rm max}).
\end{equation}
These
 formulae  
may explain the  empirical result
that the breakdown phenomenon
(the
F$\rightarrow$S transition) is most frequently
observed
at freeway bottlenecks.

\subsubsection{Double Z-Characteristics at Bottlenecks}

\begin{figure}
\begin{center}
\includegraphics*[width=.45\textwidth]{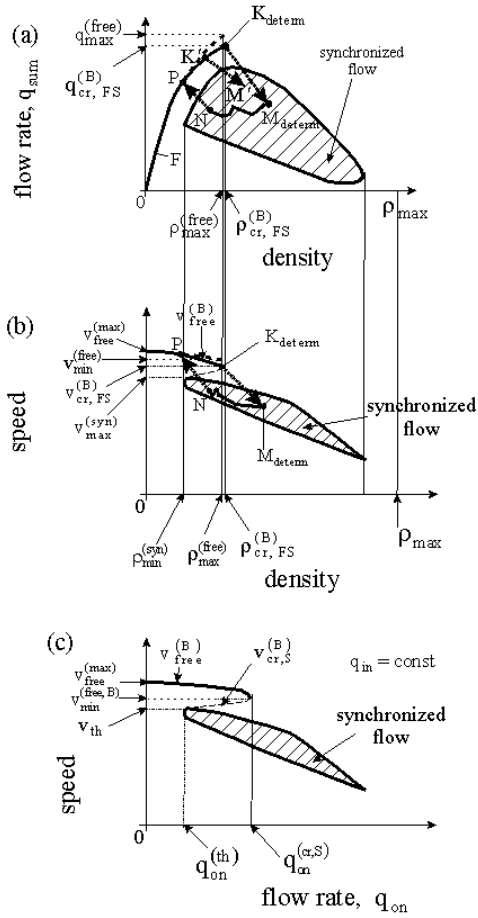}
\end{center}
\caption[]{Qualitative illustration of  the deterministic breakdown phenomenon
 (the
deterministic F$\rightarrow$S transition)
in an initial free flow at a freeway bottleneck:
(a) - States free flow (solid curves $F$ in (a) and $v^{\rm (B)}_{\rm free}$ in (b) related to
the bottleneck;
dashed branches show states of free flow on a homogeneous road) and
 steady states of
synchronized flow (dashed region) in the flow-density plane.
(b) -   related to  speed-density 
characteristics of (a). The steady states of  synchronized flow
are taken from
Fig.~\ref{Diagram} (a).
(c) - Qualitative illustration of a Z-shaped
speed-flow characteristic for the spontaneous  breakdown phenomenon
 (the
spontaneous F$\rightarrow$S transition)
in an initial free flow at a freeway bottleneck
due to an on-ramp.
States free flow (curve $v^{\rm (B)}_{\rm free}$) and
steady states of
synchronized flow (dashed region) in the flow-density plane
together with the critical branch $v^{\rm (B)}_{\rm cr, \ FS}$
(dashed curve) 
which gives the speed inside the random component of  the critical perturbation.
}
\label{Elementary_H_FS_Deter}
\end{figure}

Real {\it random}
perturbations in free flow in the bottleneck
vicinity   can  lead to the occurrence
of the
spontaneous 
F$\rightarrow$S transition
even if the critical point of free flow 
has not yet been achieved, i.e., at
\begin{equation}
q_{\rm sum}<q^{\rm (B)}_{\rm cr, \ FS} \quad
(\rho^{\rm (B)}_{\rm free}(q_{\rm sum}) <\rho^{\rm (B)}_{\rm cr, \ FS}).
\end{equation}
This spontaneous 
F$\rightarrow$S transition is symbolically marked by 
dotted arrows $K^{\prime}\rightarrow M^{\prime}$ in Fig.~\ref{Elementary_H_FS_Deter} (a).
Thus, as well as in the case of a hypothetical homogeneous road 
(Sect.~\ref{Z_FS}) we
come to a Z-shape characteristic of the speed on the density
for the onset of congestion, i.e., for the F$\rightarrow$S transition
at a bottleneck due to the on-ramp
(Fig.~\ref{Elementary_H_FS_Deter} (b)).
In this case, however, the density at the bottleneck is a function of two flow rates
$q_{\rm on}$
and $q_{\rm in}$.
Thus, it  is more convenient to use another Z-characteristic which is related to the dependence of
the speed at the bottleneck on the flow rate to the on-ramp $q_{\rm on}$
at a given flow rate $q_{\rm in}$ (Fig.~\ref{Elementary_H_FS_Deter} (c)).
In this Z-characteristic, the branch $v^{\rm (B)}_{\rm free}$ gives the speed inside
the deterministic perturbation in free flow
at the bottleneck whereas the branch $v^{\rm (B)}_{\rm cr, \ S}$  gives the speed inside
the critical perturbation with the critical amplitude
\begin{equation}
\Delta v^{\rm (B)}_{\rm cr, \ S}=v^{\rm (B)}_{\rm free}-v^{\rm (B)}_{\rm cr, \ S}.
\label{delta_v_cr_d}
\end{equation}
These Z-characteristics at the on-ramp are confirmed in numerical simulations~\cite{KKl2003A}.

It is well-known that the  breakdown phenomenon
at the bottleneck possesses a probabilistic nature~\cite{El,Persaud}.
This  also follows from the Z-characteristic for
the F$\rightarrow$S transition
(the breakdown phenomenon) at the bottleneck. Indeed,
at the critical point $q_{\rm sum}=q^{\rm (B)}_{\rm cr, \ FS}$ (Fig.~\ref{Elementary_H_FS_Deter} (a))
the critical amplitude of the local perturbation needed  
for
the F$\rightarrow$S transition is zero. Thus,
the probability of the F$\rightarrow$S transition at the bottleneck, $P^{\rm (B)}_{\rm FS}=1$.
$P^{\rm (B)}_{\rm FS}$ is defined similar to the probability $P_{\rm FS}$ (\ref{Prob_Def}).
However, rather than a  highway
section of length $L_{\rm ob}$ 
which has been considered
for the definition of $P_{\rm FS}$ on the homogeneous road, $P^{\rm (B)}_{\rm FS}$  is  the mean probability of the
 F$\rightarrow$S transition 
 for a given time interval $T_{\rm ob}$ on the main road in the bottleneck vicinity.  
Because, the critical amplitude of the local perturbation
$\Delta v^{\rm (B)}_{\rm cr, \ S}$ (\ref{delta_v_cr_d})
increases when the flow rate $q_{\rm on}$ and consequently 
the flow rate $q_{\rm sum}$ (\ref{Sum_Flow_For})
 decrease
(Fig.~\ref{Elementary_H_FS_Deter} (c)),
the probability $P^{\rm (B)}_{\rm FS}$ should decrease.
Such dependence  of the mean probability of the
 F$\rightarrow$S transition 
 of the flow rate is observed in
empirical observations~\cite{Persaud}.

After the F$\rightarrow$S transition
at a bottleneck has occurred, synchronized flow appears
at the bottleneck. The upstream front of this synchronized flow propagates
upstream, i.e., the region
of synchronized flow is widening. If the speed in the synchronized flow
becomes low enough, moving jams can emerge in this synchronized flow.
The moving jam emergence occurs
away from the bottleneck and it is independent of the reason of the synchronized flow
occurrence. This hypothesis is confirmed by empirical results~\cite{Kerner2002B}.
The S$\rightarrow$J transition in this synchronized flow should also possess
the same Z-characteristic  which has been considered for the homogeneous
road in Sect.~\ref{Elementary_2Z_Sec}
(Fig.~\ref{Elementary_2Z_SJ_Hom} (a)). This hypothesis is also confirmed 
 by numerical simulations~\cite{KKl2003A}.

\begin{figure}
\begin{center}
\includegraphics*[width=.6\textwidth]{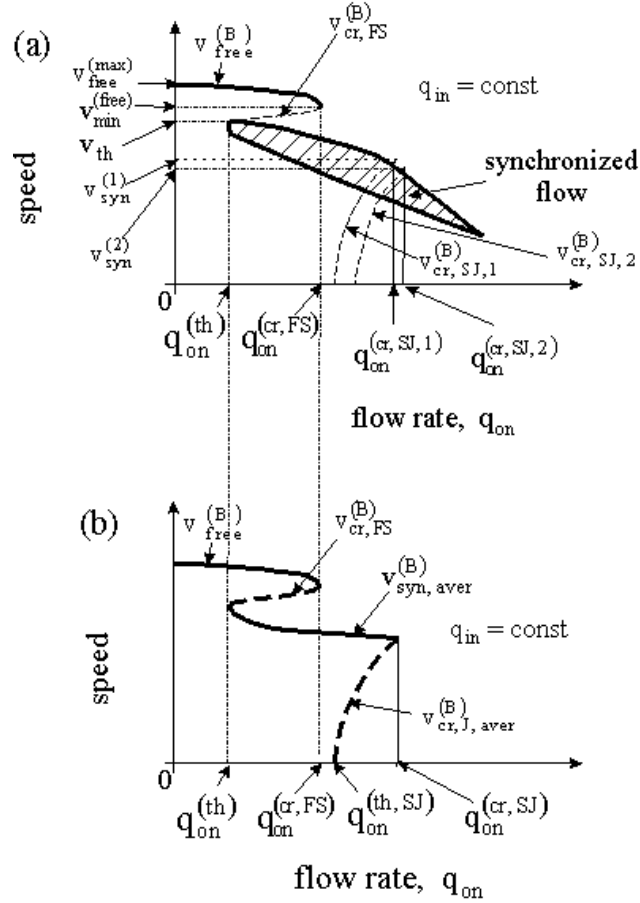}
\end{center}
\caption[]{Qualitative illustration of all possible phase transitions 
and double Z-shaped characteristics of traffic flow
at the 
 on-ramp:
(a) - States of free flow $v^{\rm (B)}_{\rm free}$, the critical branch
$v^{\rm (B)}_{\rm cr, \ FS}$
which gives the speed inside the critical 
local perturbation for the
spontaneous F$\rightarrow$S transition,
 a two-dimensional region of
steady states of synchronized flow
(dashed region),
  two critical branches
$v^{\rm (B)}_{\rm cr, \ SJ, \ i}, \ i=1,2$
which gives the speed inside the critical 
local perturbation for the
spontaneous
S$\rightarrow$J transitions for the related two speeds of synchronized flow
$v^{\rm (1)}_{\rm syn \ i}, \ i=1,2$, and the speed $v_{\rm min}=0$ inside wide
moving jams.
(b) - A simplified double Z-shaped speed-flow dependence
related to (a) 
where all
different synchronized flow speeds at a given flow rate $q_{\rm on}$ are averaged.
In (a) states of free flow
(the curve $v^{\rm (B)}_{\rm free}$),
states of synchronized flow
(dashed region), and
 the critical branch $v^{\rm (B)}_{\rm cr, \ FS}$
are taken from
Fig.~\ref{Elementary_H_FS_Deter} (c).
The critical branches $v^{\rm (SJ)}_{\rm  cr, \ 1}$ and $v^{\rm (SJ)}_{\rm cr, \ 2}$
are taken from
Fig.~\ref{Elementary_2Z_SJ_Hom}.
}
\label{E_2Z_FSJ_On}
\end{figure}

This consideration allows us also to expect  relatively complex double Z-shaped
traffic flow characteristics
at the bottleneck (Fig.~\ref{E_2Z_FSJ_On} (a)).
On this summarized figure critical branches for the F$\rightarrow$S transition and the
S$\rightarrow$J transition are shown
(see the caption to 
Fig.~\ref{E_2Z_FSJ_On}).
If we simplify this figure by an averaging of all different states
of synchronized flow to the one speed at a given flow rate $q_{\rm on}$
 we can find much more simple
double Z-shaped
traffic flow characteristics
at the bottleneck  (Fig.~\ref{E_2Z_FSJ_On} (b)). 

The hypothesis about the double Z-characteristics at the bottleneck
explains the empirical F$\rightarrow$S$\rightarrow$J transitions~\cite{Kerner2002B}.
This hypothesis is also confirmed by numerical simulations in the frame of the
three-phase traffic theory~\cite{KKl2003A}. In contrast, {\it no}
traffic flow theories and models in the
fundamental diagram approach~\cite{Sch,Helbing2001,Nagatani_R,Nagel2003A} 
can show the empirical F$\rightarrow$S$\rightarrow$J transitions
and the theoretical double Z-characteristics for traffic flow.

\subsubsection{Diagram of Congested Patterns \label{DCP}}

The qualitative difference of traffic flow theories and models in the
fundamental diagram approach~\cite{Sch,Helbing2001,Nagatani_R,Nagel2003A} and of the
three-phase traffic
theory
\cite{Kerner1999A,Kerner1998B,Kerner1998C,Kerner1999B,Kerner1999C}
can essentially clear be seen if the diagrams of congested patterns at
highway bottlenecks which should occur in these different approaches
are compared~\cite{Kerner2002B,KKl,KKW,KKl2003A}.

Recall that in the diagram of congested patterns at the on-ramp in the
fundamental diagram approach which has first been found  by Helbing
{\it et al.}  in~\cite{Helbing1999} diverse congested patterns are
possible depending on the initial flow rate $q_{\rm in}$ on a highway
upstream of the bottleneck and on the flow rate to the  on-ramp $q_{\rm on}$.

The diagram of the congested patterns at on-ramps within the
three-phase traffic theory has first been postulated by
the author~\cite{Kerner2002B,Kerner2002} based on qualitative
considerations and then derived by Kerner and Klenov based on their
microscopic traffic flow model~\cite{KKl}. It has already been shown in~\cite{KKl,KKW}
that  this
diagram is totally qualitatively different from the diagram of congested
patterns at on-ramps in the fundamental diagram
approach~\cite{Helbing1999,Lee1999,Helbing2000,Helbing2001,Nagatani_R,Nagel2003A}.

In particular, in contrast to the fundamental diagram
approach~\cite{Helbing2001,Nagatani_R}, in the diagram of congested patterns
within the three-phase traffic
theory~\cite{Kerner2002B,KKl,Kerner2002,Kerner2002D} at a high enough flow rate to
the on-ramp  moving jams always spontaneously emerge in
synchronized flow upstream of the bottleneck
rather than HCT.  At a low enough flow
rate to the on-ramp rather than moving jams synchronized flow of
higher vehicle speed can occur without an occurrence of moving
jams. These theoretical results are in agreement with the related
results of empirical observations~\cite{Kerner2002B} (see the
empirical scheme 2 and the related discussion in
Sect.~\ref{Discussion}).

A qualitative derivation of the diagram
of congested patterns at on-ramps has already been made  in~\cite{Kerner2002B,Kerner2002}.
Here 
a brief consideration of this diagram which is necessary for 
further consideration will be made. 

\begin{figure}
\begin{center}
\includegraphics*[width=.6\textwidth]{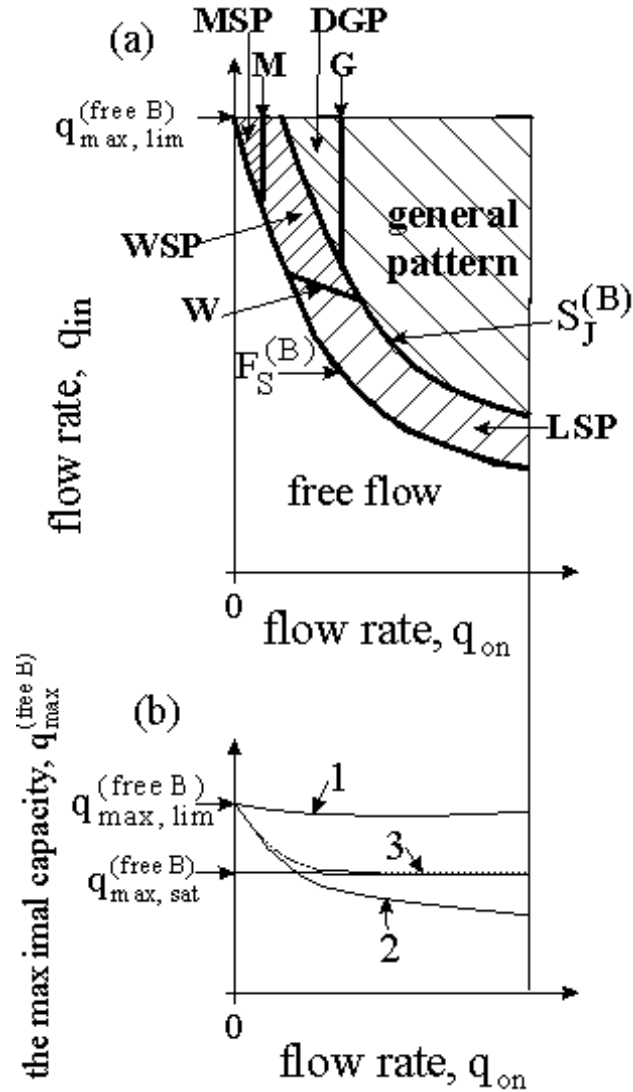}
\end{center}
\caption[]{Diagram of congested patterns at highway bottlenecks on a multi-lane highway
in
the three-phase traffic theory (a)~\protect\cite{Kerner2002B,KKl,Kerner2002,Kerner2002D}
and possible dependencies of the maximum highway capacity in free flow at the effective location of a bottleneck
$q^{\rm (free \ B)}_{\rm max}$ on the bottleneck strength $\Delta q$ (b). 
 \label{Theory} }
\end{figure}

There are two main boundaries in the diagram of congested patterns at
highway bottlenecks: $F^{\rm (B)}_{\rm S}$ and
$S^{\rm (B)}_{\rm J}$~\cite{Kerner2002B,KKl,Kerner2002}.  Below and left of
the boundary $F^{\rm (B)}_{\rm S}$ free flow is realized (Fig.~\ref{Theory} (a)).
Between the boundaries $F^{\rm (B)}_{\rm S}$ and $S^{\rm (B)}_{\rm J}$ different
SPs occur.  
Right of the boundary $S^{\rm (B)}_{\rm J}$ wide moving jams spontaneously
emerge in synchronized flow upstream of the bottleneck, i.e., a
GP occur~\cite{Kerner2002B,KKl,Kerner2002}.

To explain the boundary  $F^{\rm (B)}_{\rm S}$, let us first consider the following
hypothesis:
Let us assume that
the flow rate $q_{\rm in}$ is high enough
but the flow rate 
 $q_{\rm on}$ is extremely small, i.e.,
\begin{equation} 
q_{\rm on}\rightarrow 0 \quad
{\rm but} \quad q_{\rm on}\neq 0.
\label{OnZero_On_For}
\end{equation}
{\it Under the condition
(\ref{OnZero_On_For})
the critical flow rate $q^{\rm (free \ B)}_{\rm max, \ lim}$
for the F$\rightarrow$S transition at the bottleneck 
 is lower than 
the critical flow rate $q^{\rm (free)}_{\rm max}$ for the F$\rightarrow$S transition
on the homogeneous road:}
\begin{equation}
q^{\rm (free \ B)}_{\rm max, \ lim}<q^{\rm (free)}_{\rm max}.
\label{B_max_Max}
\end{equation}
Under the condition
(\ref{OnZero_On_For}),
the influence 
of  $q_{\rm on}$ on the flow rate  $q_{\rm sum}$ (\ref{Sum_Flow_For})
can be neglected:
 $q_{\rm sum}=q_{\rm in}$.
In the case (\ref{OnZero_On_For}), time intervals
between single vehicles squeezing onto the main road 
from the on-ramp can be  large enough. During these time intervals
 the main road can be considered as the homogeneous one, because it is
non-disturbed by the on-ramp.
However, when  single vehicles squeeze onto the main road, they
cause
time-limited {\it additional} random disturbances
in  free flow on the main road in the on-ramp vicinity. 
These additional random perturbations
can obviously cause
the F$\rightarrow$S transition at the on-ramp
at a lower flow rate $q_{\rm in}$ than  would be the case
at $q_{\rm on}= 0$, i.e., on the homogeneous road. This explains the
hypothesis (\ref{B_max_Max}) which is confirmed by
numerical simulations~\cite{KKl2003A}.

The boundary  $F^{\rm (B)}_{\rm S}$   can be found from the following 
qualitative consideration.  The critical amplitude of the critical local perturbation for
the F$\rightarrow$S transition at the bottleneck
 tends to  zero if the flow rate approaches a
critical point of free flow.
In the case of the bottleneck due to the on-ramp,
there are an infinite number of these critical points of free flow.
These critical points 
are related to an infinite number of the critical flow rates downstream
of the bottleneck $q_{\rm sum}=q_{\rm sum}(q_{\rm on}, \ q_{\rm in})\mid_{F^{\rm (B)}_{\rm S}}$
which correspond to
 the boundary $F^{\rm (B)}_{\rm S}$ in the diagram of congested patterns.
 At $q_{\rm on}\rightarrow 0$
(\ref{OnZero_On_For}) 
 the critical flow rate $q_{\rm sum}\mid_{F^{\rm (B)}_{\rm S}}$
in the related critical point of free flow is equal to
$q_{\rm in}=q^{\rm (free \ B)}_{\rm max, \ lim}$.
Thus, the F$\rightarrow$S transition
at the on-ramp must occur in this critical point.
The higher $q_{\rm on}$,  the higher  the  amplitude
of the deterministic 
local perturbation caused by the on-ramp. Therefore, the higher $q_{\rm on}$,
the lower the flow rate
 in free flow on the main road  upstream of the bottleneck $q_{\rm in}$
should be (in comparison with
$q_{\rm in}=q^{\rm (free \ B)}_{\rm max, \ lim}$)
at which
the F$\rightarrow$S-transition
at the on-ramp must occur. This explains the form of the boundary
$F^{\rm (B)}_{\rm S}$ for the spontaneous F$\rightarrow$S-transition at the bottleneck in Fig.~\ref{Theory} (a).

The boundary  $S^{\rm (B)}_{\rm J}$
 is determined by the wide moving jam emergence in synchronized flow (i.e., the 
S$\rightarrow$J
transition) upstream of the on-ramp. 
On the one hand,
between the boundaries  $F^{\rm (B)}_{\rm S}$ and $S^{\rm (B)}_{\rm J}$ 
the vehicle speed in a SP should  decrease when $q_{\rm on}$ increases. 
On the other hand, the lower the
vehicle speed in synchronized flow, the more probability of the wide moving jam emergence
in this synchronized flow. 
Thus,
in comparison with $F^{\rm (B)}_{\rm S}$,
the boundary $S^{\rm (B)}_{\rm J}$ should be shifted to the right in the flow-flow plane
(Fig.~\ref{Theory} (a)).

Between the boundaries $F^{\rm (B)}_{\rm S}$ and $S^{\rm (B)}_{\rm J}$, the higher
$q_{\rm in}$ is, the higher  the probability that the flow rate in
synchronized flow in the SP is lower than $q_{\rm in}$ and the length of the SP is
continuously increasing over time: At higher $q_{\rm in}$ a widening SP
(WSP)  and at lower $q_{\rm in}$ a localized SP
(LSP)  occurs (Fig.~\ref{Theory} (a)).
The flow rate inside a WSP is lower than $q_{\rm in}$.
Therefore, the upstream WSP front (boundary), which
separates free flow upstream and synchronized flow downstream, is
continuously widening upstream.
The mean flow rate inside the LSP is equal to $q_{\rm in}$. For this reason,
the upstream LSP front is not continuously widening upstream:
The width of the LSP is spatially limited. However, this LSP width can show oscillations
over time~\cite{KKl,KKW}.

Right of the boundary $F^{\rm (B)}_{\rm S}$ and left of the line $M$
one or a sequence of moving SPs (MSP) emerge
upstream of the on-ramp
(the region marked "MSP" in Fig.~\ref{Theory} (a)). In contrast to a wide moving jam, inside a
MSP both the
vehicle speed (40-70 km/h) and the flow rate are high. Besides, the
velocity of the downstream front of a
MSP is {\it not} a characteristic
parameter. This velocity can change in a wide range in the process of
the MSP propagation or for different MSP's. In some cases it has been
found  that after the MSP is far away from the on-ramp, the pinch
effect (the self-compression of synchronized flow) occurs inside the MSP
and a wide moving jam can be formed there~\cite{KKl}.

Right of the boundary $S^{\rm (B)}_{\rm J}$ and left of the line $G$
{\it the dissolving general pattern} or DGP
for short occurs (the
region marked "DGP" in Fig.~\ref{Theory} (a)). In the DGP, after a wide moving jam in
synchronized flow of the congested pattern has been formed,  the forming GP  dissolves over
time. As a result of this GP
dissolving process, the GP transforms into one of the SP, or free flow
occurs at the bottleneck~\cite{KKl,KKW}.

The following empirical results confirm the diagram of congested
patterns in the three-phase traffic flow theory (Fig.~\ref{Theory} (a)):
 
(i) All types of SPs and GPs are
found in empirical observations (see~\cite{Kerner2002B} and Sect.~\ref{Wide_Sec}).

(ii) In the diagram (Fig.~\ref{Theory} (a)), GP exists in the most part of the flow rates
$q_{\rm on}$ and $q_{\rm in}$ where congestion occurs. This is related to the emprirical result
that GP is the most frequent type of congested patterns at isolated highway bottlenecks~\cite{Kerner2002B}.

(iii) Corresponding to the diagram, in empirical investigations
GP transforms into a SP when the flow rate $q_{\rm on}$ decreases~\cite{Kerner2002B}.

(iv) Corresponding to the diagram, in empirical investigations
GP does not transform into another type
of congested pattern when the flow rate $q_{\rm on}$ increases~\cite{Kerner2002B}.

The hypothesis about the diagram of congested
patterns~\cite{Kerner2002B}
is also confirmed by the diagrams found in the
microscopic three-phase traffic flow theories~\cite{KKl,KKW,KKl2003A}.

In contrast to this, it has been shown
in the empirical study ~\cite{Kerner2002B}
 that no sequences of the congested pattern transformation and no theoretical congested states 
which have been predicted in
the pattern diagram in the fundamental diagram 
approach~\cite{Helbing1999,Helbing2000,Helbing2001,Nagatani_R}
    have  been observed  at isolated bottlenecks.

\section{Probabilistic Theory of Highway Capacity \label{Cap_Sec} }

The determination  of highway capacity is one of the most important applications
of any traffic theory.
Empirical observations show that the speed  breakdown 
at a bottleneck (the breakdown phenomenon) is in general accompanied by a drop in highway
capacity
(see e.g.,~\cite{Hall1992,Hall1991}). 
Here we give a qualitative theory of highway capacity
and of the capacity drop which  follows from the three-phase traffic theory.

However, firstly recall, how the breakdown phenomenon looks like in the
fundamental diagram approach.
From a numerical analysis of a macroscopic traffic flow 
model within the fundamental diagram approach Kerner and Konh\"auser
found in 1994 \cite{KK1994} that free flow is metastable with respect
to the formation of wide moving jams (F$\rightarrow$J transition), 
if the flow rate is equal to or higher than the outflow from a jam,
$q_{\rm out}$. The critical amplitude of a local perturbation in an
initial homogeneous free flow, which is needed for the
F$\rightarrow$J transition, decreases with increasing density: 
It is maximum at the threshold  
density $\rho = \rho_{\rm min}$, below which free flow is stable.
The critical amplitude becomes zero at some critical density
$\rho = \rho_{\rm cr}> \rho_{\rm min}$, above which free flow is linearly
unstable. Obviously the higher the amplitude of a random local 
perturbation the less frequent it is. Hence, the likelihood that the
F$\rightarrow$J transition occurs in a given time interval should
increase with density (or flow rate). 
The probability should tend to one at the critical density $\rho_{\rm \rm cr}$.

In 1997  Mahnke {\it et al}~\cite{Mahnke1997,Mahnke} developed
a master equation approach for calculating the
probability of the F$\rightarrow$J transition on a homogeneous road
(i.e. without bottleneck). Based on this
approach  K{\"u}hne {\it et al}~\cite{Kuehne2002} confirmed that
the probability of the F$\rightarrow$J transition
in the metastable region is increasing with the flow rate in free
flow. They applied this result to explain the
breakdown phenomenon at a highway bottleneck.  
For a recent comprehensive discussion of
the breakdown phenomenon in CA-models and in the Krau{\ss} {\it et al.} model
in the fundamental diagram
approach see also \cite{JostNa}.
The theories in~\cite{KK1994,Helbing2001,Mahnke1997,Mahnke,Kuehne2002,JostNa}
belong to the fundamental diagram approach.

In contrast to these results, in the three-phase traffic 
theory~\cite{Kerner1999A,Kerner1999B,Kerner1999C}  
it is postulated that metastable states of free flow decay into
synchronized flow (F$\rightarrow$S transition) rather than wide moving
jams (F$\rightarrow$J transition). 
In particular, even the upper limit of free flow ($q^{\rm (free \ B)}_{\rm max, \ lim}$
in Fig.~\ref{Theory})  is related to
the F$\rightarrow$S transition: In this limit point the probability of
the F$\rightarrow$S transition should be equal to one whereas the
probability of  the emergence of a moving jam (F$\rightarrow$J
transition) should be very small (Fig.~\ref{Hypote1} (c), curve $F_{\rm J}$). Thus, in this theory the breakdown
phenomenon in free traffic is related to 
the F$\rightarrow$S transition rather than to an emergence of moving jams.

Highway capacity depends on whether a homogeneous 
road (without bottlenecks) or a highway bottleneck is considered.

\subsection{Homogeneous Road \label{Hom_Road}}

On a homogeneous (without bottlenecks)
multi-lane road, 
highway capacity depends on  which traffic phase the traffic is in~\cite{Kerner1998B}: (i) The maximum
highway capacity in the traffic phase "free flow" is equal
to the maximum possible flow rate in free flow, $q^{\rm (free)}_{\rm max}$. 
(ii) The maximum highway capacity in  the traffic phase "synchronized flow" is 
equal to the maximum possible flow rate in synchronized flow, $q^{\rm (syn)}_{\rm max}$.
(iii) The maximum highway capacity downstream of the traffic phase "wide moving jam" is equal
to the flow rate in the wide moving jam  outflow, $q_{\rm out}$. 
Because of the first order phase transitions between the traffic phases each of these
maximum highway capacities has a probabilistic nature. 

In particular, the probabilistic nature of  the 
highway capacity in the traffic phase "free flow"  means the following~\cite{Kerner1999A}: 

(1) 
At the flow rate $q=q^{\rm (free)}_{\rm max}$ the probability of the spontaneous F$\rightarrow$S transition
for a given time interval $T_{\rm ob}$ and for a given highway section length $L_{\rm ob}$,
$P_{\rm FS}=1$   (\ref{PFS1}).
In other words, the maximum capacity $q=q^{\rm (free)}_{\rm max}$
depends on  $T_{\rm ob}$ (at least in some range of $T_{\rm ob}$).

(2) There is the threshold flow rate $q_{\rm th}$
for the  F$\rightarrow$S transition.
The threshold flow rate is lower than $q=q^{\rm (free)}_{\rm max}$ (Fig.~\ref{Hypote1} (a)).
At the threshold flow rate the probability of the F$\rightarrow$S transition  
$P_{\rm FS}=0\mid_{q=q_{\rm th}}$  (\ref{TH}). 

(3) If the flow rate in free flow
$q$ is within the range $[q_{\rm th}, q^{\rm (free)}_{\rm max}]$ then the higher the flow rate
$q$, the higher   the probability 
the  F$\rightarrow$S transition $P_{\rm FS}$ is. Thus, the attribute of this {\it probabilistic
highway capacity} is the probability $1-P_{\rm FS}$
that free flow remains on a road section of the length $L_{\rm ob}$
during the time interval $T_{\rm ob}$ of the observation
of this capacity in free flow. 

The hypotheses of the three-phase traffic theory are confimed by empirical findings
 of  characteristics of the  wide moving jam outflow (e.g.,~\cite{KR1996B,KR1996A}) and by numerical results of
a microscopic three-phase traffic flow theory~\cite{KKl,KKW,KKl2003A}.
 
However, if a bottleneck exist on the road, then a much more complicated non-linear phenomena determine
highway capacity. 

\subsection{Highway Capacity in Free Flow at  Bottleneck \label{Capacity_Free_B}
}

In the three-phase traffic theory, 
 the breakdown phenomenon  at  a highway bottleneck is explained by
 the F$\rightarrow$S transition at the
bottleneck~\cite{Kerner2000A}. Due to the bottleneck
the road is spatially non-homogeneous, i.e., highway capacity can depend on
a highway location~\cite{May,Manual}. We will consider highway capacity in free flow which is related
to the effective location of the bottleneck due to the on-ramp, i.e., the location where 
the probability of the F$\rightarrow$S transition for a time interval as function of 
a highway location  has a maximum
(see Fig. 1 (b) in~\cite{Kerner2000A}).

The F$\rightarrow$S transition 
occurs at the
bottleneck during a given time interval $T_{\rm ob}$  if the flow rate 
$q_{\rm in}$ and the flow rate $q_{\rm on}$ are related to
 the boundary $F^{\rm (B)}_{\rm S}$ in the diagram of congested patterns
(Fig.~\ref{Theory} (a)). 
Thus, there is {\it an infinite
multitude} of maximum freeway capacities of free flow at the bottleneck which are given by the points on
the boundary $F^{\rm (B)}_{\rm S}$.
We designate these capacities $q^{\rm (free \ B)}_{\rm max}$: 
\begin{equation}
q^{\rm (free \ B)}_{\rm max}=q_{\rm sum}\mid_{\rm F^{\rm (B)}_{\rm S}},
\label{capacity_free_For}
\end{equation} 
where $q_{\rm sum}$ (\ref{Sum_Flow_For}) is the flow rate downstream of the bottleneck.

The capacities $q^{\rm (free \ B)}_{\rm max}$ (\ref{capacity_free_For})
depend on the flow rate on the main road
upstream of the on-ramp $q_{\rm in}$ and the flow rate to the on-ramp $q_{\rm on}$, exactly on the values
$q_{\rm in}$ and 
$q_{\rm on}$ at
 the boundary $F^{\rm (B)}_{\rm S}$ in the diagram of congested 
patterns at the on-ramp (Fig.~\ref{Theory} (a)). 
The capacities (\ref{capacity_free_For})
are the maximum capacities
 related to  the time interval $T_{\rm ob}$.
This means that  the probability of the  F$\rightarrow$S transition 
at the bottleneck $P^{\rm (B)}_{\rm FS}$
 for the  time   interval $T_{\rm ob}$  
reaches one at the boundary $F^{\rm (B)}_{\rm S}$
in the diagram of congested patterns, i.e.,
corresponding to the
formula (\ref{capacity_free_For})
this occurs at the flow rate $q_{\rm sum}=q^{\rm (free \ B)}_{\rm max}$:
 \begin{equation}
P^{\rm (B)}_{\rm FS}=1\mid_{\rm q_{\rm sum}=q^{\rm (free \ B )}_{\rm max}}.
\label{capacity_free3_For}
\end{equation}

For traffic
 demand 
(values  $q_{\rm on}$ and $q_{\rm in}$)
 which is related to points  in some vicinity {\it left of} the boundary $F^{\rm (B)}_{\rm S}$
in the
diagram of congested patterns, i.e., in the free flow region of the diagram the
F$\rightarrow$S transition nevertheless
occurs at the
bottleneck during the time interval $T_{\rm ob}$ with a probability $P^{\rm (B)}_{\rm FS}<1$.
The lower this probability $P^{\rm (B)}_{\rm FS}$ is, the more  the distant  a point $(q_{\rm on},  \ q_{\rm in})$
in the
diagram of congested patterns is
from the boundary $F^{\rm (B)}_{\rm S}$ (Fig.~\ref{Theory} (a)).

The region in the diagram of congested patterns where this probabilistic
effect occurs is restricted by a threshold boundary $F^{\rm (B)}_{\rm th}$
which is left of the boundary $F^{\rm (B)}_{\rm S}$ (Fig.~\ref{KKW_Diagram} (b))
\footnote{A more detail consideration of stable and metastable free flow states and of
regions of metastability of different congested patterns in the diagram of the patterns
has recently been made in~\cite{KKl2003A}.}: 
This threshold boundary 
is related to the {\it infinite multitude} of 
threshold flow rates $q^{\rm (B)}_{\rm th}(q_{\rm on}, q_{\rm in})$. Each of these threshold
points is determined as  has been described in Sect.~\ref{Z_FS}. This means that
at the threshold boundary $F^{\rm (B)}_{\rm th}$  
the critical amplitude of the local perturbation 
for the
F$\rightarrow$S transition reaches a maximum
and the  probability $P^{\rm (B)}_{\rm FS}$ is given by the formula:
 \begin{equation}
P^{\rm (B)}_{\rm FS}=0\mid_{q_{\rm sum}=q^{\rm (B)}_{\rm th}}.
\label{FS_TH}
\end{equation} 
which is analogous to the formula (\ref{TH}). This formula is also valid
below the threshold, i.e., left of this threshold
boundary in the diagram of congested patterns $F^{\rm (B)}_{\rm th}$ we have 
$P^{\rm (B)}_{\rm FS}=0\mid_{q_{\rm sum}<q^{\rm (B)}_{\rm th}}.$

There can be a number of different dependencies of 
the maximum freeway capacity 
in free flow at the bottleneck $q^{\rm (free \ B)}_{\rm max}$ on $q_{\rm on}$
(e.g., curves $1$, $2$, $3$ in Fig.~\ref{Theory} (b)).
In particular, it can be expected that the higher the flow rate to the on-ramp $q_{\rm on}$,
 the lower  the maximum freeway capacity 
in free flow at the bottleneck $q^{\rm (free \ B)}_{\rm max}$. In this case, the maximum freeway capacity 
$q^{\rm (free \ B)}_{\rm max}$ is a decreasing function of $q_{\rm on}$ (curve $2$ in
Fig.~\ref{Theory} (b)).
A decrease of the  maximum freeway capacity 
$q^{\rm (free \ B)}_{\rm max}$ at the on-ramp when the flow rate $q_{\rm on}$ increases can have
 saturation at a high enough flow rate to the on-ramp 
$q_{\rm on}$ (Fig.~\ref{Theory} (b), dotted curve $3$). In this case,
the  maximum freeway capacity 
$q^{\rm (free \ B)}_{\rm max}$ does not reduce below some saturation value $q^{\rm (free \ B)}_{\rm max, \ sat}$
even at a very high $q_{\rm on}$. 

The highest is the capacity of free flow
at $q_{\rm on}=0$. This case is obviously related to the maximum capacity on a 
homogeneous (without bottlenecks) road (Fig.~\ref{Hypote1} (a)): 
\begin{equation}
q^{\rm (free \ B)}_{\rm max}\mid_{q_{\rm on}=0}=q^{\rm (free)}_{\rm max}.
\label{capacity_free21_For}
\end{equation}
However, at a very small flow rate $q_{\rm on}$, exactly at the limit case
(\ref{OnZero_On_For})
we have
\begin{equation}
q^{\rm (free \ B)}_{\rm max}\mid_{q_{\rm on}\rightarrow 0}=
q^{\rm (free \ B)}_{\rm max, \ lim}<q^{\rm (free)}_{\rm max} \quad
\mbox{at} \ q_{\rm on}\neq 0.
\label{capacity_free27_For}
\end{equation}
The condition (\ref{capacity_free27_For})
is linked to the effect of random perturbations which occur at the on-ramp due to 
the single vehicles
squeezing
  onto the main road from
the on-ramp as  has already been explained in Sect.~\ref{DCP}.
For this reason in 
Fig.~\ref{Theory} (b) rather than the flow rate 
$q^{\rm (free)}_{\rm max}$ the flow rate
$q^{\rm (free \ B)}_{\rm max, \ lim}$ is shown as the maximum freeway capacity at the bottleneck at the
limit case
$q_{\rm on}\rightarrow 0$.

Corresponding to (\ref{capacity_free3_For}) and (\ref{FS_TH}), if the flow rate 
$q_{\rm sum}$ (\ref{Sum_Flow_For})
at the effective location of the bottleneck
is within the range $[q^{\rm (B)}_{\rm th}, q^{\rm (free \ B)}_{\rm max, \ lim}]$ then at a given $q_{\rm on}$
the higher $q_{\rm sum}$, the higher the mean probability of the F$\rightarrow$S transition $P^{\rm (B)}_{\rm FS}$.

These hypotheses of the three-phase traffic theory are confirmed by empirical findings
where
 the probabilistic nature of freeway capacity  has been studied (e.g.,~\cite{El,Persaud})
and by numerical results~\cite{KKW,KKl2003A}.
This theory is confirmed by numerical simulations~\cite{KKl,KKW}.

\subsection{Highway Capacity in Congested Traffic at  Bottleneck. Capacity Drop
\label{Con_Road}}

In order to study the highway capacity downstream of the congested bottleneck one has to
consider the outflow from a congested bottleneck
$q^{\rm (bottle)}_{\rm out}$ (the discharge flow rate), which is measured 
downstream of the bottleneck, where free flow conditions are reached.

In the three-phase traffic theory, the discharge flow rate
$q^{\rm (bottle)}_{\rm out}$ is not just a characteristic property of the
type of bottleneck under consideration only. It also depends on the type of
congested pattern which 
actually is formed upstream of the bottleneck~\cite{Kerner2000A}. Thus,
in the three-phase traffic theory, 
highway capacity in free flow downstream of the congested bottleneck depends
on the type of congested pattern upstream of the bottleneck, the pattern characteristics and on
parameters of the bottleneck. We call this highway capacity {\it a congested pattern capacity}.

In the case of
an on-ramp, $q^{\rm (bottle)}_{\rm out}$ is expected to vary with
$(q_{\rm on}, q_{\rm in})$. Obviously, $q^{\rm (bottle)}_{\rm out}$
only limits the freeway capacity, if it is smaller than the traffic
demand upstream of the on-ramp, $q_{\rm sum}=q_{\rm in}+q_{\rm on}$, i.e. if the condition
\begin{equation}
q^{\rm (bottle)}_{\rm out}(q_{\rm on},q_{\rm in}) 
< q_{\rm sum}
\label{Cong_capacity_condition_For}
\end{equation}
is fulfilled. Note that in (\ref{Cong_capacity_condition_For})
in contrast to the discharge flow rate $q^{\rm (bottle)}_{\rm out}$,
the flow rate $q_{\rm sum}$ is related to free flow conditions 
at the bottleneck, i.e., when no congested pattern exists upstream of the bottleneck.
Under the condition (\ref{Cong_capacity_condition_For}), the 
congested pattern upstream from the on-ramp simply expands, while the
throughput remains limited by $q^{\rm (bottle)}_{\rm out}$. For example, if the general pattern (GP) is
formed at the bottleneck, an increase of $q_{\rm in}$ does not
influence the discharge flow rate $q^{\rm (bottle)}_{\rm out}$. Instead,
the width of the wide moving jam, which is mostly upstream in the
GP, simply grows.

Thus, in the case (\ref{Cong_capacity_condition_For})
the congested pattern capacity $q^{\rm (B)}_{\rm cong}$ is equal to $q^{\rm (bottle)}_{\rm out}$:
\begin{equation}
q^{\rm (B)}_{\rm cong}=q^{\rm (bottle)}_{\rm out}.
\label{capacity_condition5_For}
\end{equation}
It must be noted that  $q^{\rm (bottle)}_{\rm out}$ can strongly depend on the congested
pattern type and
congested  pattern parameters. 
The congested pattern type and
the  pattern parameters depend on initial conditions and
the flow rates $q_{\rm on}$ and $q_{\rm in}$ (Fig.~\ref{Theory} (a)).
Thus, the congested pattern capacity $q^{\rm (B)}_{\rm cong}$ 
implicitly
 depends on
the flow rates $q_{\rm on}$ and $q_{\rm in}$.

The capacity drop is the difference between 
freeway capacity in free flow at a bottleneck and in a situation, where there is
synchronized flow upstream and free flow downstream of the bottleneck (e.g.,~\cite{Hall1992}). 

Assuming that (\ref{Cong_capacity_condition_For}) is fulfilled, 
the capacity drop can be given by 
\begin{equation}
\delta q=q^{\rm (free)}_{\rm max}-q^{\rm (B)}_{\rm cong},
\label{drop2_For}
\end{equation}
where the congested pattern capacity $q^{\rm (B)}_{\rm cong}$ is given by the formula
(\ref{capacity_condition5_For});
$q^{\rm (free)}_{\rm max}$ is the maximum freeway capacity in free flow at $q_{\rm on}=0$.

If one considers all kinds of
congested patterns upstream from a bottleneck then there should be
the minimum discharge flow rate which satisfies  the condition (\ref{Cong_capacity_condition_For}).
This minimum discharge flow rate should be the
characteristic quantity for the type of bottleneck under
consideration. We denote this quantity by $q^{\rm (bottle)}_{\rm min}$.
The maximum of $q^{\rm (bottle)}_{\rm out}$ (denoted by  $q^{\rm (bottle)}_{\rm max}$) is
predicted to be the maximum flow rate, which can be realized in
synchronized flow, $q^{\rm (bottle)}_{\rm max} = q^{\rm (syn)}_{\rm max}$.
To explain the latter condition, recall that
 the downstream front of a congested pattern at the bottleneck due to
the on-ramp
 separates free flow downstream of the front
and synchronized flow upstream of the front. This downstream front  
is fixed at the bottleneck. Thus,  within the front the total
flow rate across the road (together with the on-ramp) does not depend on the  co-ordinates along 
the road.
Just downstream of the front, i.e., in free flow this flow rate is equal to  $q^{\rm (bottle)}_{\rm max}$.
Just upstream of the front it is suggested that synchronized flow occurs both on the main road and the on-ramp.
The maximum possible total flow rate in synchronized flow is equal to $q^{\rm (syn)}_{\rm max}$.
Hence, the capacity drop at a bottleneck cannot be smaller than   
\begin{equation}
\delta q_{\rm min}=q^{\rm (free)}_{\rm max}-q^{\rm (syn)}_{\rm max}.
\label{drop1}
\end{equation}

Note that
there may also be another definition of  the 
capacity drop:
\begin{equation}
\delta q=q^{\rm (free \ B)}_{\rm max}-q^{\rm (B)}_{\rm cong},
\label{drop4_For}
\end{equation}
where the congested pattern capacity $q^{\rm (B)}_{\rm cong}$ is given by the formula
(\ref{capacity_condition5_For}) and
$q^{\rm (free \ B)}_{\rm max}$ is given by (\ref{capacity_free_For}).
However, there could be a difficulty in the application of the definition (\ref{drop4_For}):
There is an infinite multitude of different maximum freeway capacities
in free flow at a bottleneck, 
$q^{\rm (free \ B)}_{\rm max}$ (see the 
formula (\ref{capacity_free_For})).

There may be one exception of the condition (\ref{Cong_capacity_condition_For}):
If a LSP 
occurs 
both on the main road and on the on-ramp 
upstream of the merge region of the on-ramp
then
the discharge flow rate is equal to traffic demand:
\begin{equation}
q^{\rm (bottle)}_{\rm out}(q_{\rm on},q_{\rm in}) 
= q_{\rm sum}.
\label{capacity_condition8}
\end{equation}
The congested pattern capacity which is related to this LSP  should be determined by 
the maximum discharge flow rate at which the LSP still exists upstream of the on-ramp.

These hypotheses of the three-phase traffic theory are confirmed by empirical findings
where
 the discharge flow rate from 
spatial-temporal congested patterns at bottlenecks  has been studied~\cite{Kerner2002B}
and by numerical results derived within the three-phase traffic theory which will be presented below.

\subsection{Numerical Study of Congested Pattern Capacity at On-Ramps}

Here we confirm and illustrate the general theory of congested pattern capacity presented above
based on a numerical simulation of a one-lane KKW cellular automata microscopic traffic flow model
within the three-phase traffic theory which has recently been proposed by Kerner, Klenov and Wolf~\cite{KKW}.

\subsubsection{KKW Cellular Automata Traffic Flow Model}

We will use for simulations the KKW-1 CA-model with a linear dependence of the synchronization distane on the 
vehicle speed~\cite{KKW}.
 This KKW CA-model consists of a dynamic part
\begin{equation}
\tilde v_{ n+1}=\max(0, \min(v_{ \rm free},  v_{{\rm s},n}, v_{{\rm
    c},n})), 
\end{equation}
\begin{equation}
v_{{\rm c},n}=\left\{\begin{array}{ll}
v_{n}+a\tau &  \textrm{for \, $g_{n} > D_{n}-d$},\\
v_{n}+a\tau {\rm sign} {(v_{\ell,n}-v_{n})} &  
\textrm{for \, $g_{n} \leq D_{n}-d$}, 
\end{array} \right.
\end{equation}
where ${\rm sign} (x)$ is 1 for $x>0$, 0 for $x=0$ and $-1$ for $x<0$,
and a stochastic (fluctuation) part
\begin{equation}
v_{n+1}=\max(0, \min({\tilde v_{n+1}+a\tau\eta_{n}, v_{n}+a\tau, 
  v_{\rm free}, v_{{\rm s},n}})),
\end{equation}
\begin{equation} 
 x_{n+1}= x_{n}+v_{n+1}\tau, 
\end{equation}
\begin{equation}
\eta_{n}=\left\{
\begin{array}{ll}
-1 &  \textrm{if $r< p_{\rm b}$}, \\
1 &  \textrm{if $p_{\rm b}\leq r<p_{\rm b}+p_{\rm a}$}, \\
0 &  \textrm{otherwise}, 
\end{array}\right. 
\end{equation}

\begin{equation}
p_{\rm b}(v_{n})=\left\{\begin{array}{ll}
                                      p_{0} & \textrm{if $v_{n}=0$}, \\
                                      p &  \textrm{if $v_{n}>0$},
                                      \end{array} \right.
\end{equation}
\begin{equation} 
                 p_{\rm a}(v_{n})=\left\{\begin{array}{ll}
                          p_{\rm a 1} & \textrm{if $v_{n} <   v_{\rm p}$}, \\
                          p_{\rm a 2} & \textrm{if $v_{n}\geq v_{\rm p}$},
                          \end{array} \right.
\end{equation} 

In this KKW CA-model,
$n=0, 1, 2, ...$ is number of time steps,
$\tau$ is the time discretization interval,
$\tilde v_{n}$ is the vehicle speed at time step $n$ without fluctuating part,
$v_{n}$ is the vehicle speed at time step $n$,
$v_{\ell,n}$ is the speed of the leading vehicle at time step $n$,
$v_{{\rm s}, n}= g_{n}/\tau$ is the safe speed at time step $n$,
$v_{\rm free}$ is the maximal speed (free flow),
$x_{n}$ is the vehicle position at the time step $n$,
$x_{\ell,n}$ is the position of the leading vehicle at time step $n$,
$d$  is the vehicle length which is the same for all vehicles,
$g_{n}=x_{\ell,n}-x_{n}-d$ is the gap (front to end distance) at time step $n$,
$D_{n}=d + kv_{n}\tau$ is the synchronization distance at time step $n$,
$\eta_{n}$ is the speed fluctuation at time step $n$, $a$ is the vehicle acceleration,
$p_{\rm a}$ is the probability of vehicle acceleration,
$p_{\rm b}$ is the probability of vehicle deceleration,
$r$ is a random number uniformly distributed between $0$ and $1$,
$k$, $p_{0}$, $p$, 
$p_{\rm a 1}$, $p_{\rm a 2}$, $v_{\rm p}$ are constant parameters.

The steady states for the model are related to a 2D region in the flow-density plane 
between the line $F$ for free flow ($v=v_{\rm free}$), the line $U$ determined by the safe speed
$v_{{\rm s}, n}$ and the line $L$ determined by the synchronization distance $D$ 
(Fig.~\ref{KKW_Diagram} (a)). 
The physics of the KKW CA-model has been considered in~\cite{KKW}.
All numerical simulations below have been performed  for a one-lane road with an 
on-ramp.
Models of the road and of the on-ramp are the same and they have
the same model parameters (in particular, the road length, the conditions for
vehicle squeezing from the on-ramp to the main road, the time and the space discretization
units, etc.) as it has been chosen
in~\cite{KKW} for the KKW-1 CA-model, parameter-set I
 (see Table III in~\cite{KKW}, the KKW-1 CA-model, parameter-set I).

\subsubsection{Transformations of Congested Patterns at On-Ramps \label{Evol_Sec}}

\begin{figure}
\begin{center}
\includegraphics*[width=.8\textwidth]{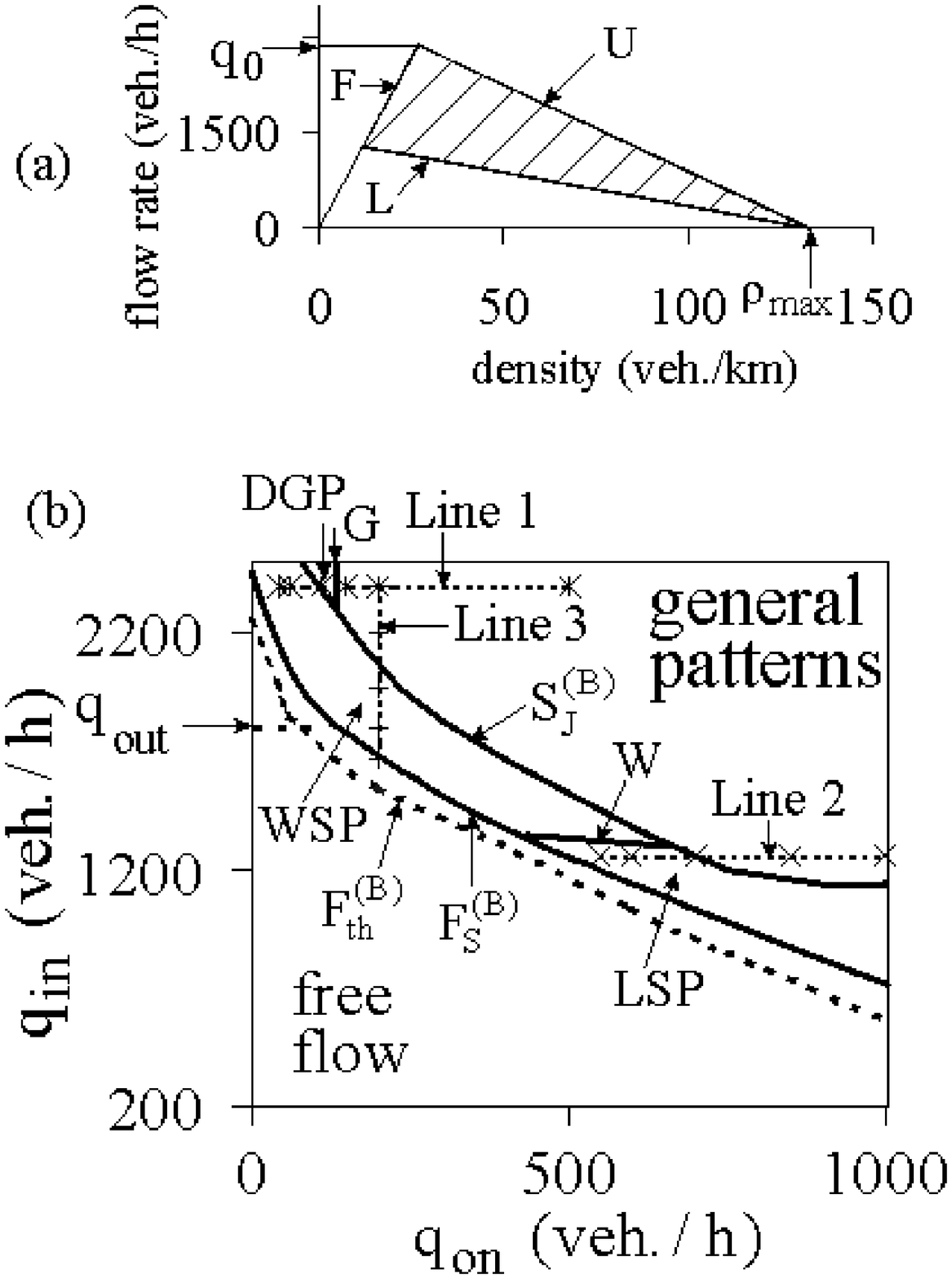}
\end{center}
\caption[]{Steady states in the flow-density plane (a)~\cite{KKW} and
the diagram of congested patterns at the on-ramp (b) for 
the KKW CA-model within
the three-phase traffic theory.
The model parameters are related to the KKW-1 CA model (parameter-set I)
in~\cite{KKW}.
 \label{KKW_Diagram} }
\end{figure}

A diagram of congested pattern at the on-ramp  
for the KKW CA-model~\cite{KKW}  is qualitatively related to the diagram in Fig.~\ref{Theory} (a)
first  predicted within the
 three-phase traffic theory~\cite{Kerner2002B} and then 
found  in the continuum model by Kerner and Klenov~\cite{KKl}
(Fig.~\ref{KKW_Diagram} (b)).
At the boundary $F^{\rm (B)}_{\rm S}$ with the probability $P^{\rm (B)}_{\rm FS}\approx 1$ the F$\rightarrow$S transition
occurs~\cite{KKW}. Between the boundaries $F^{\rm (B)}_{\rm S}$ and $S^{\rm (B)}_{\rm J}$ different SP occurs. 
Right of the boundary $S^{\rm (B)}_{\rm J}$  wide moving jams emerge in synchronized flow upstream
of the on-ramp, i.e., different GP occur.

To study the congested pattern capacity, in the diagram three lines (Line 1, Line 2 and Line 3) are  shown,
in addition
(Fig.~\ref{KKW_Diagram} (b)).
Different congested patterns have been studied when the flow rates
$q_{\rm on}$ (Line 1 and Line 2) or the flow rate $q_{\rm in}$ (Line 3)
are increasing along the related lines.

\begin{figure}
\begin{center}
\includegraphics*[width=.9\textwidth]{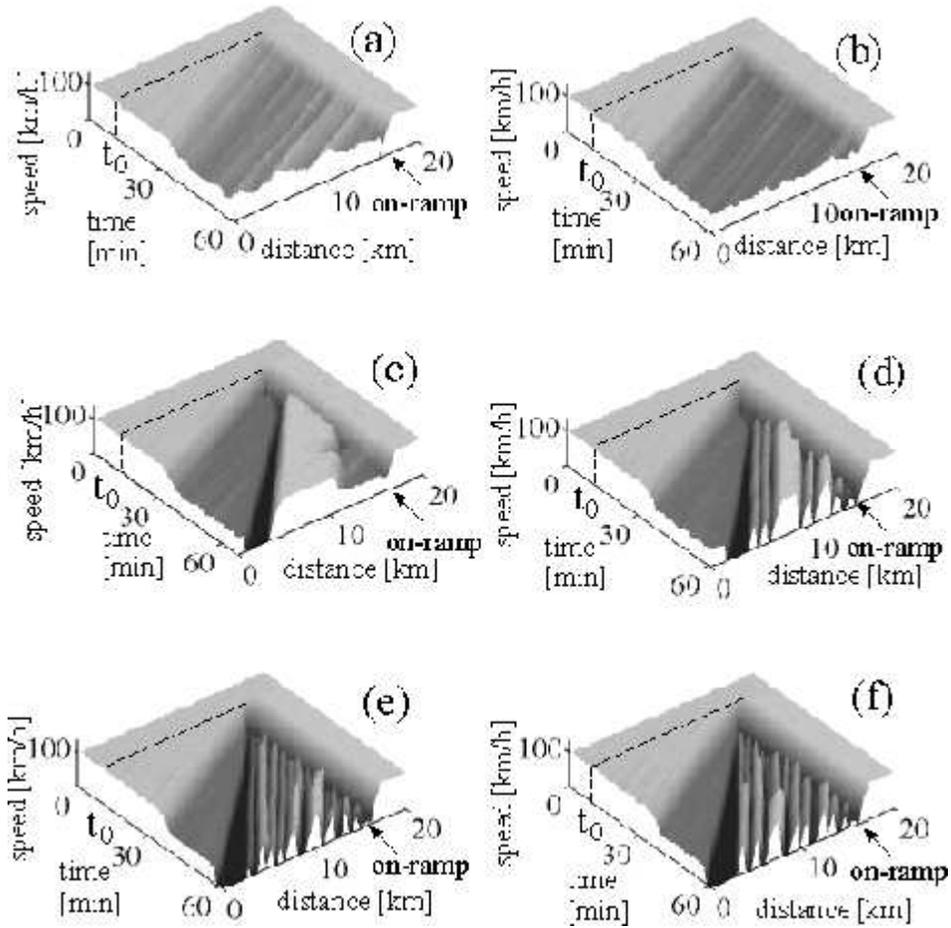}
\end{center}
\caption[]{Evolution of the vehilce speed in space and in time
at the given $q_{\rm in}= 2400 \ vehicles/h$ for different $q_{\rm on}$
related to Line 1 in Fig.~\ref{KKW_Diagram} (b):
(a, b) - widening synchronized flow patterns WSP, (c) -
the dissolving general pattern  (DGP), (d, e, f) - general patterns (GP).
The flow rate $q_{\rm on}$ is:  (a)  (40), (b) (60),
(c) (105), (d)  (150), (e) (200), (f) (500) $vehicles/h$.
The on-ramp is at the location $x=16 \ km$ (see formore detail~\cite{KKW})
 \label{Line1} }
\end{figure}

The transformation of congested patterns along Line 1 is related to 
a given {\it high} flow rate in free flow upstream of the on-ramp
$q_{\rm in}$ when 
the flow rate to the on-ramp $q_{\rm on}$
is increased from lower to higher values
(Fig.~\ref{Line1}). Right of the boundary $F^{\rm (B)}_{\rm S}$
a WSP occurs (Fig.~\ref{Line1} (a)). If the flow rate to the on-ramp is only slightly increased then
a WSP remains (Fig.~\ref{Line1} (b)) however the vehicle speed in this WSP on  average decreases
in comparison with the speed in the WSP in Fig.~\ref{Line1} (a).

Right of  the boundary $S^{\rm (B)}_{\rm J}$ at the high flow rate $q_{\rm in}$, which is related to Line 1, a 
DGP occurs
(Fig.~\ref{Line1} (c)): After the first wide moving jam has emerged in synchronized flow,
the flow rate upstream of the on-ramp decreases because it is now determined by the jam outflow.
The maximum flow rate  
in the wide moving jam outflow is reached when free flow is formed
downstream of the jam, $q_{\rm out}=1810 \ vehicles/h$. When due to the jam upstream propagation
the wide moving jam
 is far away from the on-ramp
an effective flow rate upstream of the on-ramp is determined by $q_{\rm out}$, i.e., the effective flow rate
$q^{\rm (eff)}_{\rm in}= q_{\rm out}=1810 \ vehicles/h$. This flow rate is
however considerably lower than
the initial flow rate $q_{\rm in}= 2400 \ vehicles/h$. As a result, no wide moving jams 
can emerge upstream of the on-ramp any more: the DGP occurs which consists of the only one wide
moving jam propagating upstream and a SP at the on-ramp. At the mentioned
effective flow rate $q^{\rm (eff)}_{\rm in}= 1810 \ vehicles/h$
and the flow rate to the on-ramp
$q_{\rm on}= 105 \ vehicles/h$ 
this SP at the on-ramp is a WSP 
(Figs.~\ref{KKW_Diagram} (b) and~\ref{Line1} (c)).
If the flow rate $q_{\rm on}$
is further increased then the pinch region in synchronized flow upstream of the on-ramp appears
where narrow moving jams continuously emerge (Fig.~\ref{Line1} (d)). Some of these jams transform into
wide moving jams leading to the GP formation. This occurs
right of the boundary $G$   where GPs are realized. 
When the flow rate $q_{\rm on}$ is further increased,
the GP does not transform into another pattern: It remains to be a GP at any possible flow rate $q_{\rm on}$
(Fig.~\ref{Line1} (e, f)).

\begin{figure}
\begin{center}
\includegraphics*[width=.9\textwidth]{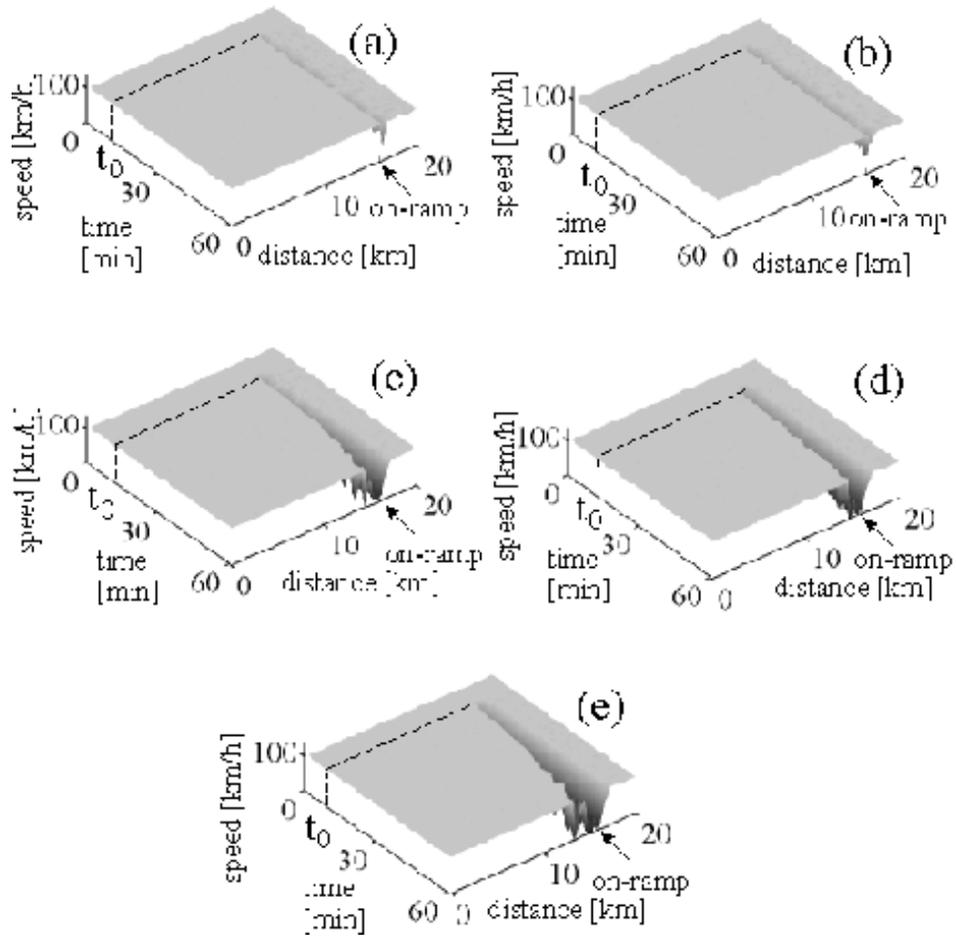}
\end{center}
\caption[]{Evolution of the vehilce speed in space and in time
at the given $q_{\rm in}= 1255 \ vehicles/h$ for different $q_{\rm on}$
related to Line 1 in Fig.~\ref{KKW_Diagram} (b):
(a, b) - localized synchronized flow patterns LSP, 
(c - e) - general patterns (GP).
The flow rate $q_{\rm on}$ is:  (a)  (550), 
(b) (630),
(c) (700), (d)  (850), (e) (1000) $vehicles/h$. 
\label{Line2} }
\end{figure}

The transformation of congested patterns along Line 2 in Fig.~\ref{KKW_Diagram} (b)
is related to 
a given {\it low} flow rate in free flow upstream of the on-ramp
$q_{\rm in}$ when 
the flow rate to the on-ramp $q_{\rm on}$
is increased beginning from a  relative high initial value
(Fig.~\ref{Line2}). 
In this case, first a LSP occurs
(Fig.~\ref{Line2} (a, b)).
If the flow rate $q_{\rm on}$ increases then the LSP transforms into a GP
(Fig.~\ref{Line2}
(c-e)).
However, whereas for the GP related to Line 1  (Fig.~\ref{Line1} (d-f))
 the condition $q_{\rm in}>q_{\rm out}$ is fulfilled, in the case of Line 2 we have
$q_{\rm in}<q_{\rm out}$
(Fig.~\ref{KKW_Diagram} (b)). This case has already been considered in~\cite{KKW} where we could see  that
the most upstream wide moving jam in the GP
 dissolves over time. This indeed occurs for the GP related to Line 2.
For this reason, the width of the GP  in Fig.~\ref{Line2}
(c-e))
increases much slower over time in comparison
with the width of the GP in Fig.~\ref{Line1} (d-f).

\begin{figure}
\begin{center}
\includegraphics*[width=.9\textwidth]{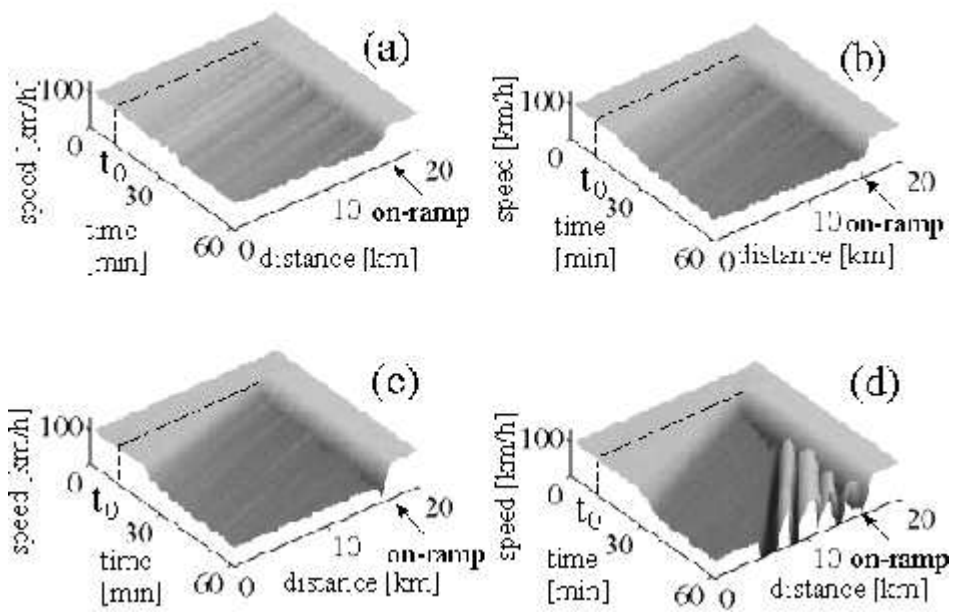}
\end{center}
\caption[]{Evolution of the vehilce speed in space and in time
at the given $q_{\rm on}= 200 \ vehicles/h$ for different $q_{\rm in}$
related to Line 3 in Fig.~\ref{KKW_Diagram} (b):
(a-c) - WSP, (d) -
GP.
The flow rate $q_{\rm in}$ is:  (a)  (1660), (b) (1800),
(c) (1960), (d)  (2200) $vehicles/h$.
 \label{Line3} }
\end{figure}

The transformation of congested patterns along Line 3 in Fig.~\ref{KKW_Diagram} (b)
is related to 
a given  flow rate to the on-ramp $q_{\rm on}$ 
 when 
the flow rate in free flow upstream of the on-ramp
$q_{\rm in}$
is increased beginning from a  relatively low initial value
(Fig.~\ref{Line3}). 
Because at this flow rate the boundary $F^{\rm (B)}_{\rm S}$
is intersected above the boundary $W$ (which separates the WSP and the LSP in the diagram
of congested patterns in Fig.~\ref{KKW_Diagram} (b)),
 a WSP occurs (Fig.~\ref{Line3} (a)). The WSP remains in a wide 
range of the flow rate $q_{\rm in}$ when this flow rate increases (Fig.~\ref{Line3} (b, c)). However,
the average speed in the WSP decreases whereas the flow rate $q_{\rm in}$  increases.
Finally, if the flow rate $q_{\rm in}$  increases
and the boundary $S^{\rm (B)}_{\rm J}$ is intersected, a GP occurs  
(Fig.~\ref{Line3} (d)). This GP transforms in a GP shown in Fig.~\ref{Line1} (e)
when the flow rate  $q_{\rm in}$ is further increased.

\subsubsection{Time-Evolution of Discharge Flow Rate}

\begin{figure}
\begin{center}
\includegraphics*[width=.8\textwidth]{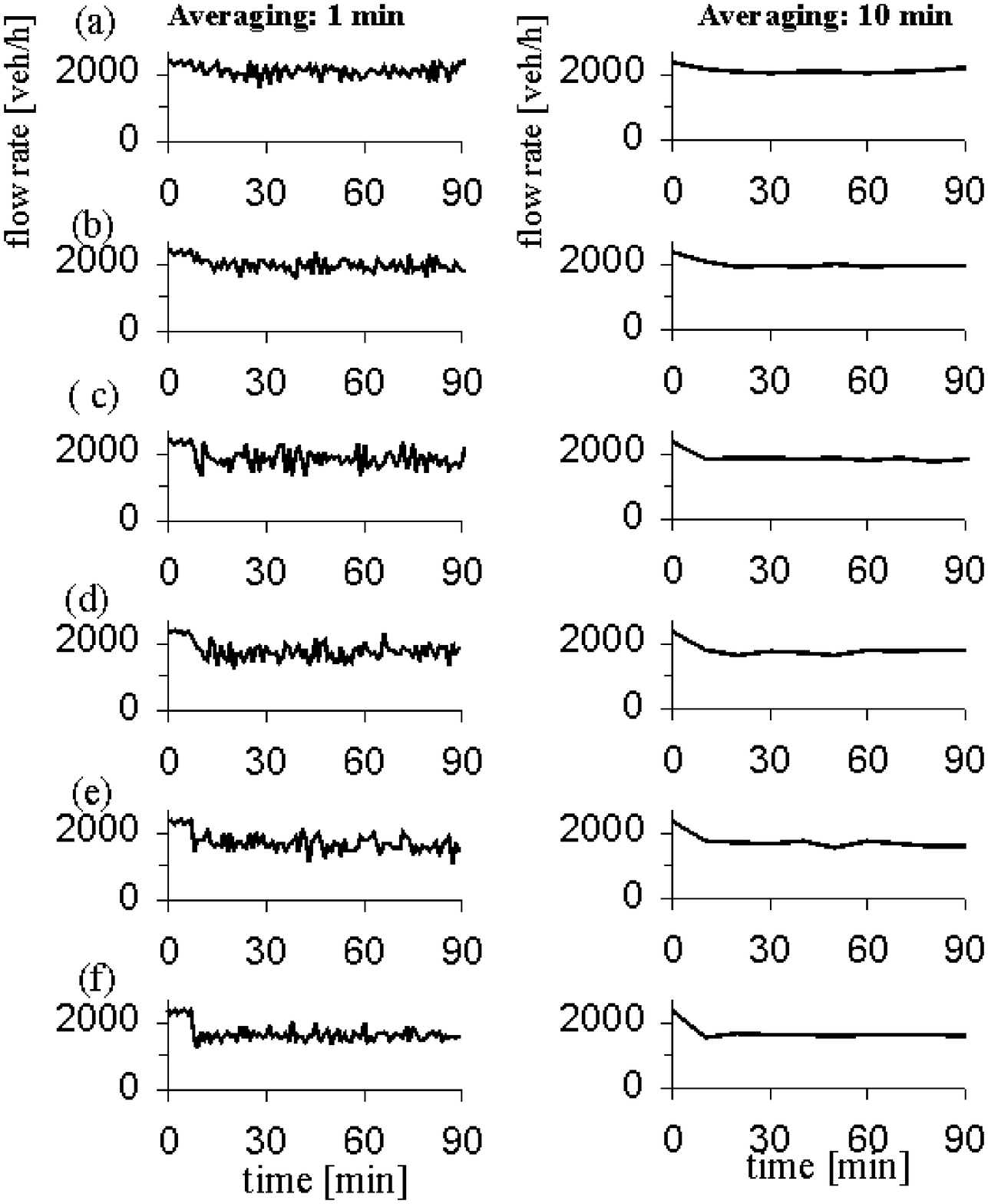}
\end{center}
\caption[]{Time-evolution of the  flow rate downstream of the on-ramp $q_{\rm down}$
during the pattern formation at the on-ramp for different congested patterns
related to Line 1 in Figs.~\ref{KKW_Diagram} (b) and~\ref{Line1}: 
The  flow rate $q_{\rm down}$
averaged during 1 min (left)
and the  flow rate $q_{\rm down}$
averaged 10 min (right). Figures (a)-(f) are related to
the congested patterns with the same letters (a)-(f) in Fig.~\ref{Line1}.
The  data from a virtual detector located at $x= 17 \ km$ in free flow 
downstream of the on-ramp (16 km). 
 \label{Line1_E} }
\end{figure}

\begin{figure}
\begin{center}
\includegraphics*[width=.9\textwidth]{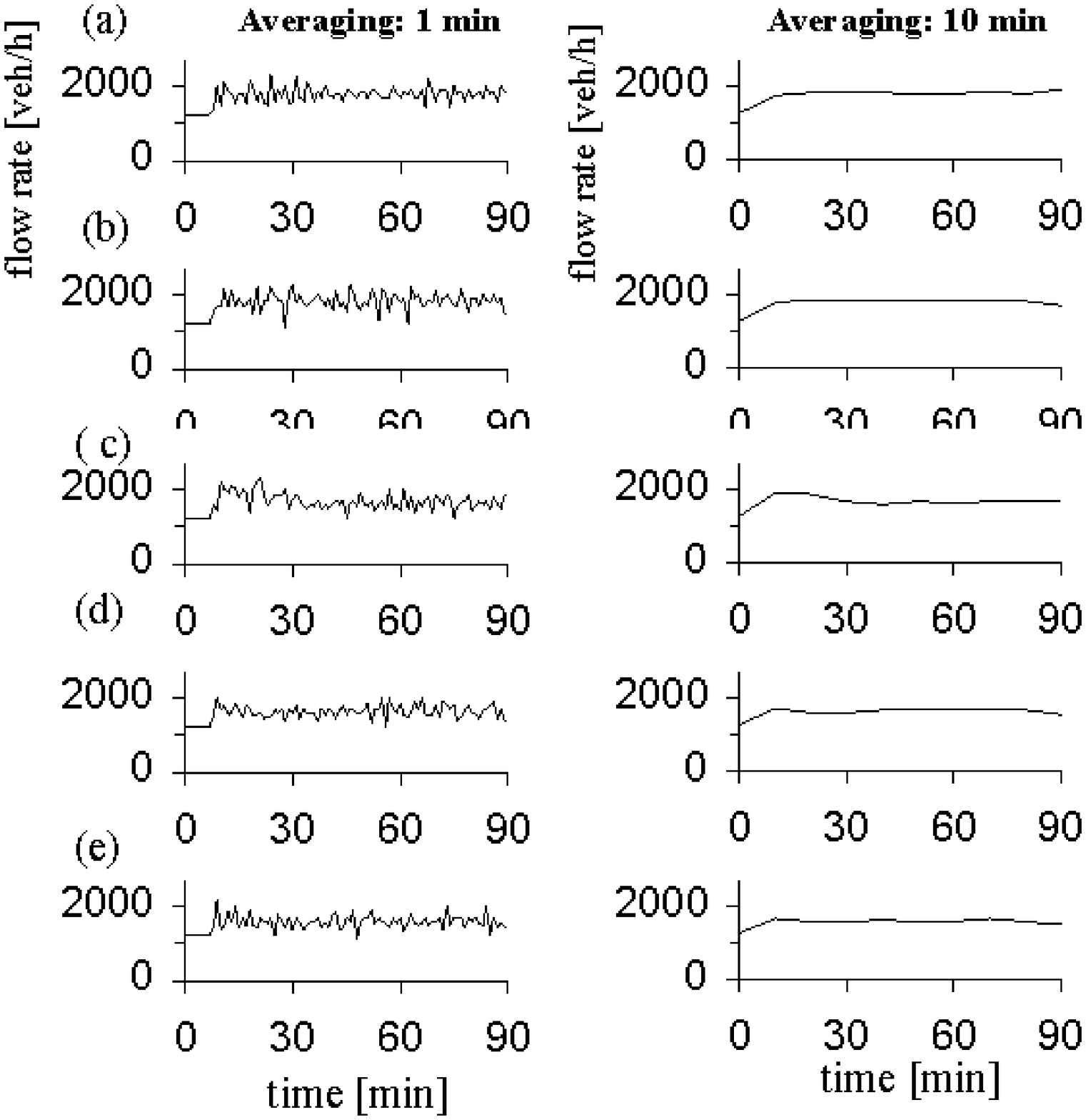}
\end{center}
\caption[]{Time-evolution of the  flow rate downstream of the on-ramp $q_{\rm down}$
during the pattern formation at the on-ramp for different congested patterns
related to Line 2 in Figs.~\ref{KKW_Diagram} (b) and~\ref{Line2}: 
The  flow rate $q_{\rm down}$
averaged during 1 min (left)
and the  flow rate $q_{\rm down}$
averaged 10 min (right). Figures (a)-(e) are related to
the congested patterns with the same letters (a)-(e) in Fig.~\ref{Line2}.
The  data from a virtual detector located at $x= 17 \ km$ in free flow 
downstream of the on-ramp (16 km). 
 \label{Line2_E} }
\end{figure}

\begin{figure}
\begin{center}
\includegraphics*[width=.9\textwidth]{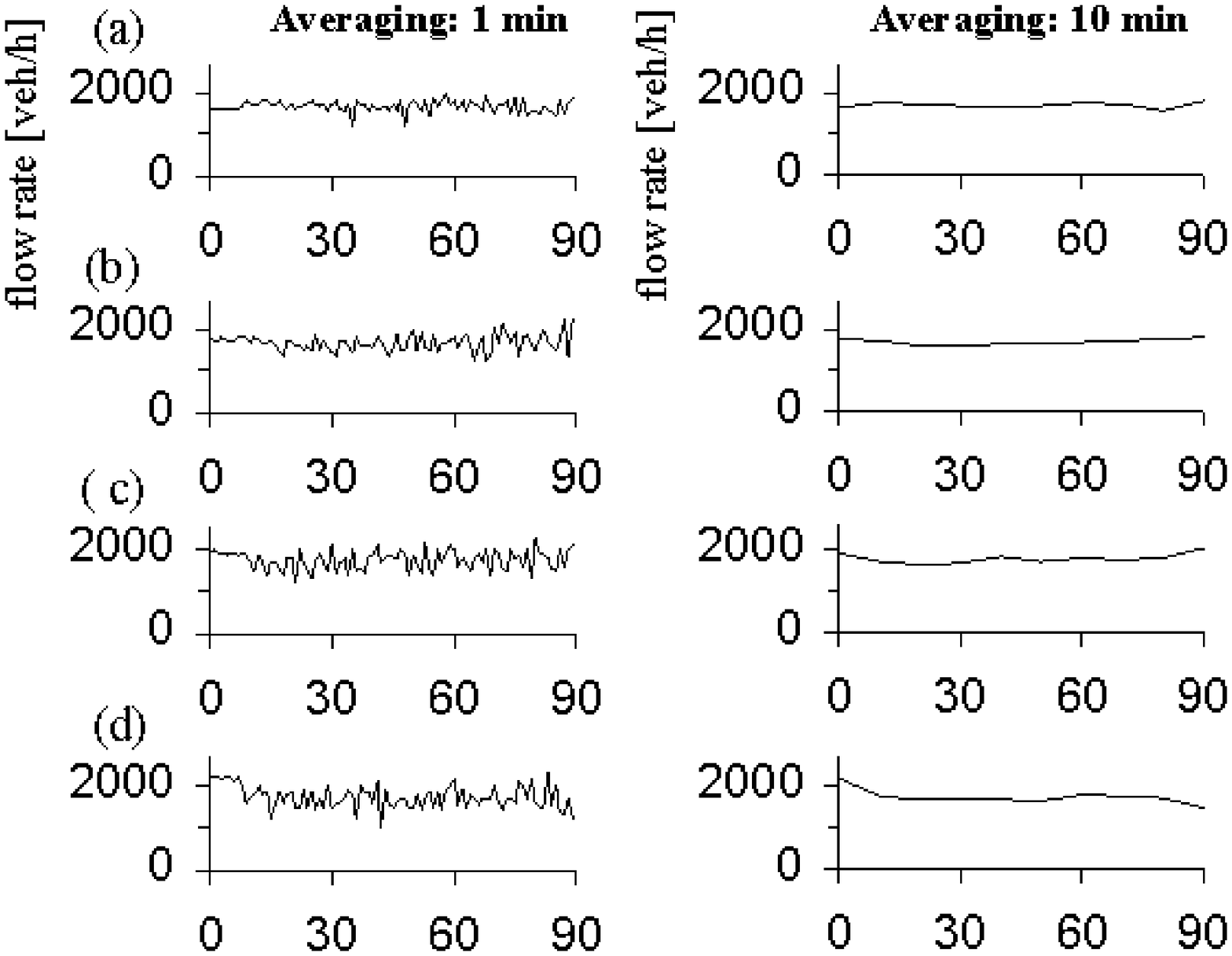}
\end{center}
\caption[]{Time-evolution of the  flow rate downstream of the on-ramp $q_{\rm down}$
during the pattern formation at the on-ramp for different congested patterns
related to the  
Line 3 
in Figs.~\ref{KKW_Diagram} (b) and~\ref{Line3}: 
The  flow rate $q_{\rm down}$
averaged during 1 min (left)
and the  flow rate $q_{\rm down}$
averaged 10 min (right). Figures (a)-(d) are related to
the congested patterns with the same letters (a)-(d) in Fig.~\ref{Line3}.
The  data from a virtual detector located at $x= 17 \ km$ in free flow 
downstream of the on-ramp (16 km). 
 \label{Line3_E} }
\end{figure}

By the simulation of  congested patterns considered above the  flow rate to the on ramp
$q_{\rm on}$ has been switched on only after the time $t=t_{\rm 0}$. 
Let us designate the flow rate far enough downstream of the on-ramp where free flow occurs as
$q_{\rm down}$. During the time $0\leq t< t_{\rm 0}$
free flow occurs at the on-ramp and $q_{\rm down}=q_{\rm sum}=
q_{\rm in}$. At $t> t_{\rm 0}$, i.e., after the on-ramp has been switched on,
the flow rate $q_{\rm down}$ should  increase because $q_{\rm sum}=q_{\rm in}+q_{\rm on}$ at $t\geq t_{\rm 0}$.
 However, at $t\geq t_{\rm 0}$
a
congested pattern has
begun to form upstream of the on-ramp. 
Thus, at $t> t_{\rm 0}$ the flow rate $q_{\rm down}$ is determined by the discharge flow rate 
$q^{\rm (bottle)}_{\rm out}$: $q_{\rm down}=q^{\rm (bottle)}_{\rm out}$. The numerical simulation
allows us to study the time-evolution  of the  flow rate $q_{\rm down}$ by the formation of each of
the congested patterns upstream of the on-ramp
 (Fig.~\ref{Line1_E}-Fig.~\ref{Line3_E}).

This evolution for congested patterns which appear along Line 1
is shown in Fig.~\ref{Line1_E}.
It can be seen that although at $t=t_{\rm 0}$ the inflow from the on-ramp is switched on, i.e., as additional vehicles
enter the main road from the on-ramp,
the discharge flow rate is lower than the flow rate $q_{\rm sum}=q_{\rm in}$ at $0\leq t< t_{\rm 0}$.
This is the result of the congested pattern formation: Each of the congested patterns
along Line 1 leads to a decrease in the flow rate on the main road just upstream of the on-ramp.
This decrease is higher than the increase of the flow rate due to the vehicles
squeezing to the main road from the on-ramp.

A different situation is realized  for the congested patterns which appear along Line 2
(Fig.~\ref{Line2_E}). In this case,
the discharge flow rate is higher than the flow rate $q_{\rm sum}=q_{\rm in}$ at $0\leq t< t_{\rm 0}$:
Each of the congested patterns
along Line 2 leads also to a decrease in the flow rate on the main road just upstream of the on-ramp.
However this decrease
is lower than the increase of the flow rate due to the vehicles
squeezing onto the main road from the on-ramp.

An intermediate case has been found 
for the congested patterns which appear along Line 3
(Fig.~\ref{Line3_E}), i.e., when the initial flow rate upstream of the on-ramp $q_{\rm in}$
is changing.
First, there is a slight increase in the flow rate $q_{\rm down}=q^{\rm (bottle)}_{\rm out}$ due to the congested pattern formation
in comparison with the initial flow rate $q_{\rm down}=q_{\rm in}$ (Fig.~\ref{Line3_E} (a)). 
For a higher flow rate $q_{\rm in}$
there is almost no change in the flow rate $q_{\rm down}$ after congested patterns have been formed
 (Fig.~\ref{Line3_E} (b, c)).
When the flow rate $q_{\rm in}$ is further increased,
the discharge flow rate is lower than the initial flow rate $q_{\rm down}=q_{\rm in}$ at $0\leq t< t_{\rm 0}$ as it was
for the congested patterns along Line 1
 (Fig.~\ref{Line3_E} (d)), i.e., the flow rate $q_{\rm down}$ decreases during the congested pattern formation.

\subsubsection{Congested Pattern Capacity}

The congested pattern capacity $q^{\rm (B)}_{\rm cong}$
(\ref{capacity_condition5_For}) has been calculated through the 60 min averaging of
the discharge flow rate $q^{\rm (bottle)}_{\rm out}$ after the related congested pattern has been formed. 
It has been found  that the congested pattern capacity $q^{\rm (B)}_{\rm cong}$
depends
 on the type of congested patterns and on the pattern parameters noticeably.

\begin{figure}[]
\begin{center}
\includegraphics[width=.9\textwidth]{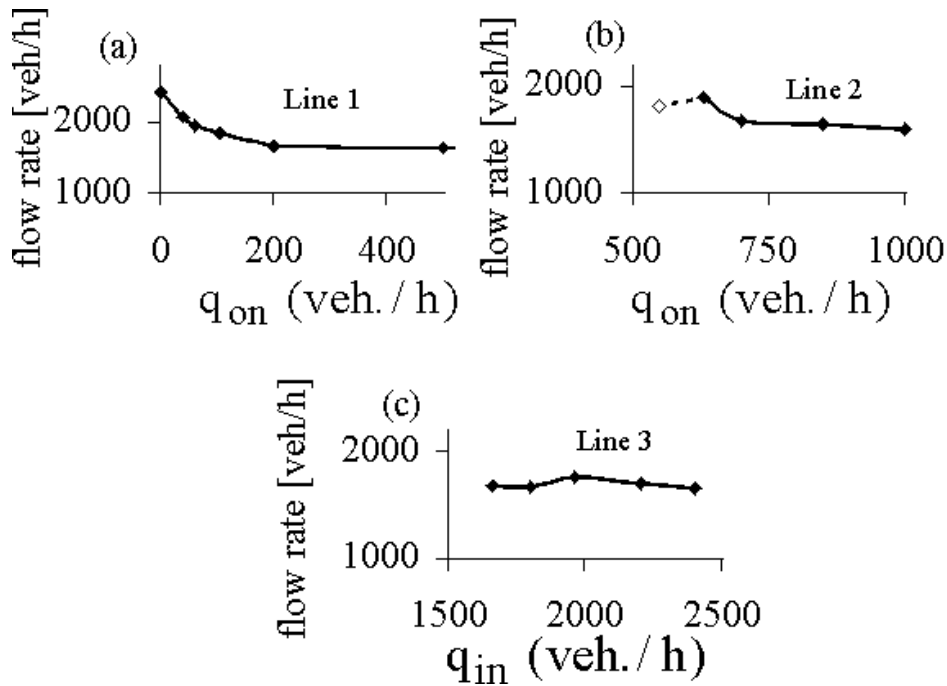}
\end{center}
\caption[]{Dependence of the average congested pattern capacity
on the patern type and pattern parameters: (a, b)
The congested pattern capacity as function on the flow rate
$q_{\rm on}$ related to Line 1(a) and Line 2 (b) in Fig.~\ref{KKW_Diagram} (b).
(c) - The congested pattern capacity as function on the flow rate
$q_{\rm in}$ related to Line 3  in Fig.~\ref{KKW_Diagram} (b).
The black points  correspond to 
the congested patterns in Figs.~\ref{Line1}-~\ref{Line3} for each of Lines 1-3, respectively. 
The averaging  of the
discharge fllow rate during 60 min beginning from the time moment $t=20 \ min$ (after
congested patterns have been formed) is performed.
The  data from a virtual detector located at $x= 17 \ km$ in free flow 
downstream of the on-ramp (16 km). 
 \label{Con_Fig} }
\end{figure}

Indeed, along Line 1 (Fig.~\ref{Con_Fig} (a))
the congested pattern capacity $q^{\rm (B)}_{\rm cong}$ is a
decreasing function of the flow rate to the on-ramp $q_{\rm on}$.
This is linked to the pinch effect in synchronized flow which occurs when $q_{\rm on}$
increases. Due to the pinch effect (a compression of synchronized flow with the narrow moving jam formation),
a relatively high vehicle speed in synchronized flow of the WSP 
 decreases. This leads to a decrease in the flow rate in synchronized flow upstream of the on-ramp.
The decrease in the latter flow rate is noticeable higher than
an increase in the flow rate to
the on-ramp. For this reason, the congested pattern capacity $q^{\rm (B)}_{\rm cong}$  decreases although
$q_{\rm on}$
increases.
The more the decrease in the speed and in the flow rate in synchronized flow upstream of the on-ramp
 is the higher  the 
flow rate $q_{\rm on}$ is.
However, there is a saturation of the decrease in the 
congested pattern capacity $q^{\rm (B)}_{\rm cong}$ when $q_{\rm on}$ is further increased.

By  calculation of the congested pattern capacity the condition (\ref{Cong_capacity_condition_For})
has been fulfilled. The only exception is the calculation of the capacity along
Line 2:
In this case, first a LSP appears (Fig.~\ref{Line2} (a, b)) and therefore the condition (\ref{capacity_condition8})
has been used. 
The congested pattern capacity which is related to this LSP  has been determined by 
the maximum discharge flow rate at which the LSP still exists upstream of the on-ramp.
It has been found  that the latter condition is only fulfilled for the  LSP 
at 
$q_{\rm on} \approx 650 \ vehicles/h$.
This has been taken into account in Fig.~\ref{Con_Fig} (b): The dashed curve is related to the
discharge flow rate from the LSP when the capacity is not reached and the solid curve 
corresponds to the congested pattern capacity.
It can been seen that the congested pattern capacity decreases when the flow rate
$q_{\rm on}$ increases. However, in comparison with Line 1 the decrease in the
capacity along Line 2
is considerably lower at the same increase in the flow rate $q_{\rm on}$. 

Along Line 3  (Fig.~\ref{Con_Fig} (c)) the
congested pattern capacity $q^{\rm (B)}_{\rm cong}$
is a non-monotonous 
 function of the flow rate to the on-ramp $q_{\rm on}$: 
There is 
a
maximum point on this dependence. It occurs that at the same flow rate to the on-ramp $q_{\rm on}$
the flow rate in synchronized flow in the WSP upstream of the on-ramp 
increases
during  the evolution of the WSP (from  the WSP in Fig.~\ref{Line3} (a)
to the WSP in Fig.~\ref{Line3} (c))
with the increase in
the flow rate $q_{\rm in}$. 
The flow rate in synchronized flow of the WSP in the maximum point in Fig.~\ref{Con_Fig} (c)
(the WSP in Fig.~\ref{Line3} (c))
is higher than the flow rate in the wide moving jam outflow $q_{\rm out}$
(see Fig.~\ref{KKW_Diagram} (a)).
When the flow rate $q_{\rm in}$ is further increased and therefore the boundary $S^{\rm (B)}_{\rm J}$ is 
intersected, then wide moving jams begin to form in synchronized flow of the initial WSP:
a GP is forming.
The flow rate in the wide moving jam outflow cannot exceed $q_{\rm out}$.
As a result, the flow rate through synchronized flow of the GP occurs to be noticeably lower than
in the WSP in Fig.~\ref{Line3} (c) which is related to the maximum point of the congested pattern
capacity in Fig.~\ref{Con_Fig} (c). Thus, the congested pattern capacity decreases when the WSP transforms into
the 
GP. This explains the maximum point in the congested pattern capacity as function of the flow rate
$q_{\rm in}$.

\section{Conclusions}

The 
three-phase traffic theory by the author 
describes phase transitions and a diverse variety of spatial-temporal
congested patterns both on homogeneous roads and at highway
bottlenecks which are related to results of empirical observations~\cite{Kerner1998B,Kerner2002B}. 
The features of these phase
transitions and of the spatial-temporal congested patterns at
bottlenecks in the three-phase traffic
theory \cite{Kerner1998B,Kerner2002B,Kerner1999A,Kerner1999B,Kerner1999C,KKl,Kerner2002}
are qualitatively  different in comparison with the related
results which have been derived in the fundamental diagram
approach \cite{Sch,Helbing2001,Nagatani_R,Nagel2003A}.  An exception is only the propagation
of wide moving jams whose characteristic parameters and features
appear to play an important role (in particular, the flow rate in the
outflow from the jam, $q_{\rm out}$) in both the three-phase traffic
theory and  traffic flow theories in the fundamental diagram approach.

Recent empirical results of a study of the congested patterns at on-
and off-ramps and of their evolution when the bottleneck strength is
gradually changing~\cite{Kerner2002B} confirm the discussed results
and conclusions of the three-phase-traffic-theory
\cite{Kerner1998B,Kerner2002B,Kerner1999A,Kerner1999B,Kerner1999C,KKl,Kerner2002}
rather than the related results and conclusions of traffic theories in
the fundamental diagram
approach~\cite{Sch,Helbing2001,Nagatani_R,Nagel2003A,Helbing1999,Lee1999,Lee2000A,Lee2000B,Helbing2000}.

The three-phase traffic
theory~\cite{Kerner1998C,Kerner1998B,Kerner1999B,Kerner1999C,KKl,Kerner2002}
has also been confirmed by the on-line application in the traffic
center of the State Hessen of some recent models "ASDA" (Automatische
Staudynamikanalyse: Automatic Tracing of Moving Traffic
Jams)~\cite{ASDA,ASDA2} and "FOTO" (Forecasting of Traffic
Objects)~\cite{FOTO,FOTO1} which are based on this traffic flow
theory. These models allow reconstruction, tracing and prediction of
spatial-temporal traffic dynamics based on local measurements of
traffic. The models ASDA and FOTO perform without validation of model
parameters at different traffic conditions (see a recent review about
the models ASDA and FOTO in~\cite{Kerner2002E}).

A mathematical microscopic traffic flow model in the frame of the
three-phase traffic theory which has recently been proposed by Kerner
and Klenov~\cite{KKl} shows a considerable potential both for the
development of qualitatively new mathematical traffic models and for
the traffic flow theory development on the basis of the discussed
hypotheses of the three-phase traffic theory. This is also confirmed
by results of a numerical study of several new cellular automata
traffic flow models which have recently been developed by Kerner,
Klenov and Wolf~\cite{KKW}.

Based on the three-phase traffic theory, a general 
probabilistic theory of highway capacity has been developed and presented in this paper.
It is shown that already in free traffic at the on-ramp there may be  infinite multitudes of
highway capacities which depend on the flow rate to the on-ramp.
When a bottleneck is congested then a much more complicated
picture
of congested pattern highway capacity can be realized: Congested pattern highway capacity
strongly depends on the type of the congested pattern and the pattern parameters.
The theory of
highway capacity and of the capacity drop
presented in the paper is confirmed by both results of 
empirical observations~\cite{KR1997,Persaud,Kerner2000A,Kerner2002B}
and by the presented numerical results of the simulations of the KKW CA-model
presented in the article and in~\cite{KKW}.

{\bf Acknowledgements}
 I would like to thank Hubert Rehborn and Sergey Klenov for their
help and acknowledges
funding by BMBF within project DAISY.

\clearpage


\clearpage


\clearpage

\clearpage

\end{document}